\documentclass[journal]{IEEEtran}
\usepackage{amsmath,amsfonts,amssymb}
\usepackage{algorithmic}
\usepackage{algorithm}
\usepackage{array}
\usepackage[caption=false,font=normalsize,labelfont=sf,textfont=sf]{subfig}
\usepackage{textcomp}
\usepackage{stfloats}
\usepackage{url}
\usepackage{verbatim}
\usepackage{graphicx}
\usepackage{cite}
\usepackage[capitalize]{cleveref}
\usepackage{bm,bbm}
\newtheorem{theorem}{Theorem}
\newtheorem{corollary}[theorem]{Corollary}
\newtheorem{lemma}[theorem]{Lemma}
\newtheorem{proposition}[theorem]{Proposition}
\newtheorem{definition}[theorem]{Definition}
\crefname{theorem}{Theorem}{Theorems}
\Crefname{theorem}{Theorem}{Theorems}
\crefname{corollary}{Corollary}{Corollaries}
\Crefname{corollary}{Corollary}{Corollaries}
\crefname{lemma}{Lemma}{Lemmas}
\Crefname{lemma}{Lemma}{Lemmas}
\crefname{proposition}{Proposition}{Propositions}
\Crefname{proposition}{Proposition}{Propositions}
\crefname{appendix}{Appendix}{Appendices}
\Crefname{appendix}{Appendix}{Appendices}

\newenvironment{remark}
  {\par\smallskip\noindent
   \hspace*{\parindent}\textit{Remark.}\ \ignorespaces}
  {\par\smallskip\ignorespacesafterend}

\newenvironment{regime}
  {\par\smallskip\noindent
   \hspace*{\parindent}\textit{Regime.}\ \ignorespaces}
  {\par\smallskip\ignorespacesafterend}

\DeclareMathOperator*{\E}{\mathbb{E}}
\newcommand{\dir}[1]{\mathsf{Dir}(#1)}

\begin{document}

\title{Batch Universal Prediction\\ via Conditional Additive Smoothing}

\author{Marco Bondaschi,~\IEEEmembership{Graduate Student Member,~IEEE,}
	and~Michael~Gastpar,~\IEEEmembership{Fellow,~IEEE}%
\thanks{The authors are with the School of Computer and Communication Sciences, \'{E}cole Polytechnique F\'{e}d\'{e}rale de Lausanne, CH-1015 Lausanne, Switzerland (e-mail: \{marco.bondaschi, michael.gastpar\}@epfl.ch).}%
\thanks{This work was presented in part at the 2024 IEEE International Symposium on Information Theory, Athens, Greece, Jul. 2024, and at the 2025 IEEE Information Theory Workshop, Sydney, Australia, Sep.-Oct. 2025.}}

\maketitle

\begin{abstract}
Modern language models motivate renewed interest in universal prediction problems in which a predictor can exploit previously observed data before predicting a fresh sequence. We study an idealized version of this setting by introducing \emph{batch regret}: a conditional analogue of classical average regret in which a predictor observes a batch of $n$ training sequences and is then evaluated on an independent test sequence generated by the same unknown source. Additive smoothing estimators, which add a bias to the observed token frequencies before predicting the next token, played a fundamental role as a simple prediction strategy with almost-optimal average regret for memoryless and Markov sources. Their equivalent under batch regret are the conditional additive smoothing estimators, which estimate the conditional distribution of the test sequence given the training batch. In this work, we study their asymptotical batch regret for binary memoryless and first-order Markov sources. For binary memoryless sources, we show how the optimal bias of the estimator, and consequently the corresponding batch regret, change as a function of the training batch size $n$, by extending the Poisson-based analysis of Krichevsky. For binary Markov sources with positive transitions, we characterize the leading coefficient of batch regret and its dependence on whether the initial law is known, free, or stationary. Finally, we introduce a R\'{e}nyi analogue of batch regret and identify the corresponding optimal predictors as conditional $\alpha$-NML estimators, thereby giving an operational interpretation to a conditional version of Sibson's $\alpha$-mutual information.
\end{abstract}


\section{Introduction}
\IEEEPARstart{I}{n} information theory, \emph{prediction} concerns the problem of estimating the continuation of a sequence of symbols, or \emph{tokens}, based on the sequence's past and on any additional information available about the data distribution. The estimate is produced by a \emph{predictor}, namely a probability distribution on the possible continuations. Under logarithmic loss, the best predictor for a known source is the source distribution itself, and the excess loss incurred by a universal predictor is called \emph{regret}. Universal prediction studies predictors whose regret is small uniformly over a prescribed class of sources.

Universal prediction is a classical topic, with roots that can be traced at least to Laplace's rule of succession \cite{laplace1814essaiprob}. Its connection to modern machine learning is nevertheless direct: language models are trained as next-token predictors, and their performance is usually evaluated through variants of logarithmic loss. However, in the classical formulation of universal prediction, one usually predicts a single long sequence generated by one unknown source. Instead, in many data-rich prediction problems such as large language model training, information about the source is available through a collection of independent training sequences, and performance is then measured on an independent test sequence. This connection motivates revisiting universal prediction in settings where a predictor has access to many previously observed examples before being evaluated on a fresh test sequence. In this work, we therefore introduce \emph{batch regret}, a conditional regret measure in which the predictor is evaluated on a test sequence of $\ell$ tokens, \emph{conditioned} on the observation of a batch of $n$ training sequences.

The main focus of this paper is the single-source batch setting, where all training sequences and the test sequence are drawn independently from the same unknown distribution $p_{\theta}$. This setting captures an idealized form of in-distribution prediction: the training batch provides information about the unknown source, and the predictor uses this information when assigning probabilities to the test sequence. More specifically, we will focus on additive smoothing estimators, or add-$\beta$ estimators, in which the predictive probability at any given moment is obtained by adding a constant bias $\beta > 0$ to the empirical frequencies of the tokens. This class includes the Laplacian \cite{laplace1814essaiprob} and the Krichevsky-Trofimov \cite{krichevsky1} estimators and has been shown to achieve near-optimal asymptotic average regret on a single sequence. Here we will study their conditional extension, in which the training batch is used to improve the estimates. In particular, we will analyze their batch regret when the target classes are binary memoryless and first-order Markov sources. Under this setting, the theoretical limits derived in this work complement a more practical line of research that studies large language models' capability to optimally predict tokens generated from controlled data distributions such as Markov sources \cite{bietti2023birth,edelman2024evolution,makkuva2024attention,makkuva2024local,rajaraman2024transformersmarkovdataconstant,bondaschi2025markov}.

\subsection{Main contributions}
Our contributions are organized as follows. In \cref{sec:ch3_definition} we formulate single-source batch regret and relate it to the Kullback-Leibler divergence between the test sequence's parametric distribution and a predictor, conditioned on the training batch. In \cref{sec:rc-theorem} we prove a conditional regret-capacity theorem, which equates minimax batch regret to a maximal conditional mutual information between the distribution parameter and the test sequence given the training batch. As the maximal mutual information can be lower-bounded by choosing any prior distribution on the parameter space, this result can be used to lower bounds on minimax batch regret, which we do in the subsequent sections. In \cref{sec:ch3_memoryless}, we introduce conditional additive smoothing estimators, or add-$\beta$ estimators, for binary memoryless sources and we prove general estimates of their batch regret in the interior and at the border of the simplex, valid whenever $n\ell\to\infty$. The overall behavior over the entire simplex depends on the regime in which the batch size $n$ and the sequence length $\ell$ interact, and is discussed in the following two sections: in \cref{sec:large-n-regime} we consider the large-$n$ regime, while in \cref{sec:fixed-n-regime} we study the complementary fixed-$n$, large-$\ell$ regime. In \cref{sec:optimal-batch-regret} we discuss the optimal batch regret over the class of all predictors instead. In \cref{sec:markov}, we extend our analysis to first-order binary Markov sources with uniformly positive transitions. Under this setting, assumptions on the initial distribution of the Markov source are fundamental: in fact, we show how the leading coefficient of batch regret changes whenever the initial law is known, free, or stationary. Finally, in \cref{sec:alpha-regret} we consider a generalized batch regret measure based on conditional R\'{e}nyi divergence, which we call batch $\alpha$-regret, and we prove its relation with a conditional extension of Sibson's $\alpha$-mutual information, thus providing an operational interpretation to it. A preliminary version of some of the results of this paper was first presented in \cite{bondaschi2024batch,bondaschi2025conditional}.

\subsection{Related work}
\label{subsec:related}
Universal prediction under cross-entropy loss and universal lossless compression are essentially mathematically equivalent \cite{coveruniversal,merhav1}, so that literature on either topic is relevant to the other. Universal compression-prediction for memoryless sources was studied by Xie and Barron under both average \cite{xie2} and worst-case regrets \cite{xie1}. Their asymptotic techniques are closely related to the compact-interior expansion in \cref{thm:inner}. The worst-case regret for Markov sources was studied in \cite{xie1,ryabko1,ryabko2}. The average case was more recently characterized in \cite{falahatgar1,hao1}. The techniques used in \cite{hao1}, in particular Paulin's concentration inequalities for Markov chains \cite{paulin2015}, are instrumental to proving our \cref{thm:markov-regret}.

A batch regret with one test sample is the average next-token regret defined in \cref{eq:NTRegret} and first studied by Krichevsky \cite{krichevsky2}. This regret measure has since been studied in several works \cite{braess2004187,komaki20113705,komaki1,fogel2}; in particular, \cite{fogel2} predicts one test symbol after a training sample and therefore belongs to this next-token setting. A form of conditional average regret for more than one token was introduced in \cite{liang1}, where it is studied for location families. The same metric is also studied by \cite{komaki20113705} and \cite{fogel3}. In particular, \cite{fogel3} focuses on memoryless sources, under which their metric is equivalent to our definition of batch regret. However, in this work we study a different regime of $n$ and $\ell$ and a different class of predictors. The equivalence between the two definitions of conditional regret does not extend to other classes of distributions, e.g., the Markov class studied in \cref{sec:markov}: according to our definition, the training batch is composed of $n$ independent Markov sequences of length $\ell$, instead of one sequence of $n\ell$ i.i.d. samples. A closer setting is the one studied in \cite{komaki20113705}, whose general formulation includes independent sequences; however, it does not study the regret as a function of the batch size $n$ and the sequence length $\ell$, which is the focus of this work. Nonetheless, both works include a conditional version of the regret-capacity theorem, similar to our \cref{thm:cond-rc}.

Sibson's $\alpha$-mutual information was introduced by Sibson in \cite{Sibson:69} as an information radius. A revisitation of it as a generalization of Shannon's mutual information was proposed in \cite{verdu1}, which also reviews many of its properties. A more recent discussion can be found in \cite{esposito2024sibsonsalphamutualinformationvariational}. Proposals for conditional versions of Sibson's mutual information, including the one used in \cref{eq:cond-sibson}, appear in \cite{tomamichel1,esposito1}. The maximization of Sibson's mutual information and related measures is studied in \cite{cai1}, including a proof of a regret-capacity theorem for $\alpha$-mutual information. The work \cite{yagli1} discusses Sibson's mutual information in a universal-compression context while studying a closely related redundancy measure called Sundaresan divergence. The predictors introduced in \cite{bondaschi1} are based on Sibson's mutual information; in \cref{thm:alpha-cond}, we obtain conditional analogues of these predictors as optimal predictors under $\alpha$-batch regret.

\section{Background}
\label{sec:background}

In this section we recall basic information-theoretic definitions and results that will be useful in the sequel. The alphabets are finite and all logarithms are in base $e$, unless stated otherwise.

\subsection{Conditional and R\'{e}nyi information measures}
Given two conditional distributions $p_{U|Z}$ and $q_{U|Z}$ and a distribution $p_Z$, we use the conditional Kullback-Leibler (KL) divergence
\begin{align}
\label{eq:cond-kl}
D(p_{U|Z}\|q_{U|Z}\mid p_Z)
&\triangleq \sum_{z\in\mathcal Z}p_Z(z)
\sum_{u\in\mathcal U}p_{U|Z}(u\mid z) \notag\\
&\qquad\cdot \log\frac{p_{U|Z}(u\mid z)}{q_{U|Z}(u\mid z)}.
\end{align}
Equivalently, when $p_{UZ}$ and $q_{UZ}$ have the same $Z$-marginal $p_Z$, this quantity is equal to the classical KL divergence $D(p_{UZ}\|q_{UZ})$. We also write it as $D(U\,\|\,\hat U\mid Z)$ when the underlying conditional laws are clear. Conditional mutual information is the corresponding information radius,
\begin{equation}
I(X;Y\mid Z)=D(p_{XY|Z}\,\|\,p_{X|Z}\, p_{Y|Z}\mid p_Z).
\end{equation}
In cases where the distribution of $X$ varies, we may also write $I_P(X;Y\mid Z)$ to highlight the fact that $X$ is distributed according to $P$.

For $\alpha>0$, $\alpha\neq1$, the R\'{e}nyi divergence \cite{renyi1} is
\begin{align}
D_{\alpha}(p_X\|q_X)
&\triangleq \frac{1}{\alpha-1}\log\sum_{x\in\mathcal X}p_X(x)^\alpha q_X(x)^{1-\alpha}.
\end{align}
Its limiting cases include $D_1(p_X\|q_X)=D(p_X\|q_X)$ and
\begin{equation}
D_{\infty}(p_X\|q_X)=\sup_{x\in\mathcal X}\log\frac{p_X(x)}{q_X(x)}.
\end{equation}
We will also use the conditional version
\begin{align}
D_{\alpha}(p_{U|Z}\|q_{U|Z}\mid p_Z)
&\triangleq \frac{1}{\alpha-1}\log \sum_{z\in\mathcal Z}p_Z(z) \notag\\
&\quad\cdot \sum_{u\in\mathcal U}p_{U|Z}(u\mid z)^\alpha
q_{U|Z}(u\mid z)^{1-\alpha}.
\end{align}

Finally, Sibson's $\alpha$-mutual information \cite{Sibson:69} is
\begin{equation}
I_{\alpha}(X;Y) \triangleq \frac{\alpha}{\alpha-1}\log \sum_{y\in\mathcal Y}\left(\sum_{x\in\mathcal X}p_X(x) p_{Y|X}(y\mid x)^\alpha\right)^{1/\alpha}.
\end{equation}
It admits the variational representation
\begin{equation}
I_{\alpha}(X;Y)=\min_{\hat Y: X\perp \hat Y}D_{\alpha}(Y\|\hat Y\mid X),
\end{equation}
for the $\alpha$-orders considered in this paper. Conditional analogues are introduced in \cref{sec:alpha-regret}.

\subsection{Classical universal prediction}\label{sec:universal_prediction}
Fix a sequence length $\ell$. A predictor assigns probabilities $\hat p(x^\ell)$ to sequences $x^\ell=(x_1,\ldots,x_\ell)\in\mathcal X^\ell$. When the predictor is used sequentially, it is specified by one-step conditionals and induces the joint distribution
\begin{equation}
\label{eq:online}
\hat{p}(x^\ell) = \prod_{i=0}^{\ell-1} \hat{p}(x_{i+1}\mid x^i).
\end{equation}
Universal prediction benchmarks $\hat p$ against a class $\mathcal P=\{p_\theta:\theta\in\Theta\}$. Under logarithmic loss,
\begin{equation}
L(\hat p,x^\ell)\triangleq -\log \hat p(x^\ell),
\end{equation}
the pointwise regret relative to $p_\theta$ is
\begin{equation}
\label{eq:regret}
R_{n,\ell}(\hat p,\theta,x^\ell)\triangleq L(\hat p,x^\ell)-L(p_\theta,x^\ell)
=\log\frac{p_\theta(x^\ell)}{\hat p(x^\ell)}.
\end{equation}
The two classical aggregate criteria are the average regret
\begin{align}
R_{\mathrm{av},\ell}(\hat p,\theta)
&\triangleq \E_{X^\ell\sim p_\theta} R_{n,\ell}(\hat p,\theta,X^\ell) \label{eq:AvR}\\
&= \sum_{x^\ell \in \mathcal X^\ell}p_\theta(x^\ell)\log\frac{p_\theta(x^\ell)}{\hat p(x^\ell)} \\
&=D(p_\theta\|\hat p), \label{RavDiv}
\end{align}
and the worst-case regret
\begin{align}
R_{\max,\ell}(\hat p,\theta)
&\triangleq \sup_{x^\ell\in\mathcal X^\ell}R_{n,\ell}(\hat p,\theta,x^\ell) \label{eq:WCR}\\
&=\sup_{x^\ell\in\mathcal X^\ell}\log\frac{p_\theta(x^\ell)}{\hat p(x^\ell)}\\
&=D_\infty(p_\theta\|\hat p). \label{RmaxDiv}
\end{align}
Equivalently,
\begin{equation}
\label{eq:RavDiff}
R_{\mathrm{av},\ell}(\hat p,\theta)=\mathbb{E}_{\theta}\,L(\hat p,X^\ell)-\mathbb{E}_{\theta}\,L(p_\theta,X^\ell),
\end{equation}
so average regret is the excess expected logarithmic loss over the source-matched predictor.

Since the parameter is unknown, the usual minimax criteria maximize over the class:
\begin{gather}
R_{\mathrm{av},\ell}(\hat p)\triangleq \sup_{\theta\in\Theta}R_{\mathrm{av},\ell}(\hat p,\theta),\label{eq:maximal-average}\\
R_{\max,\ell}(\hat p)\triangleq \sup_{\theta\in\Theta}R_{\max,\ell}(\hat p,\theta).\label{eq:maximal-wc}
\end{gather}
Whenever it exists, the normalized maximum-likelihood (NML) predictor minimizes $R_{\max}$ \cite{shtarkov1}:
\begin{equation}
\label{eq:NML}
\hat p_{\rm NML}(x^\ell)=\frac{\sup_{\theta\in\Theta}p_\theta(x^\ell)}{\sum_{x^\ell}\sup_{\theta\in\Theta}p_\theta(x^\ell)},
\end{equation}
with minimax value
\begin{equation}
\label{eq:NMLWCRegret}
\inf_{\hat p}R_{\max,\ell}(\hat p)=\log\sum_{x^\ell}\sup_{\theta\in\Theta}p_\theta(x^\ell).
\end{equation}
For average regret, the corresponding result is the redundancy- or regret-capacity theorem. Let $\mathcal W(\Theta)$ be the set of all prior distributions on $\Theta$. Then,
\begin{equation}
\label{eq:regret-capacity}
\inf_{\hat p}R_{\mathrm{av},\ell}(\hat p)=\sup_{W\in\mathcal W(\Theta)} I_W(\theta;X^\ell),
\end{equation}
where $(\theta,X^\ell)\sim W(\theta)p_\theta(X^\ell)$. For a fixed prior $W$, the Bayes-optimal predictor under average regret is the mixture
\begin{equation}
\label{eq:mixture}
\hat p_W(x^\ell)=\int_\Theta p_\theta(x^\ell)\,dW(\theta).
\end{equation}
When a capacity-achieving prior exists, its mixture is minimax optimal. Such priors are rarely available in closed form, but they can often be approximated numerically by Blahut--Arimoto-type algorithms \cite{arimoto1,blahut1}.

\subsection{Memoryless sources}
For an alphabet $\mathcal X=\{0,1,\ldots,M-1\}$, the memoryless class consists of i.i.d. distributions $p_{\theta}$ with $p_{\theta}(X_i=x)=\theta_x$, where
\begin{equation}
\bm\theta=(\theta_0,\ldots,\theta_{M-1}),\qquad \theta_x\ge0,
\qquad \sum_{x=0}^{M-1}\theta_x=1.
\end{equation}
A central online predictor for this class is the additive smoothing estimator
\begin{equation}
\label{eq:add-constant}
\hat p_\beta(X_{\ell +1}=x\mid x^\ell)=\frac{n_x(x^\ell)+\beta}{\ell+M\beta},\qquad \beta>0,
\end{equation}
where $n_x(x^\ell)$ is the number of occurrences of symbol $x$ in $x^\ell$. The special case $\beta=1/2$ is the Krichevsky--Trofimov (KT) predictor,
\begin{equation}
\label{eq:KTCondEx}
\hat p_{\rm KT}(X_{\ell+1}=x\mid x^\ell)=\frac{n_x(x^\ell)+\frac12}{\ell+\frac M2}.
\end{equation}
The add-$\beta$ predictor is the posterior predictive distribution under the symmetric Dirichlet prior $W_\beta\sim\dir\beta$; equivalently, the joint distribution induced by \cref{eq:online} satisfies
\begin{equation}
\label{eq:mixture-beta}
\hat p_\beta(x^\ell)=\int_\Theta p_\theta(x^\ell)\,dW_{\beta}(\theta).
\end{equation}
The KT predictor is asymptotically minimax for memoryless sources, up to an $O(1)$ term, under both average and worst-case regret \cite{xie1,xie2}.

We will also refer to the average next-token regret studied by Krichevsky \cite{krichevsky2},
\begin{equation}
\label{eq:NTRegret}
R_{\mathrm{nt},\ell}(\hat p)=\sup_{\theta\in\Theta}\,\mathbb{E}_{\theta}\,\log\frac{p_\theta(X_{\ell+1}\mid X^\ell)}{\hat p(X_{\ell+1}\mid X^\ell)}.
\end{equation}
For binary memoryless sources, the best add-$\beta$ predictor for this next-token criterion has $\beta = \beta_0\approx0.50922$, and asymptotic optimality among all next-token predictors is achieved by the rule of \cite[Eq. (5.18)]{braess2004187}.

\section{Batch regret}
\label{sec:ch3_definition}
In the universal prediction literature, it is customary to consider the regret for the prediction of an entire sequence of $n$ symbols, generated by a single source. That is the setting that we described in \cref{sec:universal_prediction}.
However, many prediction settings distinguish between samples available before prediction and samples revealed during prediction. This distinction is especially natural when modeling in-distribution learning: a predictor is given a finite training batch from an unknown source and must then predict a fresh sequence from the same source. We therefore introduce a new form of regret, \emph{batch regret}, which measures the excess logarithmic loss on the test sequence after conditioning on the training batch.

In the sequel, we compactly denote each sequence of the training batch as $\bm{x}_i \in\mathcal{X}^{\ell}$, so that $\bm{x}^n = (\bm{x}_1, \bm{x}_2, \dots, \bm{x}_n)$ is a set of $n$ training sequences, each consisting of $\ell$ samples. Similarly, the test sequence is denoted by $\bm{y}\in\mathcal{X}^{\ell}$. Each sequence is independent of the others, each possibly generated from a different distribution. Each distribution is determined by a parameter belonging to a parameter set $\Theta$, i.e., each distribution belongs to a certain parametric class, which we denote by $\mathcal{P}=\{p_{\theta}: \theta\in\Theta\}$. More specifically, let $\phi_i \in \Theta$ be the parameter of the distribution of the sequence $\bm{x}_i$, so that $\bm{x}_i \sim p_{\phi_i}$. We compactly denote by $\bm{\phi} = (\phi_1, \phi_2, \dots, \phi_n)$ the set of parameters of the training sequences. We also let $\theta\in\Theta$ be the parameter of the test sequence $\bm{y}$, i.e., $\bm{y} \sim p_{\theta}$. We also compactly denote, with some abuse of notation, $p_{\bm{\phi}}(\bm{x}^n) = \prod_{i=1}^n p_{\phi_i}(\bm{x}_i)$. We now define batch regret as follows.

\begin{definition}[Batch regret]
Let $\bm{x}^n = (\bm{x}_1,\bm{x}_2,\dots,\bm{x}_n)$ be a batch of $n$ training sequences, where each $\bm{x}_i \in\mathcal{X}^{\ell}$ is distributed according to $p_{\phi_i}$ independently of the others. Let $\hat{p}(\bm{y} | \bm{x}^n)$ be a predictor that, given $\bm{x}^n$, estimates the probability of a test sequence $\bm{y} \in\mathcal{X}^{\ell}$, generated independently of $\bm{x}^n$ according to $p_{\theta}$. \emph{Batch regret} is then defined as
\begin{equation}
\label{eq:batch-regret-general}
R_{n,\ell}(\hat{p}, \bm{\phi}, \theta) \triangleq \E_{\substack{\bm{X}^n \sim p_{\bm{\phi}}\\ \bm{Y} \sim p_{\theta}}}\log \frac{p_{\theta}(\bm{Y})}{\hat{p}(\bm{Y} | \bm{X}^n)}.
\end{equation}
\end{definition}

In a more general multi-source version, the training and test distributions could be different but related, for instance by belonging to a common subset $\tilde{\Theta}\subset\Theta$. In the present paper we restrict attention to the single-source case, in which all training sequences and the test sequence are drawn independently from the same distribution $p_{\theta}$. We study this setting for binary memoryless sources and first-order binary Markov sources.

When all sequences are drawn independently from the same distribution $p_{\theta}$, the batch regret takes the form
\begin{equation}
\label{eq:batch-regret}
R_{n,\ell}(\hat{p}, \theta) \triangleq \E_{\substack{\bm{X}^n \sim p_{\theta}\\ \bm{Y} \sim p_{\theta}}}\log \frac{p_{\theta}(\bm{Y})}{\hat{p}(\bm{Y} | \bm{X}^n)}.
\end{equation}
Similarly to the classical average regret analyzed in \cref{sec:universal_prediction}, the objective is to find a predictor $\hat{p}$ that is \emph{universal}, i.e., that performs well for all distributions in a given set $\mathcal{P}=\{p_{\theta}: \theta\in\Theta\}$. Hence, we are interested in the maximal batch regret
\begin{equation}
\label{eq:maximal-batch-regret}
R_{n,\ell}(\hat{p}) \triangleq \sup_{\theta\in\Theta} R_{n,\ell}(\hat{p}, \theta).
\end{equation}
The objective of this section is to study theoretical limits of the single-distribution maximal batch regret when $\mathcal{P}$ is either the set of all Bernoulli distributions, or the set of all first-order binary Markov distributions, asymptotically as $\ell$ becomes large.

\begin{remark}
We note here that if $n=0$, then we fall back into the classical full-sequence average regret setting discussed in \cref{sec:universal_prediction}. If instead $\ell=1$, we fall into the next-token average regret of \cref{eq:NTRegret} studied by Krichevsky in \cite{krichevsky2}. Our setting can be thought as unifying and generalizing those two perspectives.
\end{remark}

\begin{regime}
The interplay between $\ell$ and $n$ is fundamental, and different results below use different limits. For the class of memoryless sources, we consider both the regime $n\to\infty$ with arbitrary $\ell$, and the complementary regime where $n$ is fixed and $\ell\to\infty$. For the class of first-order Markov sources, we consider the regime where $n,\ell\to\infty$ simultaneously. Additional assumptions are stated explicitly when relevant.
\end{regime}

\section{A conditional regret-capacity theorem}
\label{sec:rc-theorem}
In \cref{sec:background} we noted that the minimax value of average regret is equal to a maximal mutual information, via the regret-capacity theorem in \cref{eq:regret-capacity}. An equivalent result also holds for batch regret, in the single-distribution case introduced in \cref{eq:batch-regret}. More specifically, batch regret can be interpreted as a conditional generalization of average regret; in this section we prove a conditional regret-capacity theorem, connecting minimax batch regret to a maximal \emph{conditional} mutual information. We first state the result in its general form, and later specialize it to the batch case introduced in \cref{sec:ch3_definition}.

Let $\mathsf X$ and $\mathsf Y$ be two finite alphabets, and let $\mathcal Q(\mathsf Y\mid\mathsf X)$ denote the set of all stochastic kernels from $\mathsf X$ to $\mathsf Y$. For any $Q \in \mathcal Q(\mathsf Y\mid\mathsf X)$, define the conditional Kullback-Leibler divergence
\begin{equation}
\label{eq:general-conditional-regret}
r_\theta(Q)
\triangleq
D\!\left(P_\theta^{Y\mid X}\,\middle\|\,Q\,\middle|\,P_\theta^X\right),
\end{equation}
where $P^{X,Y}_\theta$ is a joint probability distribution on $\mathsf X \times \mathsf Y$ parametrized by $\theta$ belonging to a parameter set $\Theta$. Aslo assume that $\Theta$ is equipped with a $\sigma$-algebra and that $\theta\mapsto P_\theta^{X,Y}(x,y)$ is measurable for every $(x,y)$.
First note the following useful conditional Bayes decomposition for the conditional KL divergence in \cref{eq:general-conditional-regret}. For any measurable prior $W$ for which the mixture is defined, let
\begin{equation}
\label{eq:mixture-w}
P_W(x,y)=\int_{\Theta}P^{X,Y}_\theta(x,y)\,dW(\theta).
\end{equation}
Then, for every stochastic kernel $Q$,
\begin{align}
\int_{\Theta} r_\theta(Q)\,dW(\theta)
&=I_W(\Theta;Y\mid X)\notag\\
&\quad+D\!\left(P_W^{Y\mid X}\middle\|Q\middle|P_W^X\right).
\label{eq:conditional-bayes-decomp}
\end{align}
With the help of this general identity, we can prove the main result of this section.

\begin{theorem}[Conditional regret-capacity theorem]
\label{thm:cond-rc}
Under the assumptions on $\Theta$ and $P_\theta^{X,Y}$ above, we have
\begin{equation}
\label{eq:cond-rc-general-2}
\min_{Q\in\mathcal Q(\mathcal Y\mid\mathcal X)}
\sup_{\theta\in\Theta} r_\theta(Q)
=
\sup_{W\in\mathcal W(\Theta)}
I_W(\Theta;Y\mid X).
\end{equation}
Furthermore, if there exists a prior $W^\star$ attaining the right-hand side supremum and a minimax $Q^\star$, then $Q^\star$ is the conditional distribution of $Y$ given $X$ induced by a mixture of the form in \cref{eq:mixture-w} with $W = W^\star$, i.e.,
\begin{equation}
\label{eq:mixture-completion}
Q^\star(\cdot\mid x)=P_{W^\star}^{Y\mid X}(\cdot\mid x),
\end{equation}
and $r_\theta(Q^\star)$ is constant $W^\star$-almost surely. Furthermore, $W^\star$ may be chosen to have finite support satisfying
\begin{equation}
\label{eq:generic-support-bound}
|\operatorname{supp}(W^\star)|
\leq
1+|\mathsf X|\bigl(|\mathsf Y|-1\bigr).
\end{equation}
\end{theorem}

The proof of the regret-capacity equivalence above is essentially a conditional generalization of \cite{haussler1997minimax}. The proof of the existence of a capacity-achieving input distribution with finite support is a conditional generalization of the cardinality bound in \cite{witsenhausen1980convexity}, which is retrieved when $\lvert\mathsf X\rvert = 1$. We defer the details to Appendix~\ref{app:cond-rc-proof}. Other versions of this general result were also proved in \cite{komaki20113705} and \cite{fogel3} under different assumptions.

We now show how this general result applies to batch universal prediction. Let $\mathcal X$ be the common alphabet of the source sequences, let
\begin{equation}
\bm{X}^n=(\bm{X}_1,\ldots,\bm{X}_n)\in(\mathcal X^\ell)^n
\qquad\text{and}\qquad
\bm{Y}\in\mathcal X^\ell
\end{equation}
be the training batch and the test sequence, respectively, and set $X=\bm{X}^n$ in \cref{thm:cond-rc}. In the single-source model, conditionally on $\theta$, the $n+1$ sequences are independent and have common law $p_\theta$ on $\mathcal X^\ell$, so
\begin{equation}
\label{eq:batch-probability-form}
P_\theta^{\bm{X}^n,\bm{Y}}(\bm{x}^n,\bm{y})
=\left(\prod_{i=1}^n p_\theta(\bm{x}_i)\right)p_\theta(\bm{y}).
\end{equation}
A batch predictor $\hat p(\bm{y}\mid \bm{x}^n)$ is precisely a stochastic kernel from $(\mathcal X^\ell)^n$ to $\mathcal X^\ell$, and its conditional information radius is the batch regret:
\begin{align}
r_\theta(\hat p)
&=D\!\left(P_\theta^{\bm{Y}\mid \bm{X}^n}\,\middle\| \ \hat p\ \middle| \ P_\theta^{\bm{X}^n}\right)\notag\\
&=\sum_{\bm{x}^n,\bm{y}}p_\theta(\bm{x}^n)p_\theta(\bm{y})
\log\frac{p_\theta(\bm{y})}{\hat p(\bm{y}\mid \bm{x}^n)}
=R_{n,\ell}(\hat p,\theta).
\label{eq:batch-specialization-cond-radius}
\end{align}
Consequently, \cref{thm:cond-rc} specializes to the following result.
\begin{corollary}[Conditional regret-capacity theorem for batch regret]
Under the same assumptions of \cref{thm:cond-rc}, the minimax value of single-distribution batch regret as defined in \cref{eq:batch-regret} satisfies
\begin{equation}
\label{eq:batch-cond-capacity}
\min_{\hat p}\sup_{\theta\in\Theta}R_{n,\ell}(\hat p,\theta)
=
\sup_{W\in\mathcal W(\Theta)}
I_W(\Theta;\bm{Y}\mid \bm{X}^n),
\end{equation}
where the infimum is taken over all predictors ${\hat{p}\in\mathcal Q(\mathcal X^\ell\mid (\mathcal X^\ell)^n)}$.
\end{corollary}

In particular, every chosen prior $W$ gives the lower bound
\begin{equation}
\label{eq:batch-prior-lower-bound}
\min_{\hat p}\sup_{\theta\in\Theta}R_{n,\ell}(\hat p,\theta)
\geq I_W(\Theta;\bm{Y}\mid \bm{X}^n).
\end{equation}
This is how the theorem will be used later: a convenient prior turns finding a lower bound to the minimax batch regret into the computation or estimation of a conditional mutual information.

\section{Conditional add-$\beta$ estimators for binary memoryless sources} \label{sec:ch3_memoryless}
We now focus on the class of memoryless sources that we introduced in \cref{sec:background}, in particular with binary alphabet $\mathcal X = \{0,1\}$, i.e., the class of Bernoulli distributions. Under any fixed distribution in this class, a batch of $n$ sequences $X^n$, each consisting of $\ell$ samples, is equivalent to one single sequence of length $n\ell$, due to the i.i.d. nature of the data. Hence, under this setting the average batch regret can be written as the sum of $\ell$ one-step conditional regrets. Let the one-step regret of the next bit given $m$ previous samples be
\begin{align}
\rho_m(\hat{p},\theta) &\triangleq \mathbb{E}_{\theta} \log \frac{p_{\theta}(X_{m+1})}{\hat{p}(X_{m+1}\mid X^m)} \\
&= \mathbb{E}_{\theta} \bigg[\theta\log\frac{\theta}{\hat{p}(X_{m+1}=1\mid X^m)} \notag\\
&\hspace{3.8em}+ (1-\theta)\log\frac{1-\theta}{\hat{p}(X_{m+1}=0 \mid X^m)}\bigg]. \label{eq:rho-general}
\end{align}
Then, batch regret is equal to the cumulative regret
\begin{equation}
R_{n,\ell}(\hat{p},\theta) = \sum_{m=t}^{t +\ell-1} \rho_m(\hat{p},\theta), \qquad t=n\ell.
\end{equation}

We now introduce an important predictor, that can be considered a conditional version of the classical additive smoothing estimator for the batch regret case. Crucially, in the batch case the predictor needs to account for both the training sequences and, in an online fashion, the test sequence as well. Formally, the predictor is defined as
\begin{equation}
\label{eq:pred-def}
\hat{p}_{\beta}(y^{\ell} \mid \bm{x}^n) = \prod_{i=0}^{\ell -1} \hat{p}_{\beta}(y_{i+1} \mid \bm{x}^n, y^i),
\end{equation}
where
\begin{equation}
\hat{p}_{\beta}(Y_{i+1}=1 \mid \bm{x}^n, y^i) = \frac{n_1(\bm{x}^n) + n_1(y^i) + \beta}{t + i + 2\beta}
\end{equation}
for a chosen additive smoothing constant $\beta>0$, and $n_1(\bm{z})$ is the number of ones in the sequence $\bm{z}$. We call this predictor the \emph{conditional additive smoothing estimator}, or \emph{conditional add-$\beta$ estimator}. As we shall see, this predictor has the correct leading-order regret on compact subsets of the Bernoulli parameter interval, even though it is not asymptotically optimal over the full interval, just like the next-token additive smoothing estimator studied by Krichevsky \cite{krichevsky2}.

\begin{remark}
It is useful to compare the conditional add-$\beta$ estimator above with possible alternatives. One such alternative is an additive smoothing predictor that estimates the probability of the test sequence $y^{\ell}$ ignoring the past. This predictor is defined, for a chosen parameter $\beta>0$, by
\begin{equation} 
\hat{p}_{\beta}(y^{\ell} \mid \bm{x}^n) = \prod_{i=0}^{\ell -1} \hat{p}_{\beta}(y_{i+1} \mid \bm{x}^n, y^i),
\end{equation}
where
\begin{equation}
\hat{p}_{\beta}(Y_{i+1}=1 \mid \bm{x}^n, y^i) = \frac{n_1(y^i) + \beta}{i + 2\beta}.
\end{equation}
Its corresponding batch regret is just the average regret for the full-sequence of length $\ell$. In this case, $\beta = 1/2$ asymptotically achieves a regret with the optimal leading coefficient, $R_{n,\ell}(\hat{p}_{1/2}) = \frac{1}{2}\log \ell + O(1)$. A second alternative estimates the probability of a sequence only considering the training batch $x^t$ and none of the past symbols of $y^{\ell}$. Such a predictor is
\begin{equation}
\label{eq:naive}
\hat{p}_{\beta}(y^{\ell}\mid\bm{x}^n) = \left(\frac{n_1(x^t) + \beta}{t + 2\beta}\right)^{n_1(y^\ell)} \left(\frac{n_0(x^t) + \beta}{t + 2\beta}\right)^{n_0(y^\ell)},
\end{equation}
where $n_0(\bm{z})$ is the number of zeros in $\bm{z}$.
In this case batch regret is equal to $\ell$ times the Krichevsky next-token regret. The best constant is then $\beta = \beta_0 \approx 0.50922\dots$, and the asymptotic batch regret equals $R_{n,\ell}(\hat{p}_{\beta_0}) = \frac{\beta_0}{n} + o(\frac{1}{n})$. One can think of the conditional add-$\beta$ predictor in \cref{eq:pred-def} as a refined version of the one in \cref{eq:naive}, where the additive smoothing constant is updated every time a new token from $y^{\ell}$ is revealed. The simpler predictor in \cref{eq:naive} will be useful in proving results on the optimal batch regret for Markov distributions in \cref{sec:markov}.
\end{remark}

Importantly, the predictor in \cref{eq:pred-def} is also the posterior predictive distribution associated with a $\mathsf{Beta}(\beta,\beta)$ prior after observing the $t=n\ell$ training symbols. That is, the predictor can be represented via the integral formula
\begin{equation}
\hat{p}_{\beta}(y^{\ell} \mid \bm{x}^n) = \int_0^1 p_{\theta}(y^{\ell})\, dW_{\beta}(\theta\mid \bm{x}^n),
\end{equation}
where 
\begin{equation}
W_{\beta}(\theta \mid \bm{x}^n) \propto \theta^{n_1(\bm{x}^n)+\beta-1} (1-\theta)^{n_0(\bm{x}^n)+\beta-1}
\end{equation}
is a prior (Beta) distribution on the parameter space $\Theta$ that depends on the training sequences $\bm{x}^n$.
This last formulation reveals how the conditional additive smoothing predictor can also be interpreted as a \emph{mixture predictor}, such that the resulting probability of the test sequence $y^{\ell}$ is an average of the probabilities assigned to the sequence by the distributions in the class $\mathcal{P}$. The particular feature of this predictor is that the weighting function $w$ changes depending on the training set, i.e., all the information obtained from the training set is encoded into the weighting function.

We now turn our attention to determining asymptotic properties of batch regret for conditional additive smoothing estimators over the class of binary memoryless sources. The next two theorems give an asymptotic expansion of batch regret under two specific parameter regimes: when the parameter belongs to the interior of the simplex, i.e., $\theta\in[\delta, 1-\delta]$ for some $\delta\in(0,\frac12)$, and when the parameter lies at the endpoints, i.e., $\theta\in\{0,1\}$. Interestingly, both results hold at the same time for two complementary regimes: the large-batch-size regime, when $n\to\infty$ and $\ell$ may or may not go to infinity, and the finite-batch-size regime, when $n$ is a fixed integer and $\ell\to\infty$. Their limitation is that they are not, by themselves, sufficient to determine the asymptotic batch regret over the full interval $[0,1]$. As we will see in the following sections, this requires a more sophisticated Poisson analysis, that depends on whether the batch size $n$ is finite or not. That is, while the two complementary regimes above share the same asymptotics in the interior and endpoint parameter regimes, this is no longer true when the entire parameter space is taken into account. 

We first present the interior regime asymptotics, which shows that, when only compact subsets of the parameter space are considered, all add-$\beta$ estimators asymptotically have the same batch regret.
\begin{theorem}[Simplex interior expansion]
\label{thm:inner}
Let $\beta>0$, let $0<\delta<\frac12$, and let $t=n\ell$. Then
\begin{equation}
\label{eq:compact-interior-sum}
\sup_{\theta\in[\delta,1-\delta]}
\left|R_{n,\ell}(\hat p_{\beta},\theta)-\frac12\sum_{m=t}^{t+\ell-1}\frac1m\right|
=O_{\beta,\delta}\!\left(\sum_{m=t}^{t+\ell-1}\frac1{m^2}\right).
\end{equation}
Consequently,
\begin{equation}
\label{eq:compact-interior-log}
\sup_{\theta\in[\delta,1-\delta]}
\left|R_{n,\ell}(\hat p_{\beta},\theta)-\frac12\log\left(1+\frac1n\right)\right|
=O_{\beta,\delta}\!\left(\frac1{n^2\ell}\right).
\end{equation}
\end{theorem}

We defer the proof to \cref{app:inner-proof}.
The next theorem, instead, concerns the interval endpoints.
\begin{theorem}[Simplex endpoints expansion]
\label{thm:boundary}
Let $\beta>0$, let $\theta \in \{0,1\}$, and let $t=n\ell$. Then, as $t\to\infty$,
\begin{equation}
\label{eq:endpoint-expansion}
R_{n,\ell}(\hat{p}_{\beta},\theta)
=\beta\log \left(1+\frac{1}{n}\right)+O_{\beta}\!\left(\frac{1}{n^2\ell}\right).
\end{equation}
\end{theorem}
\begin{IEEEproof}
We prove the case $\theta = 1$. The case $\theta = 0$ follows by symmetry. Define
\begin{equation}
 f_{\beta}(x)=\log\Gamma(x+2\beta)-\log\Gamma(x+\beta),\qquad x>0.
\end{equation}
Then
\begin{equation}
R_{n,\ell}(\hat{p}_{\beta},\theta=1)=f_{\beta}(t+\ell)-f_{\beta}(t).
\end{equation}
The standard digamma expansion, uniformly for $x\geq1$ and fixed $c>0$, gives
\begin{equation}
\psi(x+c)=\log x+\frac{c-1/2}{x}+O_c\!\left(\frac1{x^2}\right),
\end{equation}
where $\psi$ is the digamma function. Hence
\begin{equation}
 f_{\beta}'(x)=\psi(x+2\beta)-\psi(x+\beta)
 =\frac{\beta}{x}+O_{\beta}\!\left(\frac1{x^2}\right).
\end{equation}
Therefore,
\begin{align}
R_{n,\ell}(\hat{p}_{\beta},\theta=1)
&=\int_t^{t+\ell}f_{\beta}'(x)\,dx \\
&=\beta\log\frac{t+\ell}{t}+O_{\beta}\!\left(\int_t^{t+\ell}\frac{dx}{x^2}\right) \\
&=\beta\log\left(1+\frac1n\right)+O_{\beta}\!\left(\frac{\ell}{t^2}\right).
\end{align}
Since $t=n\ell$, the last error term is $O_{\beta}(1/(n^2\ell))$, completing the proof.
\end{IEEEproof}

As the next sections will clarify, it turns out that in the large-$n$ regime, the regret at the interval endpoints is actually the dominant term of the regret over the full interval whenever $\beta \geq \beta_0$. When $\beta < \beta_0$, neither the interior regret nor the endpoint regret are dominant: the largest regret is achieved for parameters in the so-called \emph{Poisson regime}, where $\theta$ is of order $1/t$, with $t=n\ell$. The phenomenon is the same that also occurs for next-token regret, which was first observed and analyzed by Krichevsky \cite{krichevsky2}. Krichevsky's one-step proof already establishes the two uniform ingredients needed for the proof of the batch case: a Poisson approximation for bounded normalized counts \cite[Lemma~5]{krichevsky2} and a large-count estimate for unbounded normalized counts \cite[Lemma~2]{krichevsky2}. Our contribution is to insert those one-step estimates into the conditional batch setting, where the regret of the test block is a sum of one-step regrets over the $\ell$ symbols of $Y^{\ell}$. Depending on whether $n$ is finite or not, this accumulation of one-step regrets produces a different cumulative batch regret: in the large-$n$ regime, the $\ell$ regret terms are maximized by approximately the same $\theta_t$; in the finite-$n$ regime, the maximizing $\theta_t$ changes along the test sequence, and a more sophisticated analysis is required.

\section{Asymptotic batch regret in the large-batch-size regime}
\label{sec:large-n-regime}
We first analyze the case when the batch size $n$ is large. The next theorem gives an asymptotic expansion of batch regret for binary memoryless sources over the full interval $[0,1]$, valid whenever $n\to\infty$. For $m\geq 1$, let
\begin{multline}
\label{eq:one-step-rho}
\rho_m^{(\beta)}(\theta) = \E_{K\sim \mathsf{Binomial}(m,\theta)}\Bigg[\theta\log\frac{\theta}{(K+\beta)/(m+2\beta)} \\
+(1-\theta)\log\frac{1-\theta}{(m-K+\beta)/(m+2\beta)}\Bigg]
\end{multline}
be the one-step regret of the add-$\beta$ predictor after $m$ past samples. Define, for $\lambda>0$,
\begin{equation}
\label{eq:Delta-beta}
\Delta_{\beta}(\lambda)\triangleq\lambda\left[\log\lambda-\E_{Z\sim\mathsf{Pois}(\lambda)}\log(Z+\beta)\right],
\end{equation}
and set $\Delta_{\beta}(0)=0$. Finally, let
\begin{equation}
\label{eq:C-beta}
C_{\beta}\triangleq\beta + \sup_{\lambda\geq0}\Delta_{\beta}(\lambda).
\end{equation}

\begin{theorem}[Full-simplex asymptotics for add-$\beta$ estimators]
\label{thm:add-beta-full-simplex}
Let $\beta>0$, let $t=n\ell$, and let $n\to\infty$, with $\ell=\ell(n)$ any positive integer sequence. Then
\begin{align}
\max_{\theta\in[0,1]}R_{n,\ell}(\hat p_{\beta},\theta)
&= C_{\beta}\log\left(1+\frac1n\right)+o\left(\frac1n\right) \\
&= \frac{C_{\beta}}{n}+o\left(\frac1n\right).
\end{align}
\end{theorem}

We defer the proof to \cref{app:add-beta-full-simplex-proof}. The theorem shows that the full-simplex maximum may not be attained either in the simplex interior or at the endpoints. More specifically, the proof of \cref{thm:add-beta-full-simplex} distinguishes between three regimes depending on asymptotic order of $\theta_t$. If either $t\theta_t \to 0$ or $t(1-\theta_t)\to 0$, i.e., the \emph{endpoint regime}, then
\begin{equation}
R_{n,\ell}(\hat{p}_{\beta},\theta_t) = \beta \log\left(1+\frac1n\right) + o\left(\frac1n\right),
\end{equation}
that is, the leading coefficient of batch regret for $\theta_t$ is $\beta$. If instead $t\theta_t \to \lambda$ or $t(1-\theta_t)\to \lambda$, $0 < \lambda < \infty$, i.e., the \emph{Poisson regime}, then
\begin{equation}
R_{n,\ell}(\hat{p}_{\beta},\theta_t) = \left(\beta + \Delta_{\beta}(\lambda)\right) \log\left(1+\frac1n\right) + o\left(\frac1n\right).
\end{equation}
Finally if both $t\theta_t \to \infty$ and $t(1-\theta_t)\to \infty$, i.e., the \emph{interior regime}, then
\begin{equation}
R_{n,\ell}(\hat{p}_{\beta},\theta_t) = \frac12 \log\left(1+\frac1n\right) + o\left(\frac1n\right).
\end{equation}
This shows that the regimes treated by \cref{thm:inner,thm:boundary} are retrieved as special cases of \cref{thm:add-beta-full-simplex}. However, \cref{thm:inner,thm:boundary} give sharper reminders.

The preceding theorem reduces the full-simplex performance of the add-$\beta$ rule to the single function $\Delta_{\beta}$. In fact, $\Delta_{\beta}$ accounts for all the three regimes described above. For $\lambda = 0$, one has $\beta+\Delta_{\beta}(0)=\beta$, which is the leading regret coefficient in the endpoint regime. On the other hand, a Taylor expansion of $\log(Z+\beta)$ around $\lambda$ gives, for fixed $\beta>0$,
\begin{equation}
\label{eq:Delta-beta-infinity}
\E\log(Z+\beta)
=\log\lambda+\frac{\beta-1/2}{\lambda}
+O_{\beta}\!\left(\frac1{\lambda^2}\right),
\end{equation}
where $Z\sim\mathsf{Pois}(\lambda)$, and hence
\begin{equation}
\label{eq:Delta-beta-half-limit}
\beta +\Delta_{\beta}(\lambda)\to\frac12,\qquad \lambda\to\infty,
\end{equation}
which corresponds to the interior regime. The Poisson regime is obtained for $\lambda \in (0,\infty)$.

Krichevsky obtained the optimal coefficient $\beta_0$ for the next-token regret by finding the value of $\beta$ for which the endpoint coefficient and the largest coefficient in the Poisson regime were equal \cite[Fig.~1 and Lemma~6]{krichevsky2}. In our notation it can be characterized as
\begin{equation}
\label{eq:beta0-def}
\beta_0=\inf\left\{\beta>0:\Delta_{\beta}(\lambda)\leq0\text{ for every }\lambda>0\right\}.
\end{equation}
For the sake of completeness, the following result establishes that the coefficient characterized by \cref{eq:beta0-def} is indeed the same as Krichevsky's optimal $\beta_0$ \cite{krichevsky2}.

\begin{proposition}[Optimal add-$\beta$ parameter]
\label{prop:beta0-optimal}
Let $\beta_0$ be defined by \eqref{eq:beta0-def}. Then the defining set is nonempty and
\begin{equation}
\frac12\leq\beta_0\leq1.
\end{equation}
Furthermore,
\begin{equation}
\label{eq:min-C-beta-beta0}
\min_{\beta>0}C_{\beta}=C_{\beta_0}=\beta_0.
\end{equation}
\end{proposition}

\begin{figure}[t]
\centering
\includegraphics[width=\linewidth]{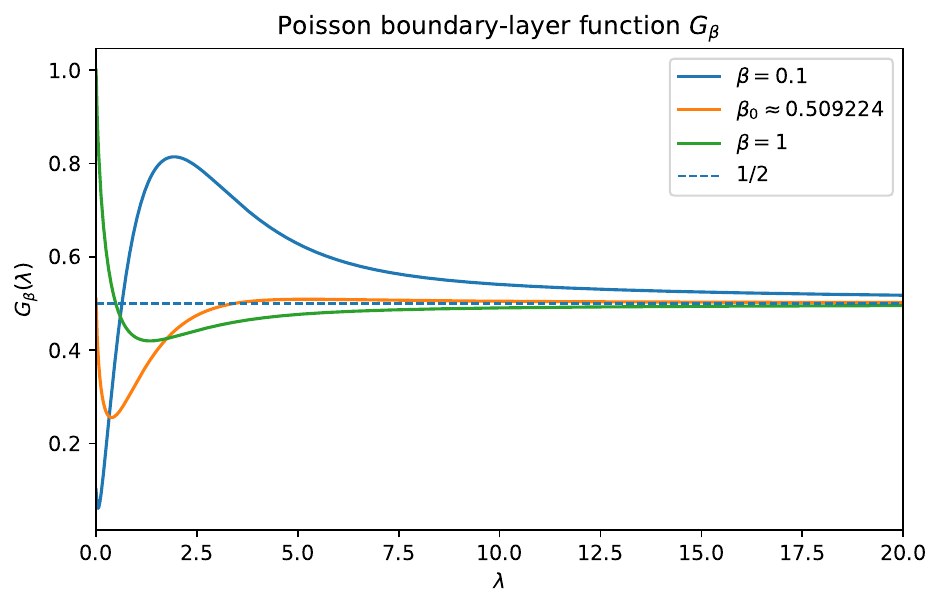}
\caption{The Poisson layer function $\beta +\Delta_{\beta}(\lambda)$ for three choices of $\beta$. The endpoint value is $\beta +\Delta_{\beta}(0)=\beta$, while all curves converge to the compact-interior value $1/2$ as $\lambda\to\infty$.}
\label{fig:G-beta-boundary-layer}
\end{figure}

\begin{figure}[t]
\centering
\includegraphics[width=\linewidth]{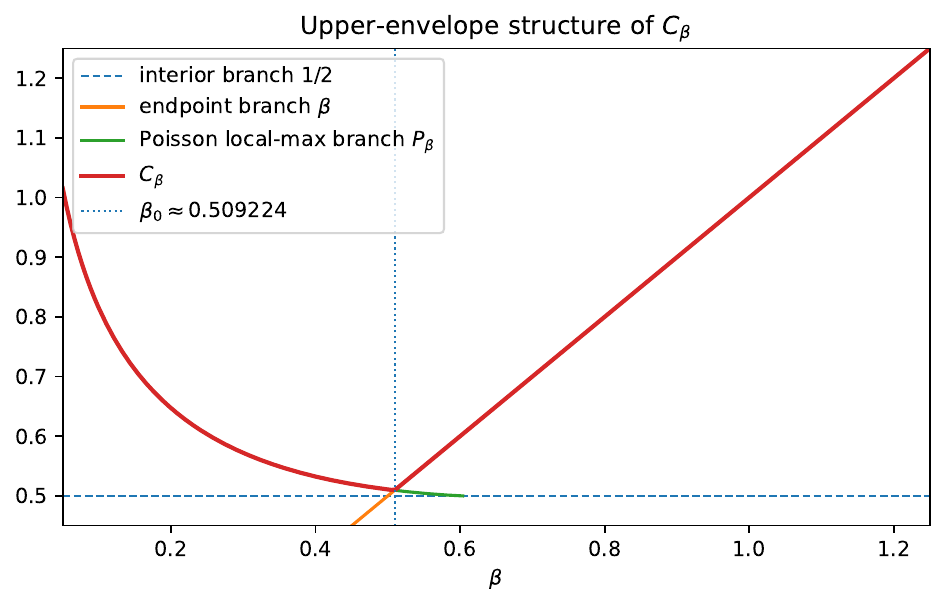}
\caption{The full-simplex add-$\beta$ coefficient $C_{\beta}$ as the upper envelope of the compact-interior branch $1/2$, the endpoint branch $\beta$, and the Poisson branch $P_{\beta}$. The crossing point is Krichevsky's value $\beta_0\approx0.509224$, where $C_{\beta_0}=\beta_0$.}
\label{fig:C-beta-envelope}
\end{figure}

We defer the proof to \cref{beta0-optimal-proof}. Numerical evaluation gives $\beta_0\approx0.509224$, in agreement with Krichevsky's analysis.
The proof of Proposition~\ref{prop:beta0-optimal} also clarifies that the optimal bias $\beta_0$ is the threshold point after which the endpoint regret becomes dominant over the largest Poisson regret. In fact, the function $C_\beta$ can be represented as the upper envelope of three functions, each corresponding to the largest regret in the interior, Poisson, and endpoint regimes. If we define the Poisson branch as
\begin{equation}
\label{eq:P-beta-stationary-branch}
P_{\beta}=\beta + \sup_{\lambda\in\mathcal M_{\beta}}\Delta_{\beta}(\lambda),
\end{equation}
where
\begin{equation}
\label{eq:local-max-set}
\mathcal M_{\beta}=\{\lambda>0:\lambda\text{ is a local maximum of }\Delta_{\beta}(\lambda)\},
\end{equation}
with the convention that $P_{\beta}=-\infty$ if $\mathcal M_{\beta}$ is empty, then we have that the optimal regret coefficient satisfies
\begin{equation}
\label{eq:C-beta-envelope}
C_{\beta}=\max\left\{\frac12,\beta,P_{\beta}\right\}.
\end{equation}
Indeed, if the supremum of $\Delta_{\beta}$ over $(0,\infty)$ exceeds both boundary values $0$ and $1/2-\beta$, then, because $\Delta_{\beta}(0) = 0$ and $\Delta_{\beta}(\lambda)\to1/2 - \beta$ as $\lambda\to\infty$, the supremum is attained at some finite positive point, which must belong to $\mathcal M_{\beta}$. The structure is illustrated numerically in Figs.~\ref{fig:G-beta-boundary-layer} and~\ref{fig:C-beta-envelope}. Figure~\ref{fig:G-beta-boundary-layer} shows representative shapes of $\Delta_{\beta}$, while Fig.~\ref{fig:C-beta-envelope} plots the three branches and their upper envelope.

\section{Asymptotic batch regret in the finite-batch-size regime}
\label{sec:fixed-n-regime}
\cref{thm:add-beta-full-simplex} assumes that the number of training sequences grows, so that the test block is short relative to the training data. There is also a complementary fixed-$n$ limit. In this case the optimal parameter $\theta_t$ changes by a non-negligible multiplicative factor over the test block, and the one-step Poisson boundary function has to be integrated. For a fixed positive integer $n$ and $\lambda\geq0$, define the \emph{normalized finite-$n$ boundary function}
\begin{align}
\label{eq:B-beta-n}
\Delta_{\beta,n}(\lambda)
&=\frac1{\log(1+1/n)}\int_1^{1+1/n}\frac{\Delta_{\beta}(u\lambda)}{u}\,du \\
&=\frac1{\log(1+1/n)}\int_{\lambda}^{(1+1/n)\lambda}
\frac{\Delta_{\beta}(\mu)}{\mu}\,d\mu,
\end{align}
with the value at $\lambda=0$ defined by continuity. Geometrically, $\Delta_{\beta,n}$ can be interpreted as a logarithmic moving average of $\Delta_{\beta}$ over the interval from $\lambda$ to $(1+1/n)\lambda$, as it is clear by substituting $u=e^s$,
\begin{equation}
\label{eq:B-log-moving-average}
\Delta_{\beta,n}(\lambda)
=\frac1{\log(1+1/n)}\int_0^{\log(1+1/n)}\Delta_{\beta}(\lambda e^s)\,ds.
\end{equation}
This function is the equivalent of $\Delta_{\beta}$ for the finite-$n$ regime, in that it is the leading coefficient of the batch regret of conditional add-$\beta$ estimators for binary memoryless sources when the batch size $n$ is fixed and the sequence length $\ell$ is large. The following theorem formalizes this fact.

\begin{theorem}[Fixed-$n$ full-simplex limit for add-$\beta$ estimators]
\label{thm:add-beta-fixed-n}
Let $\beta>0$ and fix a positive integer $n$. Then, as $\ell\to\infty$,
\begin{equation}
\label{eq:fixed-n-limit}
\max_{\theta\in[0,1]}R_{n,\ell}(\hat p_{\beta},\theta)
\longrightarrow C_{\beta,n} \log\left(1+\frac1n\right),
\end{equation}
where
\begin{equation}
\label{eq:D-beta-n}
C_{\beta,n}\triangleq \beta + \sup_{\lambda\geq0}\Delta_{\beta,n}(\lambda).
\end{equation}
\end{theorem}

The proof is given in \cref{app:add-beta-fixed-n-proof}. Similar to the large-$n$ case, $\Delta_{\beta,n}(\lambda)$ includes the interior and endpoint regimes as its extreme values: $\beta+\Delta_{\beta,n}(0) = \beta$ is the regret coefficient at the endpoints, while $\beta + \lim_{\lambda\to\infty}\Delta_{\beta,n}(\lambda) = 1/2$ is the common coefficient in the interior of the simplex.
Also notice that as $n\to\infty$ the average in \cref{eq:B-log-moving-average} converges to the single point $\Delta_{\beta}(\lambda)$, retrieving the large-$n$ coefficient; it can be shown that the convergence can be made uniform, so that $\lim_{n\to\infty} C_{\beta,n} = C_{\beta}$.

Parallel to \cref{eq:C-beta-envelope}, also the fixed-$n$ coefficient $C_{\beta,n}$ has an upper-envelope structure. If we let
\begin{equation}
\label{eq:fixed-n-local-max-set}
\mathcal M_{\beta,n}=\{\lambda>0:\lambda\text{ is a local maximum of }\Delta_{\beta,n}\}
\end{equation}
and define the Poisson branch by
\begin{equation}
\label{eq:P-beta-n}
P_{\beta,n}=\beta + \sup_{\lambda\in\mathcal M_{\beta,n}}\Delta_{\beta,n}(\lambda),
\end{equation}
with the convention $P_{\beta,n}=-\infty$ if $\mathcal M_{\beta,n}$ is empty, then we have that
\begin{equation}
\label{eq:D-beta-n-envelope}
C_{\beta,n}= \max\left\{\frac12,\beta,P_{\beta,n}\right\}.
\end{equation}
The natural finite-$n$ analogue of Krichevsky's threshold $\beta_0$ is
\begin{equation}
\label{eq:beta-n-star-def}
\beta_n^\star=
\inf\left\{\beta>0:\Delta_{\beta,n}(\lambda)\leq0
\text{ for every }\lambda>0\right\}.
\end{equation}
Thus $\beta_n^\star$ is the smallest bias for which the endpoint branch dominates the Poisson branch. The following theorem shows that $\beta^\star_n$ is indeed the optimal bias for the finite-$n$ regime and, just like in the large-$n$ regime, the optimal regret coefficient and the optimal bias are equal. Furthermore, $\beta^\star_n$ is strictly increasing with $n$ and converges to $\beta_0$ as $n\to\infty$.

\begin{theorem}[Optimal fixed-$n$ add-$\beta$ parameters]
\label{thm:fixed-n-optimal-beta}
For every fixed positive integer $n$, there exist finite $\beta_n^\star > 1/2$ and $\lambda_n^\star>1$ such that
\begin{equation}
\label{eq:fixed-n-contact}
\Delta_{\beta_n^\star,n}(\lambda_n^\star)=0
\end{equation}
and $\beta_n^\star$ satisfies
\begin{equation}
\label{eq:fixed-n-optimum}
\min_{\beta>0}C_{\beta,n}
=C_{\beta_n^\star,n}
=\beta_n^\star.
\end{equation}

Furthermore, as function of $n$, the optimal bias satisfies
\begin{equation}
\label{eq:beta-n-star-monotone}
\frac12<\beta_1^\star<\beta_2^\star<\cdots<\beta_0 \quad\text{and}\quad
\lim_{n\to\infty}\beta_n^\star=\beta_0.
\end{equation}
\end{theorem}

\begin{figure}[t]
\centering
\includegraphics[width=\columnwidth]{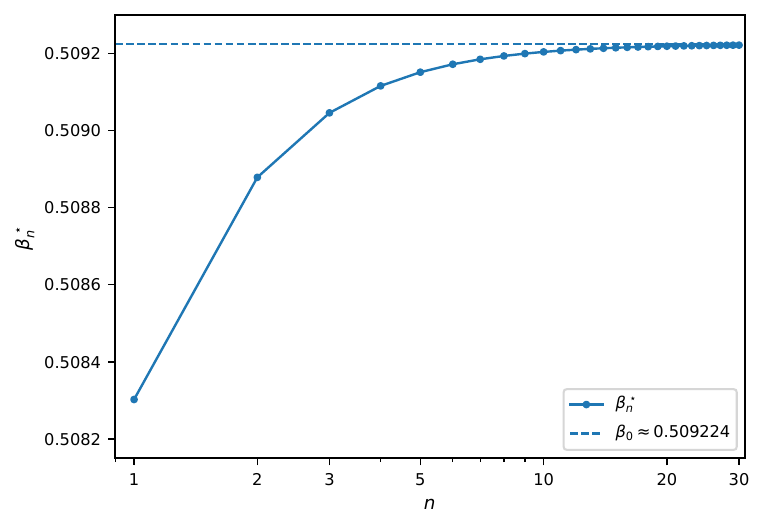}
\caption{Numerically computed optimal fixed-$n$ parameters $\beta_n^\star$, obtained from \cref{eq:fixed-n-contact-numerical}. Theorem~\ref{thm:fixed-n-optimal-beta} proves that they increase strictly to the one-step Krichevsky value $\beta_0\approx0.509224$, shown by the dashed line.}
\label{fig:beta-n-star-convergence}
\end{figure}

\begin{figure}[t]
\centering
\includegraphics[width=0.96\columnwidth]{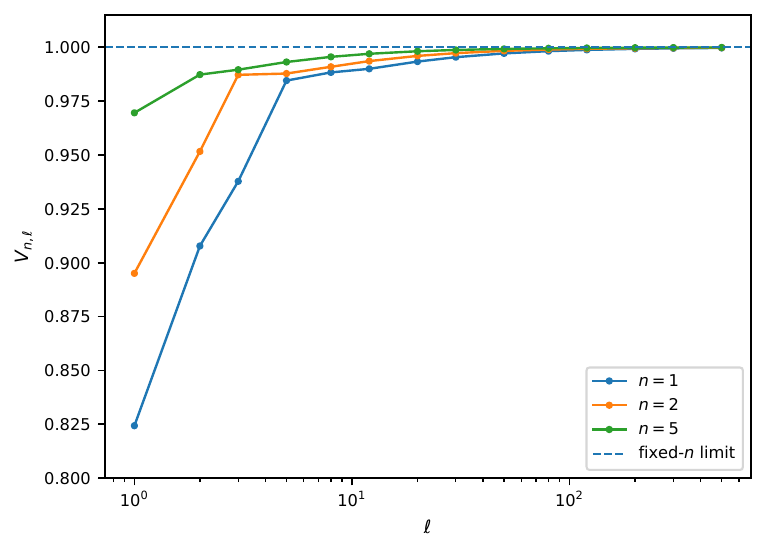}
\caption{Finite-$\ell$ validation of the asymptotic fixed-$n$ batch regret limit. The vertical axis is the worst-case regret divided by its fixed-$n$ asymptotic limit, namely $V_{n,\ell}$ from \cref{eq:finite-ell-validation-ratio}.}
\label{fig:finite-ell-validation}
\end{figure}

The proof is given in \cref{app:fixed-n-optimal-beta-proof}.
The equations
\begin{equation}
\label{eq:fixed-n-contact-numerical}
\Delta_{\beta,n}(\lambda)=0,
\qquad
\frac{\partial}{\partial\lambda}\Delta_{\beta,n}(\lambda)=0
\end{equation}
allow us to numerically characterize $(\beta_n^\star,\lambda_n^\star)$. Solving them for increasing $n$ gives \cref{fig:beta-n-star-convergence}, in agreement with \cref{thm:fixed-n-optimal-beta}. Finally, \cref{fig:finite-ell-validation} compares the limiting prediction in \cref{thm:add-beta-fixed-n,thm:fixed-n-optimal-beta} with the actual batch regret for finite $\ell$, by tracking the value of
\begin{equation}
\label{eq:finite-ell-validation-ratio}
V_{n,\ell}
=\frac{\max_{\theta\in[0,1]}R_{n,\ell}(\hat p_{\beta_n^\star},\theta)}
{\beta_n^\star \log(1+1/n)}.
\end{equation}
As $\ell\to\infty$, $V_{n,\ell}$ converges to one as expected.

\section{Optimal batch regret for binary memoryless sources}
\label{sec:optimal-batch-regret}
Having studied the asymptotic batch regret of conditional additive smoothing estimators, we can now use \cref{thm:cond-rc} to sharpen the converse for binary memoryless sources over all predictors. The first-order asymptotic minimax batch regret for memoryless sources was first obtained by Fogel and Feder \cite[Theorem~2]{fogel3}, using the sequential Braess--Sauer predictor \cite{braess2004187} for achievability. Our lower bound below identifies the same leading constant while improving the converse remainder to $O(1/(n^2\ell))$.

\begin{theorem}[Minimax lower bound for Bernoulli batch prediction]
\label{thm:regret-lower}
For the class of i.i.d. Bernoulli distributions, for any positive integer sequences $n$ and $\ell$ such that $t=n\ell\to\infty$, we have that
\begin{equation}
\label{eq:sharp-lower-bound}
\min_{\hat{p}} \max_{\theta\in[0,1]} R_{n,\ell}(\hat{p},\theta) \geq \frac{1}{2} \log \left(1 + \frac{1}{n}\right) - O\left(\frac{1}{n^2\ell}\right).
\end{equation}
\end{theorem}

The proof is given in \cref{app:lower-bound-proof}. Let now
\begin{equation}
\label{eq:bernoulli-minimax-def}
R_{n,\ell}^{\star}
\triangleq
\inf_{\hat p}\sup_{\theta\in[0,1]}R_{n,\ell}(\hat p,\theta)
\end{equation}
be the unrestricted batch regret for binary memoryless sources. Then, \cite[Theorem 2]{fogel3} shows that the preceding lower bound can be combined with the optimal one-step predictor defined by Braess and Sauer \cite{braess2004187} to obtain a tight first-order characterization of $R_{n,\ell}^\star$.

\begin{corollary}[First-order minimax Bernoulli batch regret \cite{fogel3}]
\label{cor:bernoulli-minimax}
For any positive integer sequences $n$ and $\ell$ such that ${t=n\ell\to\infty}$,
\begin{equation}
\label{eq:bernoulli-minimax-first-order}
R_{n,\ell}^{\star}
=
\left(\frac12+o(1)\right)
\log\left(1+\frac1n\right).
\end{equation}
Consequently, for every fixed $n$,
\begin{equation}
\label{eq:bernoulli-minimax-fixed-n}
\lim_{\ell\to\infty}R_{n,\ell}^{\star}
=
\frac12\log\left(1+\frac1n\right),
\end{equation}
whereas, if $n\to\infty$,
\begin{equation}
\label{eq:bernoulli-minimax-large-n}
R_{n,\ell}^{\star}
=
\frac1{2n}+o\left(\frac1n\right).
\end{equation}
\end{corollary}
\begin{IEEEproof}
The proof follows the one of \cite[Theorem 2]{fogel3}. We include it here for the sake of completeness. Let $q_m^{\rm BS}$ be the binary one-step predictor of Braess and Sauer \cite[Sec.~5, Eqs.~(5.18)--(5.21)]{braess2004187}, and let
\begin{equation}
\overline\rho_m^{\rm BS}
\triangleq
\sup_{\theta\in[0,1]}
\mathbb{E}_{\theta}\!\left[
D\!\left(\mathsf{Bernoulli}(\theta)
\middle\|
q_m^{\rm BS}(\cdot\mid X^m)
\right)
\right].
\end{equation}
Their uniform one-step analysis gives
\begin{equation}
\label{eq:BS-one-step-asymptotic}
\overline\rho_m^{\rm BS}
=
\frac{1}{2m}+o\left(\frac1m\right).
\end{equation}
Write
\begin{equation}
\overline\rho_m^{\rm BS}
\leq
\frac{\frac12+\varepsilon_m}{m},
\qquad
\varepsilon_m\to0,
\end{equation}
and set $\overline\varepsilon_t=\sup_{m\geq t}|\varepsilon_m|$, so that $\overline\varepsilon_t\to0$.

Sequentially compose these one-step rules across the test block:
\begin{equation}
\widehat p_{\rm BS}(y^\ell\mid x^t)
=
\prod_{i=0}^{\ell-1}
q_{t+i}^{\rm BS}(y_{i+1}\mid x^t,y^i).
\end{equation}
The chain rule for conditional KL divergence gives
\begin{align}
\sup_{\theta}R_{n,\ell}(\widehat p_{\rm BS},\theta)
&\leq
\sum_{m=t}^{t+\ell-1}\overline\rho_m^{\rm BS} \\
&\leq
\left(\frac12+\overline\varepsilon_t\right)
\sum_{m=t}^{t+\ell-1}\frac1m.
\end{align}
Using the classical harmonic-sum estimate,
\begin{equation}
\sum_{m=t}^{t+\ell-1}\frac1m
=
\log\left(1+\frac1n\right)
+O\left(\frac1{n^2\ell}\right),
\end{equation}
and the fact that
\begin{equation}
\frac{1/(n^2\ell)}{\log(1+1/n)}
=O\left(\frac1{n\ell}\right)=o(1),
\end{equation}
we obtain the upper bound in \cref{eq:bernoulli-minimax-first-order}. The lower bound follows directly from \cref{thm:regret-lower}.
\end{IEEEproof}

For fixed $n$ and $\ell$, \cref{thm:cond-rc} shows that there is a prior with finite support achieving the optimal batch regret. However, if applied directly to the Bernoulli batch problem, it gives an exponential support upper bound. Thankfully, the following direct conditional counterpart of the classical Witsenhausen bound for the binomial channel \cite{witsenhausen1980convexity} provides a much smaller bound.

\begin{corollary}[Finite-support Bernoulli least-favorable prior]
\label{cor:bernoulli-support}
For the binary memoryless batch problem with $t=n\ell$ training symbols and a test block of length $\ell$, there exists a capacity-achieving prior on $[0,1]$ with at most
\begin{equation}
\label{eq:bernoulli-support-bound}
(t+\ell)+1=(n+1)\ell+1
\end{equation}
support points.
\end{corollary}

\begin{IEEEproof}
The parameter interval is compact and the finite-output channel law is continuous in $\theta$, so a capacity-achieving prior exists in \cref{thm:cond-rc}. Due to the i.i.d. nature of the data, a minimax predictor may be taken to depend only on the training count $K\in\{0,\ldots,t\}$
and the test count $J\in\{0,\ldots,\ell\}$.
Let $Q^\star(j\mid k)$ be such a minimax count-based kernel and let $M=t+\ell$. The vectors appearing in \cref{eq:v-theta-support} have coordinates
\begin{align}
v_\theta(k,j)
&=\binom tk\binom\ell j
\theta^{k+j}(1-\theta)^{M-k-j}\notag\\
&\quad-\binom tk\theta^k(1-\theta)^{t-k}Q^\star(j\mid k).
\label{eq:bernoulli-vtheta}
\end{align}
Every coordinate in \cref{eq:bernoulli-vtheta} is a polynomial in $\theta$ of degree at most $M$. Equivalently, the vector-valued map $\theta\mapsto v_\theta$ is an affine image of the degree-$M$ Bernstein vector
\begin{equation}
\left(\binom Ms\theta^s(1-\theta)^{M-s}\right)_{s=0}^M.
\end{equation}
Because these $M+1$ coordinates sum to one, their affine hull has dimension at most $M$. The second part of the proof of \cref{thm:cond-rc} therefore shows that the support points are at most $M+1=(n+1)\ell+1$.
\end{IEEEproof}

We believe that the previous bound can be improved further by adapting existing results such as \cite{barletta2024binomial} to the conditional-batch case.

\section{Markov sources}
\label{sec:markov}
We now shift our attention to binary first-order Markov sources, i.e., distributions of the form
\begin{equation}
\label{eq:markov-class}
p_{\nu,p,q}(x^\ell)
=
\pi_\nu(x_1)
\prod_{j=2}^{\ell}P_{p,q}(x_j\mid x_{j-1}),
\end{equation}
where $\pi_\nu(1)=\nu$, and
\begin{equation}
\label{eq:markov-matrix}
P_{p,q}
=
\begin{pmatrix}
1-p&p\\
q&1-q
\end{pmatrix}.
\end{equation}
Following \cite{hao1}, we fix $0<\delta<1/2$ and restrict the transition parameters to
\begin{equation}
\label{eq:markov-positive-rectangle}
(p,q)\in\Theta_\delta
\triangleq[\delta,1-\delta]^2.
\end{equation}
This allows us to avoid complex regret behavior near the boundary of the simplex.
Contrary to the Bernoulli case, where the entire training batch can be represented as a single i.i.d. sequence of $n\ell$ tokens, in the Markov case this is no longer true, as every training sequence in the batch is independently re-initialized from the initial distribution of the source. Because of this, in the batch case the parametrization of the initial distribution of the Markov source becomes fundamental, and the optimal batch regret is different for different parametrizations. In this work we focus on three canonical cases, namely if the initial distribution is known, if it is a free parameter, or if it is equal to the stationary distribution of the source. The three scenarios are represented by the following three classes of distributions:
\begin{align}
\mathcal M_{\delta,\nu_0}^{\rm known}
&\triangleq
\{(\nu_0,p,q):(p,q)\in\Theta_\delta\},
\label{eq:markov-known-class}\\
\mathcal M_\delta^{\rm stat}
&\triangleq
\left\{\left(\frac{p}{p+q},p,q\right):(p,q)\in\Theta_\delta\right\},
\label{eq:markov-stationary-class}\\
\mathcal M_\delta^{\rm free}
&\triangleq
\{(\nu,p,q):\nu\in[0,1],\ (p,q)\in\Theta_\delta\}.
\label{eq:markov-free-class}
\end{align}
When the result is independent of the parametrization of the initial distribution, we simply write $M_\delta$ to mean any of the classes above.

We now focus on studying the asymptotic batch regret for Markov sources under the three parametrizations above. In this section we limit ourselves to the regime where $n\to\infty$ and $\ell\to\infty$ jointly. The $n$ training sequences $\bm X_1,\ldots,\bm X_n$ and the test sequence $\bm Y$ are independent conditional on the source parameter. For a predictor factored as
\begin{equation}
\label{eq:markov-predictor-factorization}
\widehat p(y^\ell\mid\bm x^n)
=
\widehat \pi(y_1\mid\bm x^n)
\widehat P(y_2^\ell\mid\bm x^n,y_1),
\end{equation}
the chain rule for KL divergence applied to the batch regret in \cref{eq:batch-regret} gives the decomposition
\begin{equation}
\label{eq:markov-regret-decomposition}
R_{n,\ell}(\widehat p,\theta)
=
R_{n,\ell}^{(\pi)}(\widehat \pi,\theta)
+
R_{n,\ell}^{(P)}(\widehat P,\theta),
\end{equation}
where
\begin{align}
R_{n,\ell}^{(\pi)}(\widehat \pi,\theta)
&\triangleq
\mathbb{E}_{\theta}\log
\frac{\pi_\nu(Y_1)}{\widehat \pi(Y_1\mid\bm X^n)},
\label{eq:markov-initial-regret}\\
R_{n,\ell}^{(P)}(\widehat P,\theta)
&\triangleq
\mathbb{E}_{\theta}\log
\frac{P_{p,q}(Y_2^\ell\mid Y_1)}
{\widehat P(Y_2^\ell\mid\bm X^n,Y_1)},
\label{eq:markov-transition-regret}
\end{align}
which can be identified as the regrets of the initial distribution and of the transition kernel respectively. Regarding the former, the following result holds.

\begin{theorem}[Initial-state contribution]
\label{thm:markov-initial-cases}
As $n,\ell\to\infty$ jointly, the following statements on $R_{n,\ell}^{(\pi)}$ hold.
\begin{enumerate}
\item Under the class $\mathcal M_{\delta,\nu_0}^{\rm known}$, i.e., if the initial law is known, then
\begin{equation}
\inf_{\widehat \pi} \sup_{\theta\in \mathcal M_{\delta,\nu_0}^{\rm known}}R_{n,\ell}^{(\pi)} (\widehat \pi,\theta) = 0.
\end{equation}
\item Under the class $\mathcal M_\delta^{\rm stat}$, i.e., if the initial law is stationary, then there is a predictor $\widehat \pi$ such that
\begin{equation}
\label{eq:markov-stationary-initial-upper}
\sup_{\theta\in\mathcal M_\delta^{\rm stat}}
R_{n,\ell}^{(\pi)}(\widehat \pi,\theta)
=O_\delta\left(\frac1{n\ell}\right)
=o\left(\frac1n\right).
\end{equation}
\item Under the class $\mathcal M_\delta^{\rm free}$, i.e., if the initial law is a free parameter independent of the transition kernel, then
\begin{equation}
\label{eq:markov-free-initial-asymptotic}
\inf_{\widehat \pi}
\sup_{\theta\in\mathcal M_\delta^{\rm free}}
R_{n,\ell}^{(\pi)}(\widehat \pi,\theta)
=
\frac{1}{2n}+o\left(\frac1n\right).
\end{equation}
\end{enumerate}
\end{theorem}

\begin{IEEEproof}
The first statement follows directly by taking $\widehat \pi = \nu$. For the second statement, write
\begin{equation}
\pi\triangleq\mathbb{P}_\theta(X_{i,s}=1)=\frac{p}{p+q}
\in[\delta,1-\delta]
\end{equation}
and define
\begin{equation}
\overline X_{n,\ell}
\triangleq
\frac1{n\ell}\sum_{i=1}^n\sum_{s=1}^{\ell}X_{i,s},
\qquad
\widehat\pi
\triangleq
\Pi_{[\delta,1-\delta]}(\overline X_{n,\ell}),
\label{eq:markov-stationary-initial-predictor}
\end{equation}
where $\Pi_I$ denotes Euclidean projection onto $I$. Predict $Y_1=1$ with probability $\widehat\pi$. Let $\rho=1-p-q$. The centered state function $x\mapsto x-\pi$ is an eigenfunction of $P_{p,q}$ with eigenvalue $\rho$. For a stationary Markov chain, correlations of an eigenfunction with eigenvalue $\rho$ decay as $\rho^k$; see, e.g., \cite[Lemma~12.22]{levin-peres}. Hence,
\begin{equation}
\operatorname{Cov}_\theta(X_{i,s},X_{i,s+k})
=\pi(1-\pi)\rho^k.
\end{equation}
Since $|\rho|\leq1-2\delta$,
\begin{align}
\operatorname{Var}_\theta\left(\sum_{s=1}^{\ell}X_{i,s}\right)
&=\pi(1-\pi)\left\{\ell+2\sum_{k=1}^{\ell-1}(\ell-k)\rho^k\right\}\\
&\leq\frac\ell4\left\{1+2\sum_{k=1}^{\infty}(1-2\delta)^k\right\}
\leq C_\delta\ell.
\end{align}
Since projection onto $[\delta, 1-\delta]$ does not increase squared error, and since the $n$ training sequences are independent, we have
\begin{equation}
\mathbb{E}_{\theta}(\widehat\pi-\pi)^2
\leq\mathbb{E}_{\theta}(\overline X_{n,\ell}-\pi)^2
\leq\frac{C_\delta}{n\ell}.
\end{equation}
For $a,b\in[\delta,1-\delta]$, the inequality $D(\mathsf{Bern}(a)\|\mathsf{Bern}(b))\leq(a-b)^2/[b(1-b)]$ gives \cref{eq:markov-stationary-initial-upper}.

For the third statement, notice that the $n$ initial training states are i.i.d. $\mathsf{Bernoulli}(\nu)$. Applying the one-step rule of Braess and Sauer \cite[Eq.~(5.18)]{braess2004187} on the sequence of $n$ initial states gives the upper bound $1/(2n)+o(1/n)$ uniformly over $\nu\in[0,1]$ \cite[Sec.~5]{braess2004187}. For the converse, restrict the transition kernel to $p=q=1/2$. Under this submodel, all coordinates after the first one are independent fair bits and carry no information about $\nu$. Write the training data as $(S,Z)$, where $S=(X_{1,1},\ldots,X_{n,1})$, and for any full-data predictor define $\overline p(S)=\E[\widehat \pi(1\mid S,Z)\mid S]$. Since $u\mapsto D(\mathsf{Bern}(\nu)\|\mathsf{Bern}(u))$ is convex,
\begin{align}
&\E D\left(\mathsf{Bern}(\nu)
\middle\|\mathsf{Bern}(\widehat \pi(1\mid S,Z))\right)\notag\\
&\qquad\geq
\E D\left(\mathsf{Bern}(\nu)
\middle\|\mathsf{Bern}(\overline p(S))\right).
\end{align}
Thus the additional coordinates cannot improve the minimax initial-state prediction problem, and Krichevsky's arbitrary-predictor lower bound \cite[Theorem~2]{krichevsky2} gives the matching lower bound.
\end{IEEEproof}

We next analyze the regret of the transition kernel. Let $n_{hk}(\bm{x}_i)$ be the number of transitions $h\to k$ in the sequence $\bm{x}_i$, and let $n_h(\bm{x}_i)=\sum_k n_{hk}(\bm{x}_i)$ be the number of departures from $h$. Also define the cumulative counts
\begin{equation}
\label{eq:markov-training-counts}
n_{hk}(\bm{x}^n)\triangleq\sum_{i=1}^n n_{hk}(\bm{x}_i),
\qquad
n_h(\bm{x}^n)\triangleq\sum_{i=1}^n n_h(\bm{x}^n).
\end{equation}
Consider the non-updating Krichevsky--Trofimov predictor
\begin{equation}
\label{eq:pred-markov-2}
\widehat p_{\rm KT}(y_2^\ell\mid \bm x^n,y_1)
=
\prod_{h,k\in\{0,1\}}
\left(\frac{n_{hk}(\bm{x}^n)+\frac12}{n_h(\bm{x}^n)+1}\right)^{n_{hk}(\bm{y})},
\end{equation}
where $n_{hk}(\bm{y})$ is the number of transitions $h\to k$ in the test sequence. Then, the following result holds.

\begin{theorem}[Transition upper bound]
\label{thm:markov-regret}
Uniformly over the initial distribution and over $(p,q)\in\Theta_\delta$,
\begin{equation}
\label{eq:markov-transition-expansion}
\sup_{\theta\in\mathcal M_\delta}
\left|R_{n,\ell}^{(P)}(\widehat p_{\rm KT},\theta)-\frac1n\right|
=
O_{\delta}\left(\frac{1}{n\sqrt{n\ell}}\right),
\qquad n,\ell\to\infty.
\end{equation}
Consequently,
\begin{equation}
\label{eq:markov-transition-minimax-upper}
\inf_{\widehat P}
\sup_{\theta\in\mathcal M_\delta}R_{n,\ell}^{(P)}(\widehat P,\theta)
\leq
\frac1n+o\left(\frac1n\right).
\end{equation}
\end{theorem}

We defer the proof to Appendix~\ref{app:markov-regret-proof}. The proof adapts \cite[Theorem 2]{hao1} to the batch case. More specifically, we apply Paulin's bounded-differences inequality for Markov chains \cite[Corollary~2.11]{paulin2015} separately to the token counts of each independent training sequence and then we combine the resulting bounds to the full training batch counts via standard sub-Gaussianity properties.

Matching lower bounds can be obtained for the three parametrization classes, once again via the conditional regret-capacity theorem. For each of the three initial distribution parametrizations above, the following result is instrumental in bounding the maximal mutual information in \cref{thm:cond-rc}.

\begin{lemma}
\label{lem:markov-transition-information}
Under either of the three classes $\mathcal M_{\delta,\nu_0}^{\rm known}$, $\mathcal M_\delta^{\rm stat}$ or $\mathcal M_\delta^{\rm free}$, there is a sequence of priors $W_{n,\ell}$ such that
\begin{equation}
\label{eq:markov-known-transition-information}
I_{W_{n,\ell}}\left((P,Q);Y_2^\ell
\mid\bm X^n,Y_1\right)
=
\log\left(1+\frac1n\right)
+o\left(\frac1n\right).
\end{equation}
\end{lemma}

 We defer the proof to Appendix~\ref{app:markov-minimax-proof}. For each of the three classes, define the minimax batch regret as
\begin{equation}
R_{n,\ell}(\mathcal M)
\triangleq
\inf_{\widehat p}
\sup_{\theta\in\mathcal M}R_{n,\ell}(\widehat p,\theta).
\end{equation}
Then, the following tight asymptotic characterizations hold.

\begin{theorem}[Minimax binary Markov batch regret]
\label{thm:markov-minimax}
As $n,\ell\to\infty$ jointly, we have
\begin{align}
R_{n,\ell}(\mathcal M_{\delta,\nu_0}^{\rm known})
&=
\frac1n+o\left(\frac1n\right),
\label{eq:markov-known-minimax}\\
R_{n,\ell}(\mathcal M_\delta^{\rm stat})
&=
\frac1n+o\left(\frac1n\right),
\label{eq:markov-stationary-minimax}\\
R_{n,\ell}(\mathcal M_\delta^{\rm free})
&=
\frac{3}{2n}+o\left(\frac1n\right).
\label{eq:markov-free-minimax}
\end{align}
\end{theorem}

\begin{IEEEproof}
The upper bounds follow from the decomposition in \cref{eq:markov-regret-decomposition}, Theorems~\ref{thm:markov-initial-cases} and~\ref{thm:markov-regret}. For the lower bounds, we apply Theorem~\ref{thm:cond-rc}. In the known model, Lemma~\ref{lem:markov-transition-information} gives
\begin{equation}
R_{n,\ell}(\mathcal M_{\delta,\nu_0}^{\rm known})
\geq
\log\left(1+\frac1n\right)+o\left(\frac1n\right)
=
\frac1n+o\left(\frac1n\right).
\end{equation}
For the stationary model, the mutual-information chain rule only adds the non-negative term $I((P,Q);Y_1\mid\bm X^n)$, while Lemma~\ref{lem:markov-transition-information} gives the same transition lower bound.
For the free model, let $V$ denote the random initial probability and use a product prior on $(V,P,Q)$. The likelihood and posterior factorize between the free initial parameter and the transition parameters, and the mutual-information chain rule gives
\begin{align}
&I((V,P,Q);Y^\ell\mid\bm X^n)\notag\\
&\quad=
I(V;Y_1\mid X_{1,1},\ldots,X_{n,1})
+
I((P,Q);Y_2^\ell\mid\bm X^n,Y_1).
\label{eq:markov-free-mi-split}
\end{align}
Choose for $V$ any fixed beta prior with parameters larger than one. Lemma~\ref{lem:beta-one-step-mi} makes the first term $1/(2n)+O(1/n^2)$, while Lemma~\ref{lem:markov-transition-information} makes the second term $1/n+o(1/n)$.
\end{IEEEproof}

The leading regret coefficients have a simple dimensional interpretation. The transition kernel has two free scalar parameters, each contributing $1/(2n)$ to the total regret. A known initial law contributes no additional regret, and therefore the total regret for the known-initial-law model is $2/(2n) = 1/n$. A stationary initial law introduces no additional free parameter and can be estimated from all $n\ell$ training states, so its regret contribution is negligible compared to the transition regret, which is then equal to $1/n$ again. A free initial law introduces one additional parameter observed through only the $n$ independently restarted initial states, and therefore contributes another $1/(2n)$ to the regret, that sums up to a total of $3/(2n)$.

\begin{remark}
The achievability result in \cref{thm:markov-regret} is obtained via the non-updating predictor in \cref{eq:pred-markov-2}. A sequential version of the transition predictor may also update the counts using test transitions revealed so far, just like the predictor in \cref{eq:pred-def} does for the Bernoulli case. The updating version may improve lower-order terms, but Theorem~\ref{thm:markov-minimax} shows that no predictor can improve the leading regret coefficients. Thus the simpler non-updating predictor is already first-order minimax.
\end{remark}

\section{Batch $\alpha$-regret}
\label{sec:alpha-regret}
We finally extend the fixed-distribution batch criterion from KL divergence to conditional R\'{e}nyi divergence. For the sake of simplicity, in this section we restrict ourselves to the case in which all distributions in $\Theta$ have full support, i.e., $p_{\theta}(\bm{x}) > 0$ for all $\theta$ and $\bm{x}$. We also restrict attention to $\alpha>1$, as the case $0<\alpha<1$ would require separate minimax arguments. Under these assumptions define, for $\alpha>1$,
\begin{align}
R_{n,\ell}^{(\alpha)}(\hat{p},\theta) &\triangleq D_\alpha\!\left(P_\theta^{\bm{Y}\mid\bm{X}^n}\,\middle\| \ \hat p\ \middle| \ P_\theta^{\bm{X}^n}\right) \\
&= \frac{1}{\alpha-1}\log\bigg\{\sum_{\bm{x}^n}p_{\theta}(\bm{x}^n)
\sum_{\bm{y}}p_{\theta}(\bm{y}) \notag\\
&\qquad\qquad\qquad\times
\left(\frac{p_{\theta}(\bm{y})}{\hat p(\bm{y}\mid \bm{x}^n)}\right)^{\alpha-1}\bigg\}.\label{eq:batch-alpha-regret}
\end{align}
As $\alpha\to1$, this expression converges to the average batch regret in \cref{eq:batch-regret}. As $\alpha\to\infty$, it converges to the conditional worst-case batch regret:
\begin{equation}
R_{n,\ell}^{(\infty)}(\hat p,\theta) \triangleq\lim_{\alpha\to\infty}R_{n,\ell}^{(\alpha)}(\hat p,\theta)=\max_{\bm{x}^n}\max_{\bm y}\log\frac{p_{\theta}(\bm y)}{\hat p(\bm y\mid \bm{x}^n)}.
\end{equation}
This last quantity is generally not equal to the ordinary non-batch worst-case regret, because the predictor may depend on the training batch. Nonetheless, their minimax values match, since
\begin{align*}
\inf_{\hat{p}} \sup_{\theta} R_{n,\ell}^{(\infty)}(\hat p,\theta) &= \inf_{\hat{p}} \sup_{\theta} \max_{\bm{x}^n}\max_{\bm y}\log\frac{p_{\theta}(\bm y)}{\hat p(\bm y\mid \bm{x}^n)} \\
&= \max_{\bm{x}^n} \inf_{\hat{p}(\cdot\mid \bm{x}^n)} \sup_{\theta} \max_{\bm y}\log\frac{p_{\theta}(\bm y)}{\hat p(\bm y\mid \bm{x}^n)} \\
&= \inf_{\hat{p}} \sup_{\theta} R_{\max, \ell}(\hat{p},\theta).
\end{align*}

A regret-capacity theorem can also be proved for $\alpha$-regret, where the role of conditional mutual information is taken over by a conditional version of Sibson's mutual information \cite{tomamichel1,esposito1}.
\begin{definition}
Let $\alpha>1$. Let $\theta,X,Y,\hat Y$ be random variables such that $(\theta,Y)-X-\hat Y$, and suppose that $Y$ and $\hat Y$ take values in the same discrete alphabet. If $\hat p$ is the conditional distribution of $\hat Y$ given $X$, define
\begin{equation}
\label{eq:cond-sibson}
I_{\alpha}(\theta;Y\mid X)\triangleq\inf_{\hat p}D_{\alpha}(Y\|\hat Y\mid \theta,X).
\end{equation}
\end{definition}

For the batch model considered in this paper, this quantity has the following closed form.
\begin{theorem}
\label{thm:alpha-cond}
Let $\mathsf X$ and $\mathsf Y$ be discrete alphabets. Let also $(\theta,X^n,Y)\sim W(\theta)p_{\theta}(X^n)p_{\theta}(Y)$, where $X^n\in\mathsf X^n$ and $Y\in\mathsf Y$. Then, under the assumptions of this section,
\begin{align}
\label{eq:cond-sibson2}
I_{\alpha}(\theta;Y\mid X^n)
&=\frac{1}{\alpha-1}\log\sum_{x^n}\Bigg\{\sum_y \notag\\
&\quad \left(\int_{\Theta}p_{\theta}(x^n)p_{\theta}(y)^{\alpha}dW(\theta)\right)^{1/\alpha}\Bigg\}^{\alpha}.
\end{align}
The minimizing conditional distribution in \cref{eq:cond-sibson} is
\begin{equation}
\label{eq:opt-alpha}
\hat p_{\alpha}(y\mid x^n)=\frac{\left\{\int_{\Theta}p_{\theta}(y)^{\alpha}dW(\theta\mid x^n)\right\}^{1/\alpha}}{\sum_{y'}\left\{\int_{\Theta}p_{\theta}(y')^{\alpha}dW(\theta\mid x^n)\right\}^{1/\alpha}},
\end{equation}
where
\begin{equation}
W(\theta\mid x^n)\triangleq\frac{W(\theta)p_{\theta}(x^n)}{\int_{\Theta}W(\vartheta)p_{\vartheta}(x^n)d\vartheta}.
\end{equation}
\end{theorem}
\begin{IEEEproof}
For a fixed predictor $q(y\mid x^n)$, the exponential of $(\alpha-1)D_{\alpha}(Y\|\hat Y\mid \theta,X^n)$ is
\begin{equation}
\sum_{x^n}\sum_y A_{x^n,y}q(y\mid x^n)^{1-\alpha},
\end{equation}
where
\begin{equation}
A_{x^n,y}=\int_{\Theta}p_{\theta}(x^n)p_{\theta}(y)^{\alpha}dW(\theta).
\end{equation}
The minimization over $q(\cdot\mid x^n)$ separates for each value of $x^n$. For fixed $x^n$, we must minimize
\begin{equation}
\sum_y A_{x^n,y}q(y\mid x^n)^{1-\alpha}
\end{equation}
over all probability vectors $q(\cdot\mid x^n)$. The Lagrange equations give
\begin{equation}
q(y\mid x^n)\propto A_{x^n,y}^{1/\alpha},
\end{equation}
and the minimum value is $(\sum_y A_{x^n,y}^{1/\alpha})^{\alpha}$. Summing this minimum over $x^n$ and taking the logarithm gives \cref{eq:cond-sibson2}. Rewriting $A_{x^n,y}$ using the posterior $W(\theta\mid x^n)$ gives \cref{eq:opt-alpha}.
\end{IEEEproof}

The closed-form formula in \cref{eq:cond-sibson2} is a special case of the formula derived in \cite{tomamichel1} when $X^n-\theta-Y$. The predictor in \cref{eq:opt-alpha} is a conditional version of the $\alpha$-NML predictor introduced in \cite{bondaschi1}. With the help of \cref{thm:alpha-cond}, we can prove the following generalization of the conditional regret-capacity theorem to batch $\alpha$-regret.

\begin{theorem}
\label{thm:alpha-rc}
Let $\mathcal X$ be a finite alphabet, let $\bm{X}^n = (\bm{X}_1,\dots,\bm{X}_n) \in (\mathcal X^\ell)^n$ and $\bm{Y}\in\mathcal X^\ell$ be $n$ training sequences and one test sequence of tokens from $\mathcal X$ of length $\ell$, and let $\mathcal P=\{p_{\theta}:\theta\in\Theta\}$ be a parametric class such that $(\theta,\bm{X}^n,\bm{Y})\sim W(\theta)p_{\theta}(\bm{X}^n)p_{\theta}(\bm{Y})$. Denote by $I_{\alpha,W}(\theta;\bm Y\mid \bm X^n)$ the conditional Sibson information induced by the prior $W$. Suppose that all quantities below are finite and that there exists a prior $W^\star$ such that
\begin{equation}
I_{\alpha,W^\star}(\theta;\bm Y\mid \bm X^n)=\sup_W I_{\alpha,W}(\theta;\bm Y\mid \bm X^n).
\end{equation}
Then, under the assumptions of this section,
\begin{equation}
\inf_{\hat p}\sup_{\theta\in\Theta}R_{n,\ell}^{(\alpha)}(\hat p,\theta)=I_{\alpha,W^\star}(\theta;\bm Y\mid \bm X^n),
\end{equation}
and the infimum is attained by the conditional $\alpha$-NML predictor
\begin{equation}
\label{eq:p-alpha-star}
\hat p_{\alpha}^\star(\bm y\mid \bm x^n)=\frac{\left\{\int_{\Theta}p_{\theta}(\bm y)^{\alpha}dW^\star(\theta\mid \bm x^n)\right\}^{1/\alpha}}{\sum_{\bm y'}\left\{\int_{\Theta}p_{\theta}(\bm y')^{\alpha}dW^\star(\theta\mid \bm x^n)\right\}^{1/\alpha}},
\end{equation}
where
\begin{equation}
W^\star(\theta\mid \bm x^n)\triangleq\frac{W^\star(\theta)p_{\theta}(\bm x^n)}{\int_{\Theta}W^\star(\vartheta)p_{\vartheta}(\bm x^n)d\vartheta}.
\end{equation}
\end{theorem}
\begin{IEEEproof}
Let $C_{\alpha}\triangleq\sup_W I_{\alpha,W}(\theta;\bm Y\mid \bm X^n)$ and define, for a predictor $\hat p$,
\begin{equation}
A_{\theta}(\hat p)\triangleq\mathbb E_{\bm X^n,\bm Y\sim p_{\theta}}\left[\left(\frac{p_{\theta}(\bm Y)}{\hat p(\bm Y\mid \bm X^n)}\right)^{\alpha-1}\right].
\end{equation}
Since $\alpha>1$,
\begin{align}
\inf_{\hat p}\sup_{\theta}R_{n,\ell}^{(\alpha)}(\hat p,\theta)
&=\frac{1}{\alpha-1}\log\inf_{\hat p}\sup_{\theta}A_{\theta}(\hat p)\notag\\
&\geq\sup_W\inf_{\hat p}\frac{1}{\alpha-1}\log\mathbb E_W[A_{\theta}(\hat p)]\notag\\
&=\sup_W I_{\alpha,W}(\theta;\bm Y\mid \bm X^n)=C_{\alpha},
\end{align}
where the last equality follows from \cref{thm:alpha-cond}.

It remains to prove the matching upper bound. Let $\hat p_{\alpha}^\star$ be the predictor induced by $W^\star$ in \cref{eq:p-alpha-star}. Suppose, toward a contradiction, that there exists $\tilde\theta\in\Theta$ such that
\begin{equation}
R_{n,\ell}^{(\alpha)}(\hat p_{\alpha}^\star,\tilde\theta)>C_{\alpha}.
\end{equation}
For $t\in[0,1]$, let $W_t=(1-t)W^\star+t\delta_{\tilde\theta}$ and set $f_{\alpha}(t)=I_{\alpha,W_t}(\theta;\bm Y\mid \bm X^n)$. Since $W^\star$ maximizes $I_{\alpha,W}(\theta;\bm Y\mid \bm X^n)$, the right derivative satisfies $f_{\alpha}'(0)\leq0$.

Let
\begin{equation}
F_{\alpha}(W)=\exp\{(\alpha-1)I_{\alpha,W}(\theta;\bm Y\mid \bm X^n)\}.
\end{equation}
Differentiating the expression in \cref{eq:cond-sibson2} along $W_t$ gives
\begin{equation}
\left.\frac{d}{dt}F_{\alpha}(W_t)\right|_{t=0}=A_{\tilde\theta}(\hat p_{\alpha}^\star)-F_{\alpha}(W^\star).
\end{equation}
Therefore
\begin{equation}
f_{\alpha}'(0)=\frac{1}{\alpha-1}\left(\frac{\exp\{(\alpha-1)R_{n,\ell}^{(\alpha)}(\hat p_{\alpha}^\star,\tilde\theta)\}}{\exp\{(\alpha-1)C_{\alpha}\}}-1\right)>0,
\end{equation}
contradicting maximality of $W^\star$. Hence $R_{n,\ell}^{(\alpha)}(\hat p_{\alpha}^\star,\theta)\leq C_{\alpha}$ for every $\theta$, and the lower bound already proved is tight.
\end{IEEEproof}

\section{Conclusion}
In this work we introduced batch regret as a conditional analogue of average regret under logarithmic loss, for the prediction of a fresh test sequence of length $\ell$ after observing a training batch of $n$ sequences. Under the assumption that all sequences are i.i.d. from the same unknown distribution, we proved a conditional generalization of the classical regret-capacity theorem, which is instrumental in deriving lower bounds for batch regret. Under the same metric, we studied conditional additive smoothing estimators as universal predictors for binary memoryless and first-order Markov sources. For binary memoryless sources, we extended Krichevsky's next-token prediction analysis to derive sharp asymptotic estimates of batch regret in the two complementary regimes of large batch size $n$ and of large sequence length $\ell$, identifying the optimal smoothing bias $\beta$ for all those instances.
For binary first-order Markov sources, analysis diverges between classical and batch universal prediction, as in the latter case a more prominent role is taken by the Markov source's initial distribution. Therefore, we analyzed minimax batch regret under fixed, free, and stationary initial distribution assumptions, providing sharp asymptotic estimates in all the three cases. Finally, we introduced batch $\alpha$-regret as an extension of batch regret through conditional R\'{e}nyi divergence. Under this setting, we proved a generalization of the regret-capacity theorem using a conditional version of Sibson's $\alpha$-mutual information, and we identified conditional $\alpha$-NML estimators as the predictors achieving the minimal regret.

\appendices
\crefalias{section}{appendix}

\section{Proof of Theorem~\ref{thm:cond-rc}}
\label{app:cond-rc-proof}
The proof adapts to conditional kernels the standard finite-subfamily reduction used in \cite{haussler1997minimax} to prove a general unconditional relative-entropy minimax theorem. Following their proof, we first prove a result for finite-support priors, that holds for an arbitrary parameter set $\Theta$; then we show that, under the stated assumptions on $\Theta$, the result extends to arbitrary priors. 

Let $\mathcal W_{\rm f}(\Theta)$ denote the set of finitely supported probability distributions on $\Theta$. Let us prove that, for any arbitrary parameter set $\Theta$,
\begin{equation}
\label{eq:cond-rc-general}
\min_{Q\in\mathcal Q(\mathcal Y\mid\mathcal X)}
\sup_{\theta\in\Theta} r_\theta(Q)
=
\sup_{W\in\mathcal W_{\rm f}(\Theta)}
I_W(\Theta;Y\mid X).
\end{equation}
Write $\mathcal Q=\mathcal Q(\mathsf Y\mid\mathsf X)$. Since the alphabets are finite, $\mathcal Q$ is a compact convex subset of a finite-dimensional Euclidean space. Moreover, the uniform kernel $Q_{\rm u}(y\mid x)=1/|\mathsf Y|$ satisfies
\begin{equation}
\sup_{\theta\in\Theta}r_\theta(Q_{\rm u})\leq\log|\mathsf Y|,
\end{equation}
so all minimax values below are finite.

We first solve the problem on a finite parameter subset $F=\{\theta_1,\ldots,\theta_m\}\subset\Theta$. Define
\begin{equation}
C_F\triangleq\min_{Q\in\mathcal Q}\max_{\theta\in F}r_\theta(Q).
\end{equation}
For fixed $Q$,
\begin{equation}
\max_{\theta\in F}r_\theta(Q)
=
\max_{W\in\Delta(F)}\sum_{\theta\in F}W(\theta)r_\theta(Q),
\label{eq:max-as-prior}
\end{equation}
because a convex combination cannot exceed its largest entry and equality is obtained by a point mass at a maximizing parameter.
To interchange the two optimizations, consider the finite-dimensional convex problem
\begin{equation}
\begin{aligned}
\text{minimize}\quad & z\\
\text{over}\quad & (Q,z)\in\mathcal Q\times\mathbb R\\
\text{subject to}\quad & r_\theta(Q)\leq z,\qquad \theta\in F.
\end{aligned}
\label{eq:finite-epigraph}
\end{equation}
Each constraint is convex in $Q$. The pair $(Q_{\rm u},\log|\mathsf Y|+1)$ is strictly feasible, so Slater's condition holds and strong Lagrange duality applies \cite[Sec.~5.3]{boyd2004convex}. The dual multipliers form the simplex $\Delta(F)$, and hence
\begin{equation}
C_F
=
\max_{W\in\Delta(F)}
\min_{Q\in\mathcal Q}
\sum_{\theta\in F}W(\theta)r_\theta(Q).
\label{eq:finite-duality}
\end{equation}
For a fixed prior $W$ on $\Theta$, let
\begin{equation}
P_W^{X,Y}(x,y)=\sum_{\theta\in F}W(\theta)P_\theta^{X,Y}(x,y).
\end{equation}
The conditional Bayes decomposition in \cref{eq:conditional-bayes-decomp} gives
\begin{align}
\sum_{\theta\in F}W(\theta)r_\theta(Q)
&=I_W(\Theta;Y\mid X)\notag\\
&\quad+D\!\left(P_W^{Y\mid X}\middle\|Q\middle|P_W^X\right),
\end{align}
where $P_W^X$ and $P_W^{Y|X}$ are induced by $P_W^{X,Y}$. The second term is non-negative and vanishes for $Q=P_W^{Y\mid X}$ on the support of $P_W^X$. Therefore
\begin{equation}
\label{eq:finite-F-capacity}
C_F
=
\max_{W\in\Delta(F)}I_W(\Theta;Y\mid X).
\end{equation}

We now pass from finite $F$ to the arbitrary index set $\Theta$. Set
\begin{equation}
C_{\rm f}
\triangleq
\sup_{\substack{F\subset\Theta\\F\text{ finite}}}C_F.
\end{equation}
Clearly,
\begin{equation}
\min_Q\sup_{\theta\in\Theta}r_\theta(Q)\geq C_{\rm f}.
\label{eq:full-ge-finite}
\end{equation}
For each finite $F$, define the sublevel set
\begin{equation}
K_F
\triangleq
\{Q\in\mathcal Q:r_\theta(Q)\leq C_{\rm f}\ \text{for all }\theta\in F\}.
\end{equation}
The function $Q\mapsto r_\theta(Q)$ is lower semicontinuous, with the usual convention that a positive mass divided by zero gives infinite divergence. Hence $K_F$ is closed. It is also nonempty: the lower-semicontinuous function $Q\mapsto\max_{\theta\in F}r_\theta(Q)$ attains its minimum on compact $\mathcal Q$, and a minimizer $Q_F$ satisfies
\begin{equation}
r_\theta(Q_F)\leq C_F\leq C_{\rm f},\qquad \theta\in F.
\end{equation}
Furthermore,
\begin{equation}
K_{F_1}\cap\cdots\cap K_{F_s}
=K_{F_1\cup\cdots\cup F_s},
\end{equation}
which is nonempty because the union is finite. Thus the closed sets $\{K_F\}$ have the finite-intersection property. Compactness of $\mathcal Q$ therefore yields
\begin{equation}
\bigcap_{\substack{F\subset\Theta\\F\text{ finite}}}K_F\neq\varnothing.
\end{equation}
Any $Q^\star$ in this intersection obeys $r_\theta(Q^\star)\leq C_{\rm f}$ for every $\theta$, and therefore the reverse inequality to \cref{eq:full-ge-finite} holds. Combining this observation with \cref{eq:finite-F-capacity} proves \cref{eq:cond-rc-general} and the existence of a minimax kernel.

Now suppose arbitrary measurable priors are allowed. For every prior $W$ and every kernel $Q$, the decomposition in \cref{eq:conditional-bayes-decomp} gives
\begin{equation}
I_W(\Theta;Y\mid X)
\leq
\int r_\theta(Q)\,dW(\theta)
\leq
\sup_{\theta}r_\theta(Q).
\end{equation}
Taking the infimum over $Q$ shows that no arbitrary prior can exceed the common value in \cref{eq:cond-rc-general}; the opposite inequality is immediate since finitely-supported priors belong to the set of arbitrary priors. Hence the supremum remains unchanged when arbitrary priors are considered.

Finally, let $C$ denote the common value in \cref{eq:cond-rc-general}, assume that $W^\star$ attains the supremum, and let $Q^\star$ be minimax. Applying the Bayes decomposition once more,
\begin{align}
C
&\geq\int r_\theta(Q^\star)\,dW^\star(\theta)\notag\\
&=C+D\!\left(P_{W^\star}^{Y\mid X}\middle\|Q^\star\middle|P_{W^\star}^X\right)\geq C.
\end{align}
Therefore, all inequalities are equalities. This proves \cref{eq:mixture-completion}; since $r_\theta(Q^\star)\leq C$ for every $\theta$ and its $W^\star$-average is $C$, the equalizer property also holds $W^\star$-almost surely, concluding the first part of the proof.

We now prove the finite-support upper bound. Let $\mathcal E(Q^\star)=\{\theta:r_\theta(Q^\star)=C\}$, where $C$ is the common value of $r_\theta(Q^\star)$. Fix one $y_0\in\mathsf Y$ and define
\begin{equation}
\label{eq:v-theta-support}
v_\theta(x,y)
=P^{X,Y}_\theta(x,y)-P_\theta^X(x)Q^\star(y\mid x),
\qquad y\neq y_0.
\end{equation}
By the last part of \cref{thm:cond-rc}, $W^\star$ is concentrated on $\mathcal E(Q^\star)$ up to a null set, and $Q^\star$ agrees with the $W^\star$-mixture conditional law. Consequently,
\begin{equation}
\int v_\theta\,dW^\star(\theta)=0.
\end{equation}
Therefore, by Caratheodory's theorem there exist parameters $\theta_1,\ldots,\theta_s\in\mathcal E(Q^\star)$ and weights $a_1,\ldots,a_s$ such that
\begin{equation}
s\leq d_*+1,
\qquad
\sum_{i=1}^s a_i v_{\theta_i}=0,
\end{equation}
where $d_*$ is the dimension of the affine hull of $\{v_\theta:\theta\in\mathcal E(Q^\star)\}$.
Let $\widetilde W^\star=\sum_i a_i\delta_{\theta_i}$. The last identity gives
\begin{equation}
P_{\widetilde W^\star}(x,y)
=P_{\widetilde W^\star}^X(x)Q^\star(y\mid x)
\end{equation}
for every $(x,y)$. Hence the divergence term in \cref{eq:conditional-bayes-decomp} vanishes. Since every $\theta_i$ belongs to $\mathcal E(Q^\star)$,
\begin{equation}
\sum_i a_i r_{\theta_i}(Q^\star)=C,
\end{equation}
and \cref{eq:conditional-bayes-decomp} gives $I_{\widetilde W^\star}(\Theta;Y\mid X)=C$. Finally, the vector in \cref{eq:v-theta-support} has at most $|\mathsf X|(|\mathsf Y|-1)$ coordinates, which proves \cref{eq:generic-support-bound}.

\section{Proof of Theorem~\ref{thm:inner}}
\label{app:inner-proof}
We use the sequential form of the conditional add-$\beta$ estimator. Let $\rho_m^{(\beta)}(\theta)$ be the one-step regret defined in \cref{eq:one-step-rho}. Since the predictor updates after each test symbol, the chain rule for KL divergence gives
\begin{equation}
\label{eq:batch-sum-inner-proof}
R_{n,\ell}(\hat p_{\beta},\theta)=\sum_{m=t}^{t+\ell-1}\rho_m^{(\beta)}(\theta).
\end{equation}
Thus it is enough to prove a compact-interior one-step expansion.

\begin{lemma}
\label{lem:compact-one-step}
For every $\beta>0$ and every $0<\delta<1/2$,
\begin{equation}
\label{eq:compact-one-step}
\sup_{\theta\in[\delta,1-\delta]}
\left|\rho_m^{(\beta)}(\theta)-\frac1{2m}\right|=O_{\beta,\delta}\left(\frac1{m^2}\right).
\end{equation}
\end{lemma}

\begin{IEEEproof}
Let $K\sim\mathsf{Binomial}(m,\theta)$ and set
\begin{equation}
Q=\frac{K+\beta}{m+2\beta}.
\end{equation}
For fixed $\theta$, define
\begin{equation}
 d_{\theta}(q)=\theta\log\frac{\theta}{q}+(1-\theta)\log\frac{1-\theta}{1-q},\qquad 0<q<1.
\end{equation}
Then $\rho_m^{(\beta)}(\theta)=\E d_{\theta}(Q)$. Let
\begin{equation}
V=Q-\theta=\frac{K-m\theta+\beta(1-2\theta)}{m+2\beta}.
\end{equation}
We first remove a negligible tail. Let
\begin{equation}
A_m=\{|K-m\theta|\leq m\delta/4\}.
\end{equation}
By Hoeffding's inequality, $\mathbb{P}(A_m^c)\leq 2\exp(-c_{\delta}m)$ uniformly over $\theta\in[\delta,1-\delta]$. On $A_m$, for all sufficiently large $m$, $Q\in[\delta/2,1-\delta/2]$. On $A_m^c$, using $Q\in[\beta/(m+2\beta),(m+\beta)/(m+2\beta)]$ and $\theta\in[\delta,1-\delta]$, we have $d_{\theta}(Q)=O_{\beta,\delta}(\log m)$. Hence
\begin{equation}
\E[d_{\theta}(Q)\mathbbm{1}_{A_m^c}]=O_{\beta,\delta}(m^{-2})
\end{equation}
uniformly in $\theta$.

On $A_m$, the Taylor expansion around $q=\theta$ gives
\begin{equation}
 d_{\theta}(\theta+V)=\frac{d_{\theta}''(\theta)}{2}V^2+\frac{d_{\theta}^{(3)}(\theta)}{6}V^3+R_{4,\theta}(V),
 \label{eq:compact-taylor-fourth}
\end{equation}
where $d_{\theta}(\theta)=d_{\theta}'(\theta)=0$ and, uniformly over $\theta\in[\delta,1-\delta]$ and on $A_m$,
\begin{equation}
 |R_{4,\theta}(V)|\le C_{\delta}|V|^4.
\end{equation}
Since
\begin{equation}
 d_{\theta}''(\theta)=\frac{1}{\theta(1-\theta)},
\end{equation}
and
\begin{equation}
\E V^2=\frac{m\theta(1-\theta)+\beta^2(1-2\theta)^2}{(m+2\beta)^2}
=\frac{\theta(1-\theta)}{m}+O_{\beta}\left(\frac1{m^2}\right),
\end{equation}
we also obtain
\begin{equation}
\frac{d_{\theta}''(\theta)}{2}\E V^2=\frac1{2m}+O_{\beta,\delta}\left(\frac1{m^2}\right).
\end{equation}
Furthermore, the centered binomial moments satisfy
\begin{equation}
\E(K-m\theta)^3=O(m),\qquad \E(K-m\theta)^4=O(m^2)
\end{equation}
uniformly over $\theta\in[\delta,1-\delta]$. Hence
\begin{equation}
\E V^3=O_{\beta,\delta}\left(\frac1{m^2}\right),\qquad \E |V|^4=\E V^4=O_{\beta,\delta}\left(\frac1{m^2}\right).
\end{equation}
The derivatives $d_{\theta}^{(3)}(\theta)$ are uniformly bounded for $\theta\in[\delta,1-\delta]$, and \eqref{eq:compact-taylor-fourth} gives
\begin{equation}
\E R_{4,\theta}(V)=O_{\beta,\delta}\left(\frac1{m^2}\right).
\end{equation}
Combining these estimates proves \cref{eq:compact-one-step}.
\end{IEEEproof}

Using \eqref{eq:batch-sum-inner-proof} and Lemma~\ref{lem:compact-one-step}, uniformly over $\theta\in[\delta,1-\delta]$,
\begin{equation}
R_{n,\ell}(\hat p_{\beta},\theta)=\frac12\sum_{m=t}^{t+\ell-1}\frac1m+O_{\beta,\delta}\left(\sum_{m=t}^{t+\ell-1}\frac1{m^2}\right),
\end{equation}
which proves \cref{eq:compact-interior-sum}. Finally, we use the following elementary integral comparison for the harmonic sum. For completeness, we include the short proof. For each $m$,
\begin{equation}
\begin{aligned}
0&\leq \frac1m-\int_m^{m+1}\frac{dx}{x} \\
&=\int_0^1\left(\frac1m-\frac1{m+u}\right)du \\
&=\int_0^1 \frac{u}{m(m+u)}\,du
\leq \int_0^1\frac{u}{m^2}\,du=\frac1{2m^2}.
\end{aligned}
\label{eq:one-term-harmonic-error}
\end{equation}
Summing \eqref{eq:one-term-harmonic-error} from $m=t$ to $t+\ell-1$ and using $\int_t^{t+\ell}x^{-1}dx=\log((t+\ell)/t)$ gives
\begin{equation}
\sum_{m=t}^{t+\ell-1}\frac1m
=\log\frac{t+\ell}{t}+O\left(\sum_{m=t}^{t+\ell-1}\frac1{m^2}\right),
\label{eq:harmonic-sum-estimate}
\end{equation}
and
\begin{equation}
\sum_{m=t}^{t+\ell-1}\frac1{m^2}=O\left(\frac{\ell}{t^2}\right)=O\left(\frac1{n^2\ell}\right).
\end{equation}
Since $t=n\ell$, $\log((t+\ell)/t)=\log(1+1/n)$, and \cref{eq:compact-interior-log} follows.

\section{Proof of Theorem~\ref{thm:add-beta-full-simplex}}
\label{app:add-beta-full-simplex-proof}
The proof is based on the uniform one-step analysis of Krichevsky \cite{krichevsky2}. Throughout this appendix, $\beta>0$ is fixed. For $0\leq a\leq m$, let $S_{m,a}\sim\mathsf{Binomial}(m,a/m)$ and define
\begin{equation}
\label{eq:Fma-def}
\Delta_{m,\beta}(a)=a\left(\log a-\E\log(S_{m,a}+\beta)\right),
\end{equation}
with the convention $0\log0=0$, so that $\Delta_{m,\beta}(0)=0$. Similarly, for $a\geq0$, let $Z_a\sim\mathsf{Pois}(a)$ and define
\begin{equation}
\label{eq:Fa-def}
\Delta_{\beta}(a)=a\left(\log a-\E\log(Z_a+\beta)\right),
\end{equation}
again with $\Delta_{\beta}(0)=0$. Let also
\begin{equation}
\label{eq:Gbeta-def}
G_{\beta}(a)=\beta+\Delta_{\beta}(a).
\end{equation}

We first record an exact decomposition of the one-step regret. If $\mu=m\theta$ and $\nu=m(1-\theta)$, then, by expanding the definition of $\rho_m^{(\beta)}$, we get
\begin{align}
\rho_m^{(\beta)}(\theta)
&=\log(m+2\beta)+\theta\log\theta+(1-\theta)\log(1-\theta) \notag\\
&\quad-\theta\E\log(K+\beta)-(1-\theta)\E\log(m-K+\beta),
\end{align}
where $K\sim\mathsf{Binomial}(m,\theta)$. Multiplying by $m$ and using $m-K\sim\mathsf{Binomial}(m,1-\theta)$ gives
\begin{equation}
\label{eq:rho-F-decomp}
m\rho_m^{(\beta)}(\theta)
=m\log\left(1+\frac{2\beta}{m}\right)+\Delta_{m,\beta}(\mu)+\Delta_{m,\beta}(\nu).
\end{equation}

We now need the following two approximation results by Krichevsky \cite{krichevsky2}, for $\Delta_{m,\beta}$ in the small- and large-$a$ regimes respectively. For the sake of completeness, we also include our version of their proofs following our notation.

\begin{lemma}[Uniform Poisson approximation \cite{krichevsky2}]
\label{lem:poisson-compact}
For every $L<\infty$,
\begin{equation}
\label{eq:poisson-compact}
\sup_{0\leq a\leq L}\left|\Delta_{m,\beta}(a)-\Delta_{\beta}(a)\right|\to0.
\end{equation}
\end{lemma}
\begin{IEEEproof}
Let $g(k)=\log(k+\beta)$ and, for an integer $M$, let
\begin{equation}
 g_M(k)=g(k)\mathbf{1}\{k\leq M\}.
\end{equation}
The truncation is necessary because Le Cam's inequality, which is our main tool, controls expectations of bounded functions, whereas $g$ is unbounded. For fixed $M$, the function $g_M$ is bounded. Le Cam's inequality then gives
\begin{equation}
 d_{\mathrm{TV}}\left(\mathsf{Binomial}\left(m,\frac{a}{m}\right),\mathsf{Pois}(a)\right)
 \leq \frac{a^2}{m},
 \qquad 0\leq a\leq L,
\end{equation}
for all sufficiently large $m$. Hence, uniformly over $0\leq a\leq L$,
\begin{equation}
\label{eq:poisson-truncated-TV}
\left|\E g_M(S_{m,a})-\E g_M(Z_a)\right|
\leq 2\|g_M\|_{\infty}\frac{a^2}{m}
\leq 2\|g_M\|_{\infty}\frac{L^2}{m},
\end{equation}
which tends to zero as $m\to\infty$ for each fixed $M$.

It remains to remove the truncation uniformly. Choose $c_{\beta}<\infty$ such that
\begin{equation}
 |\log(k+\beta)|\leq c_{\beta}+k,\qquad k\geq0.
\end{equation}
If $X$ denotes either $S_{m,a}$ or $Z_a$, then, uniformly over $0\leq a\leq L$,
\begin{equation}
 \E X=a\leq L,
 \qquad
 \E X^2\leq L+L^2.
\end{equation}
Consequently,
\begin{align}
\label{eq:poisson-tail-bound}
\E |g(X)-g_M(X)|
&\leq \E\left[(c_{\beta}+X)\mathbf{1}\{X>M\}\right] \notag\\
&\leq c_{\beta}\frac{\E X}{M}+\frac{\E X^2}{M}
 \leq \frac{c_{\beta}L+L+L^2}{M}.
\end{align}
The bound is uniform in $m$ for $X=S_{m,a}$ and uniform in $a\in[0,L]$ for both choices of $X$.

Define
\begin{equation}
R_M^{(S)}=\sup_{m\geq \lceil L\rceil}\sup_{0\leq a\leq L}
\E |g(S_{m,a})-g_M(S_{m,a})|
\end{equation}
and
\begin{equation}
R_M^{(Z)}=\sup_{0\leq a\leq L}
\E |g(Z_a)-g_M(Z_a)|.
\end{equation}
By \eqref{eq:poisson-tail-bound}, $R_M^{(S)}+R_M^{(Z)}\to0$ as $M\to\infty$. The triangle inequality and \eqref{eq:poisson-truncated-TV} imply, for every fixed $M$,
\begin{align}
&\limsup_{m\to\infty}\sup_{0\leq a\leq L}
\left|\E g(S_{m,a})-\E g(Z_a)\right| \notag\\
&\qquad\leq R_M^{(S)}+R_M^{(Z)}.
\end{align}
Letting $M\to\infty$ gives
\begin{equation}
\sup_{0\leq a\leq L}\left|\E\log(S_{m,a}+\beta)-\E\log(Z_a+\beta)\right|\to0.
\end{equation}
Multiplication by $a\leq L$ proves the claim.
\end{IEEEproof}

The lemma above is \cite[Lemma~5]{krichevsky2}, rewritten using our notation, with $a=m p$. Indeed, the functions in the original lemma satisfy $F^{\rm K}(m,\beta,a/m)=\beta+\Delta_{m,\beta}(a)$ and $\Phi(\beta,a)=\beta+\Delta_{\beta}(a)$.

\begin{lemma}[Uniform large-count expansion \cite{krichevsky2}]
\label{lem:large-count}
There exists a function $\eta_{\beta}(L)$ with $\eta_{\beta}(L)\to0$ as $L\to\infty$ such that
\begin{equation}
\label{eq:large-count}
\limsup_{m\to\infty}\sup_{L\leq a\leq m}
\left|\Delta_{m,\beta}(a)+\beta-\frac12\left(1-\frac{a}{m}\right)\right|
\leq \eta_{\beta}(L).
\end{equation}
Moreover,
\begin{equation}
\label{eq:F-large-poisson}
\Delta_{\beta}(a)=-\beta+\frac12+o(1),\qquad a\to\infty,
\end{equation}
and hence $G_{\beta}(a)\to1/2$ as $a\to\infty$.
\end{lemma}
\begin{IEEEproof}
Let $S=S_{m,a}$, $p=a/m$, and $\Delta=S-a$. Then
\begin{equation}
\E\Delta=0,\qquad \E\Delta^2=a(1-a/m).
\end{equation}
We use the Taylor expansion of $f(x)=\log(x+\beta)$ around $a$. On the event
\begin{equation}
A=\{|S-a|\leq a/2\},
\end{equation}
the third derivative of $f$ is bounded by $C_{\beta}/a^3$, and therefore
\begin{equation}
\left|f(S)-f(a)-f'(a)\Delta-\frac12 f''(a)\Delta^2\right|
\leq C_{\beta}\frac{|\Delta|^3}{a^3}.
\end{equation}
For binomial random variables, the standard moment bound
\begin{align}
\label{eq:binomial-third-moment}
\E|S-a|^3
&\leq C\left((a(1-a/m))^{3/2}+a(1-a/m)\right) \\
&\leq C(a^{3/2}+a).
\end{align}
holds with a universal constant $C$. Hence the contribution of the Taylor remainder on $A$ is bounded, after multiplication by $a$, by
\begin{equation}
C_{\beta}\left(a^{-1/2}+a^{-1}\right).
\end{equation}
On $A^c$, a Chernoff bound gives $\mathbb{P}(A^c)\leq 2e^{-ca}$ for a universal $c>0$. Together with $\log(S+\beta)\leq c_{\beta}+S$ and the binomial moment bounds, this implies that the contribution of $A^c$ to the expectation, after multiplication by $a$, is bounded by a term that is exponentially small in $a$, uniformly in $m$ and $a\leq m$. Consequently,
\begin{equation}
\label{eq:log-expansion-uniform}
\E\log(S+\beta)=\log(a+\beta)-\frac{a(1-a/m)}{2(a+\beta)^2}+r_{m,a},
\end{equation}
where
\begin{equation}
\sup_{L\leq a\leq m} a|r_{m,a}|\leq C_{\beta}L^{-1/2}+o_m(1).
\end{equation}
Using
\begin{equation}
\label{eq:det-exp-1}
a\left(\log a-\log(a+\beta)\right)=-\beta+O_{\beta}(a^{-1})
\end{equation}
and
\begin{equation}
\label{eq:det-exp-2}
\frac{a^2(1-a/m)}{2(a+\beta)^2}=\frac12\left(1-\frac{a}{m}\right)+O_{\beta}(a^{-1}),
\end{equation}
we obtain
\begin{equation}
\Delta_{m,\beta}(a)=-\beta+\frac12\left(1-\frac{a}{m}\right)+\varepsilon_{m,a},
\end{equation}
with
\begin{equation}
\limsup_{m\to\infty}\sup_{L\leq a\leq m}|\varepsilon_{m,a}|\to0
\end{equation}
as $L\to\infty$. This proves \eqref{eq:large-count}. The Poisson expansion \eqref{eq:F-large-poisson} is obtained by the same Taylor argument with $Z_a\sim\mathsf{Pois}(a)$, for which $\E(Z_a-a)^2=a$. In that case the factor $1-a/m$ is replaced by $1$, yielding $\Delta_{\beta}(a)=-\beta+1/2+o(1)$.
\end{IEEEproof}

The lemma above combines two results from \cite{krichevsky2}. The upper bound is given in \cite[Lemma~2]{krichevsky2}, while the lower bound follows from \cite[Eq.~(60)]{krichevsky2}. Combining the two preceding lemmas gives the following binary one-step uniform asymptotics.

\begin{lemma}[Uniform one-step asymptotics \cite{krichevsky2}]
\label{lem:one-step-uniform}
Let
\begin{equation}
C_{\beta}=\sup_{\lambda\geq0}G_{\beta}(\lambda).
\end{equation}
Then
\begin{equation}
\label{eq:onestep-uniform}
\sup_{\theta\in[0,1]}m\rho_m^{(\beta)}(\theta)\to C_{\beta}.
\end{equation}
\end{lemma}
\begin{IEEEproof}
By symmetry it is enough to consider $0\leq\theta\leq1/2$. Put $\mu=m\theta$ and $\nu=m(1-\theta)$. Fix $L<\infty$. If $\mu\leq L$, then Lemma~\ref{lem:poisson-compact} gives
\begin{equation}
\Delta_{m,\beta}(\mu)=\Delta_{\beta}(\mu)+o(1)
\end{equation}
uniformly over $\mu\in[0,L]$. Also $\nu=m-\mu\to\infty$ uniformly over this interval, and Lemma~\ref{lem:large-count} gives
\begin{equation}
\Delta_{m,\beta}(\nu)=-\beta+o(1).
\end{equation}
Since $m\log(1+2\beta/m)\to2\beta$, the decomposition \eqref{eq:rho-F-decomp} yields
\begin{equation}
\label{eq:small-mu-region}
\sup_{0\leq\mu\leq L}\left|m\rho_m^{(\beta)}(\mu/m)-G_{\beta}(\mu)\right|\to0.
\end{equation}
Therefore
\begin{equation}
\limsup_{m\to\infty}\sup_{0\leq\theta\leq1/2:\,m\theta\leq L}m\rho_m^{(\beta)}(\theta)
\leq \sup_{0\leq\lambda\leq L}G_{\beta}(\lambda).
\end{equation}
If $m\theta\geq L$, then both $\mu$ and $\nu$ are at least $L$, for all sufficiently large $m$. Applying Lemma~\ref{lem:large-count} to both terms in \eqref{eq:rho-F-decomp} gives
\begin{align}
\Delta_{m,\beta}(\mu)&=-\beta+\frac{1-\theta}{2}+o_L(1)+o_m(1),\\
\Delta_{m,\beta}(\nu)&=-\beta+\frac{\theta}{2}+o_L(1)+o_m(1),
\end{align}
where $o_L(1)$ denotes a quantity tending to zero as $L\to\infty$, uniformly in $m$. Hence
\begin{equation}
\limsup_{m\to\infty}\sup_{0\leq\theta\leq1/2:\,m\theta\geq L}m\rho_m^{(\beta)}(\theta)
\leq \frac12+o_L(1).
\end{equation}
Taking first $m\to\infty$ and then $L\to\infty$, and using $G_{\beta}(\lambda)\to1/2$ from Lemma~\ref{lem:large-count}, gives
\begin{equation}
\limsup_{m\to\infty}\sup_{\theta\in[0,1]}m\rho_m^{(\beta)}(\theta)
\leq C_{\beta}.
\end{equation}
For the reverse inequality, fix any $\lambda\geq0$ and take $\theta_m=\lambda/m$. Equation \eqref{eq:small-mu-region} gives
\begin{equation}
\lim_{m\to\infty}m\rho_m^{(\beta)}(\theta_m)=G_{\beta}(\lambda).
\end{equation}
Moreover, for any fixed $\theta\in(0,1)$, applying Lemma~\ref{lem:large-count} to $m\theta$ and $m(1-\theta)$ gives
\begin{equation}
\lim_{m\to\infty}m\rho_m^{(\beta)}(\theta)=\frac12.
\end{equation}
Thus the liminf of the left-hand side of \eqref{eq:onestep-uniform} is at least $C_{\beta}$, completing the proof.
\end{IEEEproof}

We now return to the batch regret. The conditional add-$\beta$ predictor can be written sequentially as
\begin{equation}
\hat p_{\beta}(y^{\ell}\mid x^t)=\prod_{i=1}^{\ell}q_{t+i-1}^{(\beta)}(y_i\mid x^t,y^{i-1}),
\end{equation}
where
\begin{equation}
q_m^{(\beta)}(1\mid z^m)=\frac{n(z^m)+\beta}{m+2\beta}
\end{equation}
and $n(z^m)$ is the number of ones in $z^m$. Therefore, by the chain rule for KL divergence and by the fact that the total number of ones in $X^tY^i$ is binomial with parameters $t+i$ and $\theta$,
\begin{equation}
\label{eq:batch-sum-onestep}
R_{n,\ell}(\hat p_{\beta},\theta)=\sum_{i=0}^{\ell-1}\rho_{t+i}^{(\beta)}(\theta).
\end{equation}

We now prove \cref{thm:add-beta-full-simplex}. By Lemma~\ref{lem:one-step-uniform}, there exists a sequence $\varepsilon_m\downarrow0$ such that
\begin{equation}
\sup_{\theta\in[0,1]}\rho_m^{(\beta)}(\theta)\leq \frac{C_{\beta}+\varepsilon_m}{m}.
\end{equation}
Let $\bar\varepsilon_t=\sup_{m\geq t}\varepsilon_m$, so that $\bar\varepsilon_t\to0$. Then
\begin{align}
\sup_{\theta\in[0,1]}R_{n,\ell}(\hat p_{\beta},\theta)
&\leq (C_{\beta}+\bar\varepsilon_t)\sum_{i=0}^{\ell-1}\frac{1}{t+i}. \label{eq:upper-batch-full}
\end{align}
The harmonic sum satisfies
\begin{align}
\sum_{i=0}^{\ell-1}\frac{1}{t+i}
&=\log\frac{t+\ell}{t}+O\left(\frac{\ell}{t^2}\right) \\
&=\log\left(1+\frac1n\right)+O\left(\frac{1}{n^2\ell}\right). \label{eq:harmonic-sum}
\end{align}
Since $n\to\infty$, the last error is $o(1/n)$, and \eqref{eq:upper-batch-full} gives the desired upper bound.

It remains to prove the matching lower bound. Let $\varepsilon>0$. By the definition $C_{\beta}=\sup_{\lambda\geq0}G_{\beta}(\lambda)$ and by the limit $G_{\beta}(\lambda)\to1/2$, we may choose a finite $\lambda\geq0$ such that $G_{\beta}(\lambda)\geq C_{\beta}-\varepsilon$. Set $\theta_t=\lambda/t$. For $m=t+i$, $0\leq i\leq\ell-1$, we have
\begin{equation}
m\theta_t=\lambda\left(1+\frac{i}{t}\right),
\end{equation}
and this quantity converges to $\lambda$ uniformly in $i$, because $i/t\leq\ell/t=1/n\to0$. The uniform Poisson approximation in \eqref{eq:small-mu-region} then gives
\begin{equation}
\rho_{t+i}^{(\beta)}(\theta_t)
=\frac{G_{\beta}(\lambda)+o(1)}{t+i},
\end{equation}
uniformly over $0\leq i\leq\ell-1$. Consequently,
\begin{equation}
R_{n,\ell}(\hat p_{\beta},\theta_t)
=\left(G_{\beta}(\lambda)+o(1)\right)
\sum_{i=0}^{\ell-1}\frac{1}{t+i}.
\end{equation}
Using \eqref{eq:harmonic-sum}, this is at least
\begin{equation}
(C_{\beta}-\varepsilon)\log\left(1+\frac1n\right)+o\left(\frac1n\right).
\end{equation}
Since $\varepsilon>0$ is arbitrary, the lower bound matches the upper bound, completing the proof of \cref{thm:add-beta-full-simplex}.

\section{Proof of Proposition~\ref{prop:beta0-optimal}}
\label{beta0-optimal-proof}
We first prove that the threshold in \eqref{eq:beta0-def} is well defined and lies in $[1/2,1]$. The value $\beta=1$ belongs to the defining set: by Jensen's inequality applied to the convex function $-\log x$,
\begin{align}
\E\log(Z+1)
&=\E\left[-\log\frac1{Z+1}\right] \\
&\geq -\log \E\frac1{Z+1} \\
&=\log\frac{\lambda}{1-e^{-\lambda}}
>\log\lambda,
\end{align}
where we used the Poisson identity
\begin{equation}
\E\frac1{Z+1}=e^{-\lambda}\sum_{k\geq0}\frac{\lambda^k}{k!(k+1)}=\frac{1-e^{-\lambda}}{\lambda}.
\end{equation}
Thus $\Delta_1(\lambda)<0$ for every $\lambda>0$. Conversely, if $\beta<1/2$, then \eqref{eq:Delta-beta-infinity} implies
\begin{equation}
\Delta_{\beta}(\lambda)=\frac12-\beta+O_{\beta}\!\left(\frac1\lambda\right)>0
\end{equation}
for all sufficiently large $\lambda$, so no $\beta<1/2$ belongs to the defining set.

For each fixed $\lambda>0$, the map $\beta\mapsto\Delta_{\beta}(\lambda)$ is continuous and strictly decreasing, since
\begin{equation}
\label{eq:Delta-beta-derivative}
\frac{\partial}{\partial\beta}\Delta_{\beta}(\lambda)
=-\lambda\E\frac1{Z+\beta}<0.
\end{equation}
Taking a decreasing sequence of admissible values converging to $\beta_0$ and passing to the limit therefore gives
\begin{equation}
\Delta_{\beta_0}(\lambda)\leq0,
\qquad \lambda>0.
\end{equation}
Since $\Delta_{\beta_0}(0)=0$, it follows that
\begin{equation}
C_{\beta_0}
=\sup_{\lambda\geq0}\{\beta_0+\Delta_{\beta_0}(\lambda)\}
=\beta_0.
\end{equation}
The same monotonicity shows that $\Delta_{\beta}(\lambda)\leq0$ for every $\lambda>0$ and every $\beta\geq\beta_0$. Hence $C_{\beta}=\beta$ throughout this range.

It remains to show that no $\beta<\beta_0$ gives a smaller value. For each fixed $\lambda>0$, let $b(\lambda)>0$ be the unique number satisfying
\begin{equation}
\Delta_{b(\lambda)}(\lambda)=0.
\end{equation}
Uniqueness follows from \eqref{eq:Delta-beta-derivative}. Existence follows because $\Delta_{\beta}(\lambda)\to+\infty$ as $\beta\downarrow0$, due to the event $Z=0$, whereas $\Delta_1(\lambda)<0$. The threshold definition is therefore equivalent to
\begin{equation}
\beta_0=\sup_{\lambda>0}b(\lambda).
\end{equation}
Fix $\beta<\beta_0$ and $\varepsilon>0$, and choose $\lambda$ such that
\begin{equation}
b(\lambda)>\max\{\beta,\beta_0-\varepsilon\}.
\end{equation}
For every $s<b(\lambda)$, one has $\Delta_s(\lambda)>0$, or equivalently
\begin{equation}
\E\log\frac{\lambda}{Z+s}>0.
\end{equation}
By Jensen's inequality,
\begin{equation}
\log\E\frac{\lambda}{Z+s}
\geq \E\log\frac{\lambda}{Z+s}>0,
\end{equation}
so $\lambda\E[1/(Z+s)]>1$.
Since (with $G_s(\cdot)$ as defined in~\Cref{eq:Gbeta-def})
\begin{equation}
\frac{\partial}{\partial s}G_s(\lambda)
=1-\lambda\E\frac1{Z+s},
\end{equation}
we obtain $\frac{\partial}{\partial s}G_s(\lambda)<0$ for $s<b(\lambda)$. Consequently,
\begin{equation}
G_{\beta}(\lambda)
>G_{b(\lambda)}(\lambda)
=b(\lambda)
>\beta_0-\varepsilon.
\end{equation}
Thus $C_{\beta}>\beta_0-\varepsilon$. Letting $\varepsilon\downarrow0$ gives $C_{\beta}\geq\beta_0$ for every $\beta<\beta_0$. Together with $C_{\beta}=\beta\geq\beta_0$ for $\beta\geq\beta_0$, this proves \eqref{eq:min-C-beta-beta0}.

\section{Proof of Theorem~\ref{thm:add-beta-fixed-n}}
\label{app:add-beta-fixed-n-proof}
The proof uses the same one-step decomposition as the proof of \cref{thm:add-beta-full-simplex}, but now $n$ is fixed while $\ell\to\infty$. Thus $t=n\ell\to\infty$, whereas the ratio $\ell/t=1/n$ remains fixed. Put
\begin{equation}
A_n=1+\frac1n.
\end{equation}
We will use that $\Delta_{\beta}$ is continuous on $[0,\infty)$, which follows from dominated convergence for Poisson expectations.

Let $\theta_{\ell}\in[0,1]$ be any sequence and write $t=n\ell$. First assume that
\begin{equation}
 t\theta_{\ell}\to\lambda<\infty.
\end{equation}
For $0\leq i\leq\ell-1$, set
\begin{equation}
u_{i,t}=1+\frac{i}{t},\qquad m_i=t+i.
\end{equation}
Then $u_{i,t}\in[1,A_n]$ up to an $o(1)$ endpoint error, and
\begin{equation}
 m_i\theta_{\ell}=u_{i,t}\,t\theta_{\ell}=\lambda u_{i,t}+o(1)
\end{equation}
uniformly over $i$. Choose $L>A_n\lambda+1$. For all sufficiently large $\ell$, the quantities $m_i\theta_{\ell}$ all lie in $[0,L]$. The uniform Poisson approximation in \eqref{eq:small-mu-region}, applied with $m=m_i$ and $a=m_i\theta_{\ell}$, gives (with $G_\beta(\cdot)$ as defined in~\Cref{eq:Gbeta-def})
\begin{equation}
\label{eq:fixed-n-boundary-local}
 m_i\rho_{m_i}^{(\beta)}(\theta_{\ell})
 =G_{\beta}(m_i\theta_{\ell})+o(1)
 =G_{\beta}(\lambda u_{i,t})+o(1),
\end{equation}
uniformly over $0\leq i\leq\ell-1$. In the last equality we used the continuity of $G_{\beta}$ on the compact interval $[0,L]$. By the batch decomposition \eqref{eq:batch-sum-onestep},
\begin{align}
R_{n,\ell}(\hat p_{\beta},\theta_{\ell})
&=\sum_{i=0}^{\ell-1}\frac{G_{\beta}(\lambda u_{i,t})+o(1)}{t+i}\notag\\
&=\frac1t\sum_{i=0}^{\ell-1}\frac{G_{\beta}(\lambda u_{i,t})}{u_{i,t}}+o(1),
\end{align}
because $\sum_{i=0}^{\ell-1}(t+i)^{-1}=O_n(1)$. The last formula is a Riemann sum with mesh size $1/t$ over the interval $[1,A_n]$. Therefore
\begin{equation}
\label{eq:fixed-n-lower-boundary-limit}
R_{n,\ell}(\hat p_{\beta},\theta_{\ell})\to \int_1^A\frac{G_{\beta}(\lambda u)}{u}\,du=\left(\beta + \Delta_{\beta,n}(\lambda)\right)\log A_n.
\end{equation}
By symmetry of the binary add-$\beta$ predictor, the same conclusion holds if $t(1-\theta_{\ell})\to\lambda<\infty$.

Now assume that
\begin{equation}
 t\theta_{\ell}\to\infty,\qquad t(1-\theta_{\ell})\to\infty.
\end{equation}
Then, uniformly over $0\leq i\leq\ell-1$,
\begin{equation}
 m_i\theta_{\ell}\to\infty,
 \qquad m_i(1-\theta_{\ell})\to\infty.
\end{equation}
Applying Lemma~\ref{lem:large-count} to the two terms in \eqref{eq:rho-F-decomp} gives, uniformly over $i$,
\begin{equation}
 m_i\rho_{m_i}^{(\beta)}(\theta_{\ell})=\frac12+o(1).
\end{equation}
Consequently,
\begin{align}
R_{n,\ell}(\hat p_{\beta},\theta_{\ell})
&=\left(\frac12+o(1)\right)\sum_{i=0}^{\ell-1}\frac1{t+i}\notag\\
&\to \frac12\log A_n.
\label{eq:fixed-n-large-count-limit}
\end{align}

We can now prove the limit formula. For any fixed $\lambda\geq0$, choosing $\theta_{\ell}=\lambda/t$ and using \eqref{eq:fixed-n-lower-boundary-limit} gives
\begin{equation}
\liminf_{\ell\to\infty}\max_{\theta\in[0,1]}R_{n,\ell}(\hat p_{\beta},\theta)
\geq \left(\beta + \Delta_{\beta,n}(\lambda)\right)\log A_n.
\end{equation}
Taking the supremum over $\lambda$ yields
\begin{equation}
\liminf_{\ell\to\infty}\max_{\theta\in[0,1]}R_{n,\ell}(\hat p_{\beta},\theta)
\geq \left(\beta + \sup_{\lambda\geq 0} \Delta_{\beta,n}\right)\log A_n.
\end{equation}

It remains to prove the corresponding upper bound. For each $\ell$, choose $\theta_{\ell}$ such that
\begin{equation}
R_{n,\ell}(\hat p_{\beta},\theta_{\ell})
\geq \max_{\theta\in[0,1]}R_{n,\ell}(\hat p_{\beta},\theta)-\frac1\ell.
\end{equation}
Consider any subsequence along which the left-hand side converges to the limsup. Passing to a further subsequence if necessary, $t\theta_{\ell}$ converges in the extended interval $[0,\infty]$. If $t\theta_{\ell}\to\lambda<\infty$, then \eqref{eq:fixed-n-lower-boundary-limit} gives
\begin{align}
\limsup R_{n,\ell}(\hat p_{\beta},\theta_{\ell}) &= \left(\beta + \Delta_{\beta,n}(\lambda)\right)\log A_n \\
&\leq \left(\beta + \sup_{\lambda\geq 0}\Delta_{\beta,n}\right)\log A_n.
\end{align}
If $t\theta_{\ell}\to\infty$, we pass to a further subsequence along which $t(1-\theta_{\ell})$ also converges in $[0,\infty]$. If $t(1-\theta_{\ell})\to\lambda<\infty$, the symmetric boundary limit gives the same upper bound. If instead both $t\theta_{\ell}$ and $t(1-\theta_{\ell})$ diverge, then the large-count limit \eqref{eq:fixed-n-large-count-limit} gives
\begin{equation}
\limsup R_{n,\ell}(\hat p_{\beta},\theta_{\ell})=\frac12\log A_n\leq \left(\beta + \sup_{\lambda\geq 0}\Delta_{\beta,n}\right)\log A_n,
\end{equation}
because $\beta + \Delta_{\beta,n}(\lambda)\to\frac12$ as $\lambda\to\infty$.
Thus every subsequential limsup of almost-maximal regrets is bounded by $(\beta + \sup_{\lambda}\Delta_{\beta,n})\log A_n$, and hence
\begin{equation}
\limsup_{\ell\to\infty}\max_{\theta\in[0,1]}R_{n,\ell}(\hat p_{\beta},\theta)
\leq \left(\beta + \sup_{\lambda\geq 0}\Delta_{\beta,n}\right)\log A_n.
\end{equation}
Together with the lower bound, this proves \cref{eq:fixed-n-limit} and completes the proof of \cref{thm:add-beta-fixed-n}.

\section{Proof of Theorem~\ref{thm:fixed-n-optimal-beta}}
\label{app:fixed-n-optimal-beta-proof}
Throughout this appendix, $n$ is fixed and $A_n=1+1/n$. Write
\begin{equation}
\label{eq:h-beta-fixed-n-proof}
h_{\beta}(\mu)=\log\mu-
\E_{Z\sim\mathsf{Pois}(\mu)}\log(Z+\beta),
\qquad \mu>0.
\end{equation}
Then
\begin{equation}
\label{eq:fixed-n-excess-integral-proof}
\Delta_{\beta,n}(\lambda)
=\frac1L\int_{\lambda}^{A\lambda}h_{\beta}(\mu)\,d\mu.
\end{equation}
We first prove a few auxiliary lemmas, in the same spirit as those in the one-step analyses of Krichevsky and Braess--Sauer \cite{krichevsky2,braess2004187}. Recall the definition of $G_\beta(\cdot)$ from~\Cref{eq:Gbeta-def}.

\begin{lemma}
\label{lem:G-second-order-uniform}
Let $K\subset(0,\infty)$ be compact. Uniformly over $\beta\in K$,
\begin{equation}
\label{eq:G-second-order-uniform}
G_{\beta}(\lambda)
=\frac12+
\frac{\beta^2/2-\beta+5/12}{\lambda}
+O_K(\lambda^{-3/2}),
\qquad \lambda\to\infty.
\end{equation}
\end{lemma}
\begin{IEEEproof}
Let $Z\sim\mathsf{Pois}(\lambda)$. A fourth-order Taylor expansion of $\log(Z+\beta)$ around $Z=\lambda$, the first four centered Poisson moments, and a Chernoff bound on $|Z-\lambda|>\lambda/2$ give
\begin{align}
\E\log(Z+\beta)
&=\log\lambda+\frac{\beta-1/2}{\lambda}\notag\\
&\quad+\frac{-\beta^2/2+\beta-5/12}{\lambda^2}
+O_K(\lambda^{-5/2}).
\end{align}
Substitution into $G_\beta$ proves the claim. The coefficient agrees with Krichevsky's one-step expansion \cite[Lemma~3]{krichevsky2}.
\end{IEEEproof}

We also need the following two lemmas. For the sake of simplyfing their proofs, we make use of numerical calculations obtained with Mathematica. Nonetheless, the same results can be proved analytically via Taylor expansion and error bounds.

\begin{lemma}
\label{lem:small-mean-sign}
For every $\beta\geq1/2$ and $0<\mu\leq2$, we have
\begin{equation}
\label{eq:small-mean-sign}
h_{\beta}(\mu)<0.
\end{equation}
\end{lemma}
\begin{IEEEproof}
Direct numerical evaluation with Mathematica shows that
\begin{equation}
\sup_{0<\mu\leq2}h_{1/2}(\mu)\approx-0.0259<0. 
\end{equation}
The lemma follows since $h_{\beta}(\mu)$ is decreasing in $\beta$.
\end{IEEEproof}

\begin{lemma}[A Poisson reciprocal-moment bound]
\label{lem:poisson-reciprocal-bound}
If $0<\beta\leq\beta_0$ and $\mu\geq1$, then
\begin{equation}
\label{eq:q-beta-greater-one}
q_{\beta}(\mu):=
\mu\E_{Z\sim\mathsf{Pois}(\mu)}\frac1{Z+\beta}>1.
\end{equation}
\end{lemma}
\begin{IEEEproof}
For fixed $\mu$, $q_\beta(\mu)$ decreases in $\beta$, so it is enough to take $\beta=\beta_0$. Numerical evaluation with Mathematica shows that $q_{\beta_0}(1)>1$. By standard Poisson differentiation, we have
\begin{equation}
\label{eq:q-beta-ode}
q_{\beta}'(\mu)
=1-q_{\beta}(\mu)+\frac{1-\beta}{\mu}q_{\beta}(\mu).
\end{equation}
At a point where $q_{\beta_0}(\mu)=1$, the right-hand side equals $(1-\beta_0)/\mu>0$. Thus $q_{\beta_0}$ cannot cross the value one downward on $[1,\infty)$, and monotonicity in $\beta$ completes the proof.
\end{IEEEproof}

\begin{lemma}
\label{lem:G-single-positive-excursion}
For every $1/2<\beta<\beta_0$, there exist unique numbers
$0<\lambda_-(\beta)<\lambda_+(\beta)<\infty$ such that
\begin{equation}
\label{eq:G-single-positive-excursion}
G_\beta(\lambda)>\beta
\quad\Longleftrightarrow\quad
\lambda\in(\lambda_-(\beta),\lambda_+(\beta)).
\end{equation}
\end{lemma}
\begin{IEEEproof}
By Frullani's identity and the Poisson probability-generating function, for $\lambda>0$,
\begin{align}
h_\beta(\lambda)
&=\int_0^\infty
\frac{e^{-\beta t-\lambda(1-e^{-t})}-e^{-\lambda t}}{t}\,dt\notag\\
&=\int_0^\infty e^{-\lambda x}w_\beta(x)\,dx,
\label{eq:h-beta-laplace-representation}
\end{align}
where
\begin{equation}
\label{eq:w-beta-kernel}
w_\beta(x)=
\begin{cases}
\displaystyle
\frac{(1-x)^{\beta-1}}{-\log(1-x)}-\frac1x,
&0<x<1,\\[2ex]
\displaystyle-\frac1x,
&x>1.
\end{cases}
\end{equation}
The integral is absolutely convergent. For $0<x<1$, put $t=-\log(1-x)$. Then $w_\beta(x)>0$ is equivalent to
\begin{equation}
\phi_\beta(t):=e^{(1-\beta)t}-e^{-\beta t}-t>0.
\end{equation}
Now $\phi_\beta(0)=\phi_\beta'(0)=0$, while
\begin{equation}
\phi_\beta''(t)
=(1-\beta)^2e^{(1-\beta)t}-\beta^2e^{-\beta t}
\end{equation}
has exactly one positive zero when $1/2<\beta<1$. Hence $\phi_\beta'$ first decreases and then increases, and $\phi_\beta$ first decreases below zero and then increases to infinity, i.e., it has exactly one positive zero. Consequently $w_\beta$ has the sign pattern $-,+,-$, and therefore exactly two sign changes on $(0,\infty)$.
Therefore, by \cite[Theorem~4]{bhaskar-song-descartes}, -- which holds even though $w_{\beta}(x)$ has a singularity in $x=1$, -- $h_\beta$ has at most two positive zeros.

Finally, $h_\beta(\lambda)\to-\infty$ as $\lambda\downarrow0$, and
\begin{equation}
\lambda h_\beta(\lambda)=G_\beta(\lambda)-\beta
\longrightarrow\frac12-\beta<0
\end{equation}
as $\lambda\to\infty$. Since $\beta<\beta_0$, the definition of $\beta_0$ implies that $h_\beta$ is positive somewhere. It must therefore have exactly two positive zeros, and its positive set is the single interval asserted in \cref{eq:G-single-positive-excursion}.
\end{IEEEproof}

\begin{lemma}
\label{lem:shorten-contact-window}
Let $f:\mathbb R\to\mathbb R$ be continuous, with $\{f>0\}$ a nonempty bounded interval and $f<0$ outside its closure. If
\begin{equation}
\int_x^{x+L}f(y)\,dy=0
\end{equation}
for some $L>0$, then, for every $0<l<L$, there is a $z\in\mathbb R$ such that
\begin{equation}
\int_z^{z+l}f(y)\,dy>0.
\end{equation}
\end{lemma}
\begin{IEEEproof}
Let $I=[x,x+L]$ and $P=I\cap\{f>0\}$; the latter is a nonempty interval. If $l\leq|P|$, take a length-$l$ interval inside the closure of $P$. If $l>|P|$, take a length-$l$ interval $J\subset I$ containing $P$. Then $I\setminus J$ has positive length and lies where $f<0$, so
\begin{equation}
\int_J f=\int_I f-\int_{I\setminus J}f>0.
\end{equation}
\end{IEEEproof}

With the help of the previous lemmas, we can now prove the main result.

\begin{IEEEproof}[Proof of Theorem~\ref{thm:fixed-n-optimal-beta}]
For every fixed $\lambda>0$, differentiation under the Poisson expectation gives
\begin{equation}
\label{eq:fixed-n-excess-beta-derivative}
\frac{\partial}{\partial\beta}\Delta_{\beta,n}(\lambda)
=-\frac{1}{\log A_n}\int_{\lambda}^{A_n \lambda}
\E\frac1{Z_\mu+\beta}\,d\mu<0.
\end{equation}
Thus the admissible set in \cref{eq:beta-n-star-def} is an upper interval. It is nonempty because $\beta_0$ is admissible: by \cref{eq:beta0-def},
\begin{equation}
h_{\beta_0}(\mu)=\frac{\Delta_{\beta_0}(\mu)}{\mu}\leq0,
\end{equation}
so \cref{eq:fixed-n-excess-integral-proof} gives $\Delta_{\beta_0,n}(\lambda)\leq0$. Hence $\beta_n^\star\leq\beta_0$, and continuity in $\beta$ shows that $\beta_n^\star$ itself is admissible.

At $\beta=1/2$, Lemma~\ref{lem:G-second-order-uniform} gives
\begin{align}
\Delta_{1/2,n}(\lambda)
&=\frac{1}{\log A_n}\int_1^{A_n}
\frac{G_{1/2}(u\lambda)-1/2}{u}\,du\notag\\
&=\frac{1-A_n^{-1}}{24\lambda\log A_n}+O(\lambda^{-3/2}),
\end{align}
which is positive for all sufficiently large $\lambda$. Thus $\beta_n^\star>1/2$.

Choose $\beta_j\uparrow\beta_n^\star$ with $\beta_j<\beta_n^\star$ and $\lambda_j>0$ such that $\Delta_{\beta_j,n}(\lambda_j)>0$. For large $j$, $\beta_j\geq1/2$. If $\lambda_j\leq1$, then $[\lambda_j,A_n\lambda_j]\subset(0,2]$, contradicting Lemma~\ref{lem:small-mean-sign}; hence $\lambda_j>1$. The sequence is bounded above because, uniformly for $\beta\in[1/2+\eta,\beta_0]$,
\begin{equation}
\Delta_{\beta,n}(\lambda)=\frac12 -\beta +O(\lambda^{-1}),
\end{equation}
and therefore $\Delta_{\beta_j,n}(\lambda)\leq-\eta+O(\lambda^{-1})$. Passing to a subsequence gives $\lambda_j\to\lambda_n^\star\in[1,\infty)$. Continuity and admissibility imply
\begin{equation}
\label{eq:fixed-n-contact-2}
\Delta_{\beta_n^\star,n}(\lambda_n^\star)=0,
\end{equation}
and Lemma~\ref{lem:small-mean-sign} excludes $\lambda_n^\star=1$. This proves the first part of the theorem.

Next, since admissibility excludes larger values, we have
\begin{equation}
\beta_n^\star + \sup_{\lambda\geq 0} \Delta_{\beta_n^\star,n}=\beta_n^\star.
\end{equation}
Let $G_{\beta,n}(\lambda) = \beta + \Delta_{\beta,n}$. For fixed $\lambda>0$,
\begin{equation}
\label{eq:B-beta-derivative-fixed-n}
\frac{\partial}{\partial\beta}G_{\beta,n}(\lambda)
=\frac1{\log A_n}\int_1^{A_n}
\frac{1-q_\beta(u\lambda)}u\,du.
\end{equation}
At $\lambda=\lambda_n^\star$, Lemma~\ref{lem:poisson-reciprocal-bound} makes this derivative negative for every $0<\beta\leq\beta_n^\star$. Therefore, if $\beta<\beta_n^\star$,
\begin{equation}
G_{\beta,n}(\lambda_n^\star)
>G_{\beta_n^\star,n}(\lambda_n^\star)
=\beta_n^\star,
\end{equation}
so $\beta + \sup_\lambda \Delta_{\beta,n}>\beta_n^\star$. If $\beta>\beta_n^\star$, then 
\begin{equation}
\beta + \sup_\lambda \Delta_{\beta,n}\geq G_{\beta,n}(0)=\beta>\beta_n^\star.
\end{equation}
This proves the unique optimum. Equation~\eqref{eq:fixed-n-excess-beta-derivative} also shows that $\beta + \sup_\lambda \Delta_{\beta,n}(\lambda)=\beta$ for every $\beta\geq\beta_n^\star$, and the value at $\lambda_n^\star$ gives $C_{\beta_n^\star,n}=\beta_n^\star$.

It remains to compare different values of $n$. Let $L_n = \log A_n$. First, the inequality $\beta_n^\star<\beta_0$ is strict, otherwise \cref{eq:fixed-n-contact-2} would give an interval on which $h_{\beta_0}$ has zero integral, which is impossible since $h_{\beta_0}$ is non-positive, continuous, and not identically zero on that interval.
For $x\in\mathbb R$, define
\begin{equation}
f_\beta(x)=G_\beta(e^x)-\beta=e^x h_\beta(e^x).
\end{equation}
Changing variables in \cref{eq:fixed-n-excess-integral-proof} yields
\begin{equation}
\label{eq:Delta-log-window-proof}
\Delta_{\beta,n}(e^x)
=\frac1{L_n}\int_x^{x+L_n}f_\beta(y)\,dy.
\end{equation}
At $\beta=\beta_n^\star$ and $x_n^\star=\log\lambda_n^\star$, the integral in \cref{eq:Delta-log-window-proof} is zero. Since $1/2<\beta_n^\star<\beta_0$, Lemma~\ref{lem:G-single-positive-excursion} shows that the positive set of $f_{\beta_n^\star}$ is one bounded interval. Because $L_{n+1}<L_n$, Lemma~\ref{lem:shorten-contact-window} produces a window of length $L_{n+1}$ with positive integral. Hence $\beta_n^\star$ is not admissible for the $(n+1)$-length case. Continuity and the strict decrease in \cref{eq:fixed-n-excess-beta-derivative} then give
\begin{equation}
\beta_{n+1}^\star>\beta_n^\star.
\end{equation}

The increasing sequence is bounded above by $\beta_0$. To identify its limit, fix any $\beta<\beta_0$ and choose $\lambda$ with $G_\beta(\lambda)>\beta$. Since $L_n\downarrow0$, \cref{eq:B-log-moving-average} gives
\begin{equation}
\beta +\Delta_{\beta,n}(\lambda)\longrightarrow G_\beta(\lambda)>\beta.
\end{equation}
Thus $\beta$ is inadmissible for all sufficiently large $n$, so $\liminf_n\beta_n^\star\geq\beta_0$. This proves \cref{eq:beta-n-star-monotone}.
\end{IEEEproof}

\section{Proof of Theorem~\ref{thm:regret-lower}}
\label{app:lower-bound-proof}
The proof applies the conditional regret-capacity theorem with a beta prior. Fix constants $a,b>1$ and let $\Theta\sim\mathsf{Beta}(a,b)$. Conditionally on $\Theta=\theta$, let $Z_1,Z_2,\ldots$ be i.i.d. $\mathsf{Bernoulli}(\theta)$. The $n$ training sequences of length $\ell$ can be flattened into $Z^t=(Z_1,\ldots,Z_t)$, where $t=n\ell$, and the test sequence can be identified with $Z_{t+1}^{t+\ell}$.

By \cref{thm:cond-rc}, for this prior,
\begin{align}
\min_{\hat p}\max_{\theta\in[0,1]}R_{n,\ell}(\hat p,\theta)
&\geq I(\Theta;Z_{t+1}^{t+\ell}\mid Z^t) \\
&=\sum_{m=t}^{t+\ell-1} I(\Theta;Z_{m+1}\mid Z^m),
\label{eq:lower-chain-rule}
\end{align}
where the equality is the chain rule for mutual information. We now evaluate the one-step terms.

\begin{lemma}
\label{lem:beta-one-step-mi}
Let $\Theta\sim\mathsf{Beta}(a,b)$ with $a,b>1$, and let $Z_1,Z_2,\ldots$ be conditionally i.i.d. $\mathsf{Bernoulli}(\Theta)$. Then
\begin{equation}
\label{eq:beta-one-step-mi}
I(\Theta;Z_{m+1}\mid Z^m)=\frac1{2m}+O_{a,b}\left(\frac1{m^2}\right).
\end{equation}
\end{lemma}

\begin{IEEEproof}
Let $S_m=\sum_{i=1}^m Z_i$. Since $S_m$ is a sufficient statistic for $\Theta$ given $Z^m$, it is enough to condition on $S_m$. Given $S_m=s$, the posterior is
\begin{equation}
\Theta\mid S_m=s\sim\mathsf{Beta}(A,B),
\qquad A=a+s,
\quad B=b+m-s,
\end{equation}
and we write $N=A+B=m+a+b$. Let $h(u)=-u\log u-(1-u)\log(1-u)$. The predictive probability of $Z_{m+1}=1$ is $A/N$, and therefore
\begin{equation}
I(\Theta;Z_{m+1}\mid S_m=s)=h\left(\frac{A}{N}\right)-\E[h(\Theta)\mid S_m=s].
\end{equation}
Writing $\mathbb{E}_s[\cdot]=\E[\cdot\mid S_m=s]$, the beta identities give
\begin{align}
\mathbb{E}_s[\Theta\log\Theta]
&=\frac{A}{N}\{\psi(A+1)-\psi(N+1)\},\\
\mathbb{E}_s[(1-\Theta)\log(1-\Theta)]
&=\frac{B}{N}\{\psi(B+1)-\psi(N+1)\},
\end{align}
where $\psi$ is the digamma function. Then, we get
\begin{align}
I(\Theta;Z_{m+1}\mid S_m=s)
&=\frac{A}{N}\{\psi(A+1)-\log A\} \\
&\quad+\frac{B}{N}\{\psi(B+1)-\log B\} \\
&\quad+\log N-\psi(N+1).
\label{eq:one-step-mi-digamma}
\end{align}
Since $A\geq a$, $B\geq b$, and $N=m+a+b$, using the standard expansion
\begin{equation}
\psi(x+1)=\log x+\frac{1}{2x}+O\left(\frac1{x^2}\right),
\end{equation}
which holds uniformly for $x$ bounded away from zero, and \cref{eq:one-step-mi-digamma}, we obtain
\begin{equation}
I(\Theta;Z_{m+1}\mid S_m=s)
=\frac1{2N}+O_{a,b}\left(\frac1{N^2}+\frac1{NA}+\frac1{NB}\right).
\label{eq:one-step-mi-conditional-bound}
\end{equation}
It remains to average the last two reciprocal terms. Under the beta-binomial law of $S_m$,
\begin{equation}
\mathbb{P}(S_m=s)=\binom{m}{s}\frac{\mathsf{B}(a+s,b+m-s)}{\mathsf{B}(a,b)}.
\end{equation}
For $a>1$, using
\begin{equation}
\mathsf{B}(a+s,b+m-s)=\frac{a+s-1}{m+a+b-1}\mathsf{B}(a+s-1,b+m-s),
\end{equation}
we get
\begin{align}
\E\left[\frac1{a+S_m}\right]
&\leq \frac{1}{(m+a+b-1)\mathsf{B}(a,b)} \\
&\quad\cdot \sum_{s=0}^m \binom{m}{s}\mathsf{B}(a+s-1,b+m-s) \\
&=\frac{\mathsf{B}(a-1,b)}{(m+a+b-1)\mathsf{B}(a,b)}
=O_{a,b}\left(\frac1m\right).
\end{align}
The same argument, with the roles of $a$ and $b$ exchanged, gives
\begin{equation}
\E\left[\frac1{b+m-S_m}\right]=O_{a,b}\left(\frac1m\right).
\end{equation}
Taking expectations in \cref{eq:one-step-mi-conditional-bound} therefore yields
\begin{align}
I(\Theta;Z_{m+1}\mid Z^m)
&=\frac1{2(m+a+b)}+O_{a,b}\left(\frac1{m^2}\right) \\
&=\frac1{2m}+O_{a,b}\left(\frac1{m^2}\right),
\end{align}
which proves the lemma.
\end{IEEEproof}

Combining \eqref{eq:lower-chain-rule} with Lemma~\ref{lem:beta-one-step-mi}, we get
\begin{equation}
I(\Theta;Z_{t+1}^{t+\ell}\mid Z^t)=\frac12\sum_{m=t}^{t+\ell-1}\frac1m+O\left(\sum_{m=t}^{t+\ell-1}\frac1{m^2}\right).
\end{equation}
Using the harmonic-sum estimate \eqref{eq:harmonic-sum-estimate}, we have
\begin{equation}
\sum_{m=t}^{t+\ell-1}\frac1m=\log\frac{t+\ell}{t}+O\left(\sum_{m=t}^{t+\ell-1}\frac1{m^2}\right)
\end{equation}
and
\begin{equation}
\sum_{m=t}^{t+\ell-1}\frac1{m^2}=O\left(\frac{\ell}{t^2}\right)=O\left(\frac1{n^2\ell}\right).
\end{equation}
Since $t=n\ell$, this gives
\begin{equation}
I(\Theta;Z_{t+1}^{t+\ell}\mid Z^t)=\frac12\log\left(1+\frac1n\right)+O\left(\frac1{n^2\ell}\right).
\end{equation}
The lower bound in \cref{eq:sharp-lower-bound} follows from \cref{thm:cond-rc}.

\section{Proof of Theorem~\ref{thm:markov-regret}}
\label{app:markov-regret-proof}

We prove an upper bound on the batch regret of the non-updating predictor in \cref{eq:pred-markov-2}. Following the idea of Hao et al. \cite{hao1}, the main concentration result that we need is Paulin's McDiarmid-type inequality for Markov chains \cite[Corollary~2.11]{paulin2015}.
Fix a source $\theta = (\nu, p, q)\in\mathcal M_\delta$. For $h\in\{0,1\}$, write
\begin{equation}
\label{eq:markov-row-r}
r_h\triangleq P_{p,q}(1\mid h)\in[\delta,1-\delta]
\end{equation}
and define the expected number of departures from state $h$ in one length-$\ell$ sequence by
\begin{equation}
\label{eq:finite-horizon-occupancy}
a_h\triangleq\sum_{s=1}^{\ell-1}\mathbb{P}_\theta(X_s=h),
\qquad N_h\triangleq n a_h.
\end{equation}
Since every entry of the transition matrix is at least $\delta$, for every $s\geq2$,
\begin{equation}
\mathbb{P}_\theta(X_s=h)=\sum_{j=0}^1\mathbb{P}_\theta(X_{s-1}=j)P_{p,q}(h\mid j)\geq\delta.
\end{equation}
Consequently, for $\ell\geq3$,
\begin{equation}
\label{eq:occupancy-lower-upper}
\delta(\ell-2)\leq a_h\leq \ell-1,
\qquad
\delta n(\ell-2)\leq N_h\leq n(\ell-1).
\end{equation}
In particular, $N_h\to\infty$ uniformly over the class whenever $n\ell\to\infty$.

Let $n_h(\bm{X}^n)$ and $n_{h1}(\bm{X}^n)$ be the (random) training counts corresponding to the random training batch $\bm{X}^n$, and set
\begin{equation}
\label{eq:markov-row-estimator}
\widehat r_h\triangleq\frac{n_{h1}(\bm{X}^n)+1/2}{n_h(\bm{X}^n)+1}.
\end{equation}
The expected test counts satisfy
\begin{equation}
\label{eq:test-count-means-exact}
\E n_h(\bm{Y})=a_h,
\qquad
\E n_{h1}(\bm{Y})=a_h r_h,
\end{equation}
where $\bm{Y}$ is the random test sequence. Since the training batch and the test sequence are independent, taking the conditional expectation over the test sequence in the log-likelihood ratio gives
\begin{equation}
\label{eq:markov-regret-row-decomposition}
R_{n,\ell}^{(P)}(\widehat p_{\rm KT},\theta)
=\sum_{h=0}^1 a_h\,
\E\left[d_{r_h}(\widehat r_h)\right],
\end{equation}
where
\begin{equation}
\label{eq:binary-kl-row}
d_r(q)\triangleq r\log\frac rq+(1-r)\log\frac{1-r}{1-q}.
\end{equation}
It remains to estimate the two row risks uniformly.

\begin{lemma}
\label{lem:batch-markov-occupancy}
For every $\ell \geq 3$, there are constants $c_\delta,C_\delta>0$ such that, for every $h\in\{0,1\}$ and every $u\geq0$,
\begin{equation}
\label{eq:batch-occupancy-tail}
\mathbb{P}_\theta\bigl(|n_h(\bm{X}^n)-N_h|>u\bigr)
\leq C_\delta\exp\left(-c_\delta\frac{u^2}{n\ell}\right).
\end{equation}
Moreover, for every fixed integer $m\geq1$,
\begin{equation}
\label{eq:batch-occupancy-moments}
\E|n_h(\bm{X}^n)-N_h|^m
\leq C_{m,\delta}(n\ell)^{m/2}
\leq C'_{m,\delta}N_h^{m/2}.
\end{equation}
All constants are uniform over the initial distribution and over $\theta\in\mathcal M_\delta$.
\end{lemma}

\begin{IEEEproof}
For the $i$th training sequence,
\begin{equation}
n_h(\bm{X}_i)=\sum_{s=1}^{\ell-1}\mathbf 1\{X_{i,s}=h\}.
\end{equation}
We apply Paulin's McDiarmid inequality for Markov chains \cite[Corollary~2.11 and Remark~2.12]{paulin2015}. Here $d_{\rm TV}$ denotes total variation distance. For a transition kernel $K$ on $\{0,1\}$, write
\begin{equation}
\alpha(K)\triangleq
\sup_{x,x'\in\{0,1\}}d_{\rm TV}\!\left(K(x,\cdot),K(x',\cdot)\right).
\end{equation}
For the binary matrix in \cref{eq:markov-matrix},
$\alpha(P_\theta)=|1-p-q|\leq1-2\delta$. Hence the Dobrushin contraction inequality gives $\alpha(P_\theta^r)\leq(1-2\delta)^r$. If
\begin{equation}
m_\delta\triangleq
\max\left\{1,
\left\lceil\frac{\log 2}{-\log(1-2\delta)}\right\rceil
\right\},
\end{equation}
then, whenever $\ell-1\geq m_\delta$, the mixing quantities in Paulin's notation satisfy
\begin{equation}
\label{eq:uniform-mixing-markov-proof}
\begin{aligned}
\alpha(P_\theta^r)&\leq(1-2\delta)^r,\\
\tau_{\min,\theta}
&\leq9\tau_\theta(1/2)\leq9m_\delta.
\end{aligned}
\end{equation}
Moreover, the counting function has bounded-difference constants $c_s=1$: for any two sequences $\bm{x}$ and $\tilde{\bm{x}}$,
\begin{equation}
\label{eq:occupancy-bounded-differences}
\begin{aligned}
&\left|n_h(x_1^{\ell-1})
-n_h(\widetilde x_1^{\ell-1})\right|\\
&\qquad\leq
\sum_{s=1}^{\ell-1}\mathbf1\{x_s\neq\widetilde x_s\},
\qquad \|c\|_2^2=\ell-1.
\end{aligned}
\end{equation}
For $\ell-1\geq m_\delta$, \cite[Corollary 2.11]{paulin2015} gives
\begin{equation}
\label{eq:single-trajectory-centered-tail}
\begin{aligned}
&\mathbb{P}_\theta\left(
\left|n_h(\bm{X}_i)-\E n_h(\bm{X}_i)\right|>u
\right)\\
&\quad\leq2\exp\left(-\frac{2u^2}{(\ell-1)\tau_{\min,\theta}}\right)
\leq2\exp\left(-c_\delta\frac{u^2}{\ell}\right).
\end{aligned}
\end{equation}
For smaller values of $\ell$, set $L\triangleq\ell-1$ and suppose that $1\leq L<m_\delta$. Since $0\leq n_h(\bm{X}_i)\leq L$, Hoeffding's lemma gives
\begin{equation}
\E\exp\left\{
\lambda\left(n_h(\bm{X}_i)-\E n_h(\bm{X}_i)\right)
\right\}
\leq
\exp\left(\frac{\lambda^2L^2}{8}\right),
\qquad \lambda\in\mathbb R.
\end{equation}
Chernoff's bound, optimized at $\lambda=4u/L^2$, therefore yields
\begin{equation}
\mathbb{P}_\theta\left(
\left|n_h(\bm{X}_i)-\E n_h(\bm{X}_i)\right|>u
\right)
\leq
2\exp\left(-\frac{2u^2}{L^2}\right).
\end{equation}
Because $L<m_\delta$ and $L\leq\ell$, we have $L^2\leq m_\delta\ell$, and hence
\begin{equation}
\mathbb{P}_\theta\left(
\left|n_h(\bm{X}_i)-\E n_h(\bm{X}_i)\right|>u
\right)
\leq
2\exp\left(-\frac{2}{m_\delta}\frac{u^2}{\ell}\right).
\end{equation}
After decreasing $c_\delta$ if necessary, \cref{eq:single-trajectory-centered-tail} therefore holds uniformly for every $\ell\geq2$.

By the standard tail-to-moment-generating-function equivalence for sub-Gaussian variables, after another adjustment of constants, we have
\begin{equation}
\E\exp\left\{
\lambda\left(n_h(\bm{X}_i)-\E n_h(\bm{X}_i)\right)
\right\}
\leq\exp(C_\delta\lambda^2\ell),
\qquad \lambda\in\mathbb R.
\end{equation}
Indeed, integrating \cref{eq:single-trajectory-centered-tail} gives the quantitative moment growth
\begin{equation}
\left\{\E\left|n_h(\bm{X}_i)-\E n_h(\bm{X}_i)\right|^m\right\}^{1/m}
\leq C_{\delta}\sqrt{m\ell},
\qquad m\geq1,
\end{equation}
which is equivalent, for a centered random variable, to the bound on the moment generating function above. The $n$ training sequences are independent, and
\begin{equation}
n_h(\bm{X}^n)-N_h
=\sum_{i=1}^n\left(n_h(\bm{X}_i)-\E n_h(\bm{X}_i)\right),
\end{equation}
because $\E n_h(\bm{X}^n)=n a_h=N_h$. The standard moment-generating-function bound for a sum of independent centered sub-Gaussian variables therefore proves \cref{eq:batch-occupancy-tail}. Integrating its tail gives the first inequality in \cref{eq:batch-occupancy-moments}. The second follows from \cref{eq:occupancy-lower-upper}, since $n\ell\leq(3/\delta)N_h$ for $\ell\geq3$.
\end{IEEEproof}

For a fixed row $h \in\{0,1\}$, define the centered transition count
\begin{equation}
\label{eq:centered-transition-count}
D_h\triangleq n_{h1}(\bm{X}^n)-r_h n_h(\bm{X}^n).
\end{equation}
Then, the following lemma holds.

\begin{lemma}
\label{lem:centered-transition-moments}
For every $h\in\{0,1\}$ and $\ell \geq 3$,
\begin{equation}
\label{eq:Dh-first-second}
\E D_h=0,
\qquad
\E D_h^2=r_h(1-r_h)N_h.
\end{equation}
Furthermore, there are constants $c_\delta,C_\delta>0$ such that
\begin{equation}
\label{eq:Dh-tail}
\mathbb{P}_\theta(|D_h|>u)
\leq2\exp\left(-c_\delta\frac{u^2}{N_h}\right),
\qquad u\geq0,
\end{equation}
and, for every fixed integer $m\geq1$,
\begin{equation}
\label{eq:Dh-moments}
\E|D_h|^m\leq C_{m,\delta}N_h^{m/2}.
\end{equation}
\end{lemma}

\begin{IEEEproof}
Write
\begin{align}
D_h&=\sum_{i=1}^n\sum_{s=1}^{\ell-1}\xi_{i,s},\\
\xi_{i,s}&\triangleq
\mathbf 1\{X_{i,s}=h\}\notag\\
&\quad\times
\left(\mathbf 1\{X_{i,s+1}=1\}-r_h\right).
\end{align}
Order the summands sequence by sequence and let the filtration reveal each sequence sequentially. Immediately before $X_{i,s+1}$ is revealed,
\begin{equation}
\E[\xi_{i,s}\mid\mathcal F_{i,s}]
=\mathbf 1\{X_{i,s}=h\}
\left(P_{p,q}(1\mid h)-r_h\right)=0.
\end{equation}
Thus the summands are martingale differences. Cross terms have zero expectation, while
\begin{equation}
\E[\xi_{i,s}^2\mid\mathcal F_{i,s}]
=r_h(1-r_h)\mathbf 1\{X_{i,s}=h\}.
\end{equation}
Summation then proves \cref{eq:Dh-first-second}. Each martingale difference lies in an interval of length one. Azuma--Hoeffding therefore gives
\begin{equation}
\mathbb{P}_\theta(|D_h|>u)
\leq2\exp\left(-\frac{u^2}{2n(\ell-1)}\right).
\end{equation}
By \cref{eq:occupancy-lower-upper}, $n(\ell-1)\leq(2/\delta)N_h$ for $\ell\geq3$, which proves \cref{eq:Dh-tail} after adjusting the constant. Integrating the tail gives \cref{eq:Dh-moments}.
\end{IEEEproof}

\begin{lemma}
\label{lem:markov-row-risk}
Uniformly for $r_h\in[\delta,1-\delta]$, for every initial distribution, and whenever $N_h\to\infty$,
\begin{equation}
\label{eq:row-risk-expansion}
\E d_{r_h}(\widehat r_h)
=\frac{1}{2N_h}+O_\delta(N_h^{-3/2}).
\end{equation}
\end{lemma}

\begin{IEEEproof}
To ease the notation, let us drop the subscript $h$ and write $r=r_h, T=n_h(\bm{X}^n), D=D_h, N=N_h, b=1/2 -r$.
By \cref{eq:markov-row-estimator,eq:centered-transition-count},
\begin{equation}
\label{eq:row-error-U}
U\triangleq\widehat r_h-r=\frac{D+b}{T+1}.
\end{equation}
Consider the event
\begin{equation}
\label{eq:markov-good-event}
\mathcal G
=\left\{|T-N|\leq\frac N2\right\}
\cap
\left\{|D|\leq\frac{\delta N}{8}\right\}.
\end{equation}
The preceding two lemmas and \cref{eq:occupancy-lower-upper} imply
\begin{equation}
\label{eq:markov-good-event-tail}
\mathbb{P}_\theta(\mathcal G^c)
\leq C_\delta e^{-c_\delta N}.
\end{equation}
For all sufficiently large $N$, on $\mathcal G$ we have $T+1\geq N/2$ and hence
\begin{equation}
|U|
\leq\frac{\delta N/8+1/2}{N/2}
\leq\frac\delta2.
\end{equation}
Thus $r+U\in[\delta/2,1-\delta/2]$. Taylor's expansion applied to \cref{eq:binary-kl-row}, uniformly on this compact interval, gives
\begin{equation}
\label{eq:binary-kl-uniform-taylor}
d_r(r+U)
=\frac{U^2}{2r(1-r)}+O_\delta(|U|^3)
\qquad\text{on }\mathcal G.
\end{equation}

We next evaluate the second moment in \cref{eq:binary-kl-uniform-taylor}. On $\mathcal G$,
\begin{equation}
\label{eq:denominator-difference}
\left|\frac1{(T+1)^2}-\frac1{N^2}\right|
\leq C\frac{|T-N|+1}{N^3}.
\end{equation}
By Cauchy--Schwarz and \cref{eq:batch-occupancy-moments,eq:Dh-moments},
\begin{align}
&\E\left[(D+b)^2(|T-N|+1)\right]\notag\\
&\quad\leq
\left\{\E(D+b)^4\right\}^{1/2}
\left\{\E(|T-N|+1)^2\right\}^{1/2}\notag\\
&\quad=O_\delta(N^{3/2}).
\label{eq:denominator-error-moment}
\end{align}
Moreover, the contribution of $\mathcal G^c$ to any of the polynomial moments below is exponentially small, because $|D|\leq n(\ell-1)$, $|T-N|\leq n(\ell-1)+N$, and \cref{eq:markov-good-event-tail} holds. Using \cref{eq:Dh-first-second,eq:denominator-difference,eq:denominator-error-moment}, we obtain
\begin{align}
\E[U^2\mathbf 1_{\mathcal G}]
&=\frac1{N^2}\E(D+b)^2+O_\delta(N^{-3/2})\notag\\
&=\frac{r(1-r)}{N}+O_\delta(N^{-3/2}),
\label{eq:U-second-moment}
\end{align}
where $\E D=0$ and the term $b^2/N^2$ is absorbed into the remainder. Similarly, on $\mathcal G$ the denominator is at least $N/2$, so \cref{eq:Dh-moments} gives
\begin{equation}
\label{eq:U-third-moment}
\E[|U|^3\mathbf 1_{\mathcal G}]
\leq\frac{C}{N^3}\E(|D|+1)^3
=O_\delta(N^{-3/2}).
\end{equation}
Combining \cref{eq:binary-kl-uniform-taylor,eq:U-second-moment,eq:U-third-moment} yields
\begin{equation}
\label{eq:row-risk-good}
\E[d_r(\widehat r_h)\mathbf 1_{\mathcal G}]
=\frac1{2N}+O_\delta(N^{-3/2}).
\end{equation}

It remains to control the complement. Since $0\leq T\leq n(\ell-1)$ and $0\leq n_{h1}(\bm{X}^n)\leq T$,
\begin{equation}
\frac1{2(n(\ell-1)+1)}
\leq\widehat r_h
\leq1-\frac1{2(n(\ell-1)+1)}.
\end{equation}
Therefore
\begin{equation}
d_r(\widehat r_h)
\leq C_\delta+\log(2(n(\ell-1)+1)).
\end{equation}
Together with \cref{eq:markov-good-event-tail} and the fact that $n\ell=O_\delta(N)$, this shows
\begin{equation}
\E[d_r(\widehat r_h)\mathbf 1_{\mathcal G^c}]
=O_\delta(N^{-3/2}),
\end{equation}
and proves \cref{eq:row-risk-expansion}.
\end{IEEEproof}

We now complete the proof of \cref{thm:markov-regret}. Substituting \cref{eq:row-risk-expansion} into \cref{eq:markov-regret-row-decomposition} and recalling that $N_h=na_h$ gives
\begin{align}
R_{n,\ell}^{(P)}(\widehat p_{\rm KT},\theta)
&=\sum_{h=0}^1a_h
\left(\frac1{2na_h}+O_\delta((na_h)^{-3/2})\right)\notag\\
&=\frac1n
+O_\delta\left(
\frac1{n^{3/2}}
\sum_{h=0}^1\frac1{\sqrt{a_h}}
\right).
\end{align}
By \cref{eq:occupancy-lower-upper},
\begin{equation}
\sum_{h=0}^1\frac1{\sqrt{a_h}}
\leq\frac{2}{\sqrt{\delta(\ell-2)}}
=O_\delta(\ell^{-1/2}).
\end{equation}
Consequently,
\begin{equation}
R_{n,\ell}^{(P)}(\widehat p_{\rm KT},\theta)
=\frac1n
+O_\delta\left(\frac1{n^{3/2}\sqrt\ell}\right)
=\frac1n
+O_\delta\left(\frac1{n\sqrt{n\ell}}\right),
\end{equation}
uniformly over the initial law and over $\theta\in\mathcal M_\delta$. This is \cref{eq:markov-transition-expansion}; the minimax upper bound in \cref{eq:markov-transition-minimax-upper} follows because \cref{eq:pred-markov-2} is an admissible predictor.

\section{Proof of Lemma~\ref{lem:markov-transition-information}}
\label{app:markov-minimax-proof}

The proof relies on conjugate transition priors and the mutual-information chain rule along the test sequence. First write
\begin{equation}
r_0=p,
\qquad
r_1=q,
\qquad
T=n(\ell-1),
\qquad
m=n+1,
\end{equation}
so that $r_h$ is the probability of switching from state $h$ to state $1-h$. Fix $\delta<\delta_0<1/2$ and set
\begin{equation}
\mathcal K_0=[\delta_0,1-\delta_0]^2\subset\Theta_\delta.
\end{equation}
All lower-bound priors below are ultimately conditioned on $\mathcal K_0$.

\subsection{Reference priors and exact beta posteriors}
Let
\begin{equation}
a_T=A\log T,
\end{equation}
where $A>0$ will be chosen sufficiently large. For the known-initial-distribution model, and for the transition component of the free-initial-distribution model, use the reference prior
\begin{equation}
\widetilde W_T^{\rm K}
=\mathsf{Beta}(a_T,a_T)\otimes\mathsf{Beta}(a_T,a_T).
\label{eq:markov-known-reference-prior}
\end{equation}
For the stationary-initial-distribution model, set
\begin{equation}
b_T=a_T+\frac m2
\end{equation}
and use the reference density
\begin{align}
\widetilde W_T^{\rm S}(p,q)
&=\frac1{Z_T}(p+q)^m
p^{a_T-1}(1-p)^{b_T-1}\\
&\qquad\times q^{a_T-1}(1-q)^{b_T-1},
\qquad (p,q)\in(0,1)^2.
\label{eq:markov-stationary-reference-prior}
\end{align}
Then, the following lemma holds.

\begin{lemma}
\label{lem:markov-reference-prior-concentration}
For every fixed $B>0$, $A$ can be chosen so that, for $j\in\{{\rm K},{\rm S}\}$,
\begin{equation}
\widetilde W_T^j(\mathcal K_0^c)\leq C_BT^{-B}.
\label{eq:markov-reference-prior-tail}
\end{equation}
\end{lemma}

\begin{IEEEproof}
Let
\begin{gather}
F(p,q)=16p(1-p)q(1-q), \\
H(p,q)=2(p+q)\sqrt{(1-p)(1-q)}.
\end{gather}
For $j\in\{\rm K, \rm S\}$, denote by $f_T^j$ the corresponding
density divided by its value at $(1/2,1/2)$, i.e.,
\begin{gather}
f_T^K(p,q)=F(p,q)^{a_T-1},\\
f_T^S(p,q)=F(p,q)^{a_T-1}H(p,q)^m,
\end{gather}
where $F,H\le 1$ on $[0,1]^2$.

Since $(p,q)\notin \mathcal K_0$ implies that at least one
coordinate lies outside $[\delta_0,1-\delta_0]$,
\begin{equation}
F(p,q)\le \eta\triangleq
4\delta_0(1-\delta_0)<1,
\qquad (p,q)\in \mathcal K_0^c.
\end{equation}
Consequently, for all sufficiently large $T$,
\begin{equation}
\int_{\mathcal K_0^c} f_T^j(p,q)\,dp\,dq
\le \eta^{a_T-1}
=O(e^{-c a_T}),
\end{equation}
where $c=-\log\eta>0$.

It remains to lower-bound the normalizing integral.
Write $p=1/2+x$ and $q=1/2+y$. Then
\begin{equation}
F(p,q)=(1-4x^2)(1-4y^2)
\end{equation}
and
\begin{equation}
H(p,q)=(1+x+y)\sqrt{(1-2x)(1-2y)}.
\end{equation}
For $|x|,|y|\le 1/8$, the elementary inequalities
$\log(1-u)\ge-2u$ for $0\le u\le1/2$ and
$\log(1+u)\ge u-u^2$ for $|u|\le1/2$ give
\begin{equation}
\log F(p,q)\ge-8(x^2+y^2)
\end{equation}
and
\begin{equation}
\begin{aligned}
\log H&(p,q) \\
&=\log(1+x+y)
 +\frac12\log(1-2x)
 +\frac12\log(1-2y)\\
&\ge -(x+y)^2-2(x^2+y^2)\\
&\ge -4(x^2+y^2).
\end{aligned}
\end{equation}
Thus, for both $j\in\{K,S\}$, provided $a_T\ge2$,
\begin{equation}
f_T^j(1/2+x,1/2+y)
\ge \exp\{-8(a_T+m)(x^2+y^2)\}.
\end{equation}

Fix
\begin{equation}
\kappa=\min\{1/8,\,1/2-\delta_0\}>0
\end{equation}
and consider the square
\begin{equation}
\mathcal Q_T=\left\{
(1/2+x,1/2+y):
|x|,|y|\le\frac{\kappa}{\sqrt{a_T+m}}
\right\}\subseteq \mathcal K_0.
\end{equation}
For every point in $\mathcal Q_T$,
\begin{equation}
(a_T+m)(x^2+y^2)\le2\kappa^2,
\end{equation}
so $f_T^j(p,q)\ge e^{-16\kappa^2}$ throughout
$\mathcal Q_T$. Since
\begin{equation}
|\mathcal Q_T|=\frac{4\kappa^2}{a_T+m},
\end{equation}
we obtain the explicit lower bound
\begin{equation}
\begin{aligned}
\int_{(0,1)^2}f_T^j(p,q)\,dp\,dq
&\ge\int_{\mathcal Q_T}f_T^j(p,q)\,dp\,dq\\
&\ge e^{-16\kappa^2}|\mathcal Q_T|\\
&=\frac{4\kappa^2e^{-16\kappa^2}}{a_T+m}.
\end{aligned}
\end{equation}
Combining the numerator and denominator estimates,
\begin{equation}
\begin{aligned}
\widetilde W_T^j(\mathcal K_0^c)
&=\frac{\int_{\mathcal K_0^c}f_T^j(p,q)\,dp\,dq}
        {\int_{(0,1)^2}f_T^j(p,q)\,dp\,dq}\\
&\le C(a_T+m)e^{-c a_T}.
\end{aligned}
\end{equation}
Finally, $a_T=A\log T$, while $m=n+1=O(T)$
for $\ell\ge2$. Therefore,
\begin{equation}
\widetilde W_T^j(\mathcal K_0^c)
\le C_A T^{1-cA}.
\end{equation}
For any prescribed $B>0$, choosing $A$ so that
$cA\ge B+1$ establishes \cref{eq:markov-reference-prior-tail}.
\end{IEEEproof}

Let now
\begin{equation}
\mathcal F_t
=\sigma(\bm X_1,\ldots,\bm X_n,Y_1,\ldots,Y_t),
\qquad 1\leq t\leq\ell-1.
\end{equation}
For $h\in\{0,1\}$, let $N_{h,t}$ be the total number of observed departures from state $h$ in the training sequences and in the partially revealed test sequence up to time $t$, and let $S_{h,t}$ be the corresponding number of switches from $h$ to $1-h$.

Under \cref{eq:markov-known-reference-prior}, the posterior row parameters are independent and
\begin{equation}
r_h\mid\mathcal F_t
\sim
\mathsf{Beta}\left(a_T+S_{h,t},a_T+N_{h,t}-S_{h,t}\right).
\label{eq:markov-known-beta-posterior}
\end{equation}
The same identity holds for the transition posterior in the free model after conditioning on the initial states.

For the stationary model, let $K$ be the number of ones among the $n$ training initial states and $Y_1$. Their likelihood is
\begin{equation}
\left(\frac p{p+q}\right)^K
\left(\frac q{p+q}\right)^{m-K}
=\frac{p^Kq^{m-K}}{(p+q)^m}.
\label{eq:markov-stationary-initial-likelihood}
\end{equation}
Multiplication by \cref{eq:markov-stationary-reference-prior} cancels the denominator exactly. Consequently,
\begin{align}
p\mid\mathcal F_t
&\sim
\mathsf{Beta}\left(a_T+K+S_{0,t},
 b_T+N_{0,t}-S_{0,t}\right),\\
q\mid\mathcal F_t
&\sim
\mathsf{Beta}\left(a_T+m-K+S_{1,t},
 b_T+N_{1,t}-S_{1,t}\right),
\label{eq:markov-stationary-beta-posterior}
\end{align}
and the two coordinates are independent. Thus, in every case considered here, the row posterior is $\mathsf{Beta}(A_{h,t},B_{h,t})$, with
\begin{equation}
A_{h,t},B_{h,t}\geq a_T,
\qquad
A_{h,t}+B_{h,t}=N_{h,t}+C_{h,T},
\label{eq:markov-beta-shape-sum}
\end{equation}
where
\begin{equation}
2a_T\leq C_{h,T}\leq 2a_T+\frac{3m}{2}.
\label{eq:markov-beta-offset-bound}
\end{equation}
For each $h$ and $T$, the offsets $C_{h,T}$ are constant in the known and free cases and are functions only of the observed initial states in the stationary case.

\subsection{One-step beta mutual information}
For $R\sim\mathsf{Beta}(A,B)$ and $Z\mid R\sim\mathsf{Bernoulli}(R)$, write
\begin{equation}
\mathcal J(A,B)\triangleq I(R;Z).
\end{equation}

\begin{lemma}
\label{lem:markov-beta-information}
If $A,B\geq a\geq2$ and $M=A+B$, then there exists a universal constant $C > 0$ such that
\begin{equation}
\left\lvert\mathcal J(A,B) - \frac1{2M}\right\rvert \leq \frac{C}{aM}.
\label{eq:markov-beta-information-expansion}
\end{equation}
\end{lemma}

\begin{IEEEproof}
The posterior-predictive success probability is $A/M$. The beta identities used in Lemma~\ref{lem:beta-one-step-mi} give the exact expression
\begin{align}
\mathcal J(A,B)
&=h\left(\frac AM\right)
+\frac AM\psi(A+1)+\frac BM\psi(B+1)\notag\\
&\qquad-\psi(M+1),
\label{eq:markov-beta-information-exact}
\end{align}
where $h$ is binary entropy and $\psi$ is the digamma function. Using
\begin{equation}
\psi(x+1)=\log x+\frac1{2x}+O(x^{-2}),
\qquad x\geq2,
\end{equation}
we have that the logarithmic terms cancel out, the terms of order $x^{-1}$ give $1/(2M)$, while the remaining terms are
\begin{equation}
O\left(\frac1{M^2}+\frac1{MA}+\frac1{MB}\right)
=O\left(\frac1{aM}\right).
\end{equation}
\end{IEEEproof}

Conditionally on $\mathcal F_t$ and $Y_t=h$, the next transition is Bernoulli with parameter $r_h$. Hence the mutual-information chain rule gives, under either reference prior,
\begin{align}
&I_{\widetilde W_T}((P,Q);Y_2^\ell\mid\bm X^n,Y_1)\notag\\
&\quad=
\sum_{t=1}^{\ell-1}\sum_{h=0}^1
\mathbb{E}_{\widetilde W_T}\left[
\mathbf1\{Y_t=h\}\mathcal J(A_{h,t},B_{h,t})
\right].
\label{eq:markov-conjugate-chain-rule}
\end{align}

\subsection{Reciprocal counts estimate}
For a fixed parameter $\theta=(\nu,p,q)$, define
\begin{equation}
a_h(\theta)=\sum_{s=1}^{\ell-1}\mathbb{P}_\theta(Y_s=h),
\qquad
b_{h,t}(\theta)=\sum_{s=1}^{t-1}\mathbb{P}_\theta(Y_s=h).
\end{equation}
Then
\begin{equation}
b_{h,1}=0,
\qquad
b_{h,\ell}=a_h,
\qquad
b_{h,t+1}-b_{h,t}=\mathbb{P}_\theta(Y_t=h),
\label{eq:markov-conjugate-b-increments}
\end{equation}
and
\begin{equation}
m_{h,t}(\theta)\triangleq\mathbb{E}_\theta N_{h,t}=na_h(\theta)+b_{h,t}(\theta).
\label{eq:markov-conjugate-prefix-mean}
\end{equation}
Furthermore, we have that $c_\delta T\leq m_{h,t}(\theta)\leq C_\delta T$ for $(p,q)\in\mathcal K_0$ and all sufficiently large $\ell$.
Lemma~\ref{lem:batch-markov-occupancy}, applied to the training batch, and the same Paulin bound applied to the partial test sequence give
\begin{equation}
\mathbb{P}_\theta\{|N_{h,t}-m_{h,t}(\theta)|>u\}
\leq C_\delta\exp\left(-c_\delta\frac{u^2}{T}\right),
\label{eq:markov-conjugate-prefix-tail}
\end{equation}
and therefore
\begin{equation}
\mathbb{E}_\theta|N_{h,t}-m_{h,t}(\theta)|\leq C_\delta\sqrt T,
\label{eq:markov-conjugate-prefix-first-moment}
\end{equation}
uniformly over $(p,q)\in\mathcal K_0$, $h$, and $t$.

\begin{lemma}
\label{lem:markov-shifted-reciprocal-sum}
Let $C_{h,T}$ be any possibly random offset satisfying
\begin{equation}
2a_T\leq C_{h,T}\leq C(n+\log T).
\end{equation}
Then, uniformly for $(p,q)\in\mathcal K_0$ and $h\in\{0,1\}$,
\begin{align}
&\sum_{t=1}^{\ell-1}
\mathbb{E}_\theta\left[
\frac{\mathbf1\{Y_t=h\}}{N_{h,t}+C_{h,T}}
\right]\notag\\
&\quad=
\log\left(1+\frac1n\right)
+O_\delta\left(
\frac1{n^{3/2}\sqrt\ell}
+\frac1{n\ell}
+\frac{\log T}{n^2\ell}
\right).
\label{eq:markov-shifted-reciprocal-sum}
\end{align}
\end{lemma}

\begin{IEEEproof}
Let $p_{h,t}=\mathbb{P}_\theta(Y_t=h)$ and $m_{h,t}=na_h+b_{h,t}$. On the event $|N_{h,t}-m_{h,t}|\leq m_{h,t}/2$,
\begin{equation}
\left|\frac1{N_{h,t}+C_{h,T}}-
\frac1{m_{h,t}+C_{h,T}}\right|
\leq\frac{2|N_{h,t}-m_{h,t}|}{m_{h,t}^2}.
\end{equation}
Using \cref{eq:markov-conjugate-prefix-tail,eq:markov-conjugate-prefix-first-moment} and $C_{h,T}\geq2a_T$, the complementary event contributes exponentially little, and hence
\begin{align}
&\left|\mathbb{E}_\theta\left[
\frac{\mathbf1\{Y_t=h\}}{N_{h,t}+C_{h,T}}
\right]
-\mathbb{E}_\theta\left[
\frac{\mathbf1\{Y_t=h\}}{m_{h,t}+C_{h,T}}
\right]\right|\notag\\
&\qquad\leq CT^{-3/2}+Ce^{-cT}/a_T.
\end{align}
Since $m_{h,t}$ is deterministic for fixed $\theta$ and $C_{h,T}\leq C(n+\log T)$, we also have
\begin{equation}
\left|\mathbb{E}_\theta\left[
\frac{\mathbf1\{Y_t=h\}}{m_{h,t}+C_{h,T}}
\right]-\frac{p_{h,t}}{m_{h,t}}\right| \leq\frac{C(n+\log T)}{T^2}.
\end{equation}
Finally, \cref{eq:markov-conjugate-b-increments} gives
\begin{equation}
\sum_{t=1}^{\ell-1}\frac{p_{h,t}}{na_h+b_{h,t}}
=
\sum_{t=1}^{\ell-1}
\frac{b_{h,t+1}-b_{h,t}}{na_h+b_{h,t}}.
\end{equation}
Since $x\mapsto(na_h+x)^{-1}$ is decreasing and the increments lie in $[0,1]$, we have
\begin{equation}
0 \leq
\sum_{t=1}^{\ell-1}
\frac{b_{h,t+1}-b_{h,t}}{na_h+b_{h,t}}
-\int_0^{a_h}\frac{dx}{na_h+x} \leq\frac1{2n^2a_h}.
\end{equation}
The integral equals $\log(1+1/n)$ and $a_h\geq\delta(\ell-2)$. Summing the preceding errors over $t$ proves the lemma.
\end{IEEEproof}

By \cref{eq:markov-beta-shape-sum,eq:markov-beta-offset-bound,eq:markov-beta-information-expansion} and Lemma~\ref{lem:markov-shifted-reciprocal-sum}, for every fixed $(p,q)\in\mathcal K_0$ each row contributes
\begin{equation}
\frac12\log\left(1+\frac1n\right)+o(1/n)
\label{eq:markov-conjugate-row-contribution}
\end{equation}
uniformly. Indeed, the accumulated remainder from \cref{eq:markov-beta-information-expansion} is $O(1/(n a_T))=o(1/n)$.

\subsection{Conditioning onto the Markov class}
We finish by passing from the reference priors to priors supported in $\mathcal K_0$. The following elementary observation is useful.

\begin{lemma}
\label{lem:conditioning-reference-prior}
Let $\widetilde W$ be a prior distribution, let $\mathcal C$ be a parameter event with $\widetilde W(\mathcal C^c)=\varepsilon$, and let $W=\widetilde W(\cdot\mid\mathcal C)$. If $Y$ takes at most $M$ values, then
\begin{equation}
\left|(1-\varepsilon)I_W(\Theta;Y\mid X)
-I_{\widetilde W}(\Theta;Y\mid X)\right|
\leq h_2(\varepsilon)+\varepsilon\log M.
\label{eq:conditioning-reference-prior-bound}
\end{equation}
\end{lemma}

\begin{IEEEproof}
Let $C=\mathbf1\{\Theta\in\mathcal C\}$. Since $C$ is a function of $\Theta$, the chain rule gives
\begin{align}
I_{\widetilde W}(\Theta;Y\mid X)
&=I(C;Y\mid X)+(1-\varepsilon)I_W(\Theta;Y\mid X)\notag\\
&\quad+\varepsilon I_{\widetilde W(\cdot\mid\mathcal C^c)}(\Theta;Y\mid X).
\end{align}
Now $I(C;Y\mid X)\leq h_2(\varepsilon)$ and the last conditional mutual information is at most $\log M$.
\end{IEEEproof}

For $j\in\{{\rm K},{\rm S}\}$, let $\varepsilon_T^j=\widetilde W_T^j(\mathcal K_0^c)$. Under the reference joint law, the expression in \cref{eq:markov-conjugate-chain-rule} is nonnegative and is at most $(\ell-1)\log2$. Therefore \cref{eq:markov-reference-prior-tail,eq:markov-conjugate-row-contribution} imply
\begin{equation}
I_{\widetilde W_T^j}((P,Q);Y_2^\ell\mid\bm X^n,Y_1)
=
\log\left(1+\frac1n\right)+o(1/n).
\end{equation}
Lemma~\ref{lem:markov-reference-prior-concentration} with $B>1$ and Lemma~\ref{lem:conditioning-reference-prior} with $M=2^{\ell-1}$ show that the conditioned priors
\begin{equation}
W_T^j=\widetilde W_T^j(\cdot\mid\mathcal K_0)
\end{equation}
satisfy the same expansion above, because
\begin{equation}
h_2(\varepsilon_T^j)+\varepsilon_T^j(\ell-1)\log2=o(1/n).
\end{equation}
Furthermore, they are supported in the fixed compact subrectangle $\mathcal K_0\subset\Theta_\delta$. This proves the known and stationary cases of Lemma~\ref{lem:markov-transition-information}. In the free model, combine $W_T^{\rm K}$ with an independent prior on the free initial probability; after conditioning on the initial states, the transition posterior is exactly the known-initial posterior, so the same expansion above applies.

\section*{Acknowledgments}
This research was supported by the Swiss National Science Foundation Grant No. 10004381.

\bibliographystyle{IEEEtran}
\bibliography{IEEEabrv,References}

@STRING(IT="{IEEE} Transactions on Information Theory")

@STRING(ISIT="{IEEE} International Symposium on Information Theory")

@STRING(ITW="{IEEE} Information Theory Workshop")

@STRING(ITA="{IEEE} Information Theory and Applications Workshop")

@inproceedings{renyi1,
  author = {R\'enyi, Alfr\'ed},
  title = {On measures of entropy and information},
  booktitle = {Proc. 4th {B}erkeley {S}ympos. {M}ath. {S}tatist. and {P}rob.,
              {V}ol. {I}},
  year = {1960},
}

@book{coveruniversal,
  title={{Universal gambling schemes and the complexity measures
of Kolmogorov and Chaitin}},
  author={Cover, Thomas M.},
  year={1974},
  publisher={Tech. Rep. 12, Dept. Statist., Stanford
Univ., Stanford, CA}
}

@article{blahut1,
  author={Blahut, R.},
  journal={IEEE Transactions on Information Theory}, 
  title={Computation of channel capacity and rate-distortion functions}, 
  year={1972},
  volume={18},
  number={4},
  pages={460-473},
}

@article{arimoto1,
  author={Arimoto, S.},
  journal={IEEE Transactions on Information Theory}, 
  title={An algorithm for computing the capacity of arbitrary discrete memoryless channels}, 
  year={1972},
  volume={18},
  number={1},
  pages={14-20},
  }

@article{braess2004187,
title = {Bernstein polynomials and learning theory},
journal = {Journal of Approximation Theory},
volume = {128},
number = {2},
pages = {187-206},
year = {2004},
author = {Dietrich Braess and Thomas Sauer},
}

@article{komaki1,
title = {Asymptotically minimax Bayesian predictive densities for multinomial models},
journal = {Electron. J. Statist.},
volume = {6},
year = {2012},
author = {Fumiyasu Komaki},
}

@article{komaki20113705,
title = {Bayesian predictive densities based on latent information priors},
journal = {Journal of Statistical Planning and Inference},
volume = {141},
number = {12},
pages = {3705-3715},
year = {2011},
author = {Fumiyasu Komaki},
}

@inproceedings{fogel2,
  author={Fogel, Yaniv and Feder, Meir},
  booktitle={2018 IEEE International Symposium on Information Theory}, 
  title={Universal Batch Learning with Log-Loss}, 
  year={2018},
}

@inproceedings{fogel3,
  author={Fogel, Yaniv and Feder, Meir},
  booktitle={2024 IEEE International Symposium on Information Theory Workshops (ISIT-W)},
  title={Combining Batch and Online Prediction},
  year={2024},
  organization={IEEE},
  doi={10.1109/ISIT-W61686.2024.10591756}
}

@article{ryabko1,
  title={Twice-universal coding},
  author={Ryabko, Boris Y.},
  journal={Problems of information transmission},
  volume={20},
  number={3},
  pages={173--177},
  year={1984}
}

@article{ryabko2,
  title={Prediction of random sequences and universal coding},
  author={Ryabko, Boris Y.},
  journal={Problems of information transmission},
  volume={24},
  number={2},
  pages={87--96},
  year={1988}
}

@inproceedings{falahatgar1,
  title={Learning {M}arkov distributions: Does estimation trump compression?},
  author={Falahatgar, Moein and Orlitsky, Alon and Pichapati, Venkatadheeraj and Suresh, Ananda Theertha},
  booktitle={2016 IEEE International Symposium on Information Theory (ISIT)},
  year={2016},
}

@article{liang1,
  title={Exact minimax strategies for predictive density estimation, data compression, and model selection},
  author={Liang, Feng and Barron, Andrew},
  journal={IEEE Transactions on Information Theory},
  volume={50},
  number={11},
  pages={2708--2726},
  year={2004},
  publisher={IEEE}
}

@article{xie1,
  author={Qun Xie and Barron, A.R.},
  journal=IT, 
  title={Asymptotic minimax regret for data compression, gambling, and prediction}, 
  year={2000},
  volume={46},
  number={2},
  pages={431-445},
  doi={10.1109/18.825803}
}

@article{shtarkov1,
  author={Y. M. Shtarkov},
  journal="Problems Inform. Transmission", 
  title={Universal sequential coding of single messages}, 
  year={1987},
  volume={23},
  number={3},
  pages={175-186}
}

@article{krichevsky1,
  author={Krichevsky, R. and Trofimov, V.},
  journal=IT, 
  title={The performance of universal encoding}, 
  year={1981},
  volume={27},
  number={2},
  pages={199-207},
  doi={10.1109/TIT.1981.1056331}
}

@article{merhav1,
  author={Merhav, N. and Feder, M.},
  journal=IT, 
  title={Universal prediction}, 
  year={1998},
  volume={44},
  number={6},
  pages={2124-2147},
  doi={10.1109/18.720534}
}

@article{xie2,
  author={Qun Xie and Barron, A.R.},
  journal=IT, 
  title={Minimax redundancy for the class of memoryless sources}, 
  year={1997},
  volume={43},
  number={2},
  pages={646-657},
  doi={10.1109/18.556120}
}

@article{krichevsky2,
  author={Krichevsky, R.E.},
  journal=IT, 
  title={Laplace's law of succession and universal encoding}, 
  year={1998},
  volume={44},
  number={1},
  pages={296-303},
  doi={10.1109/18.651051}
}

@article{hao1,
  title={On Learning Markov Chains},
  author={Hao, Yi and Orlitsky, Alon and Pichapati, Venkatadheeraj},
  journal={Advances in Neural Information Processing Systems},
  volume={31},
  year={2018},
}

@article{paulin2015,
  author={Paulin, Daniel},
  title={Concentration inequalities for Markov chains by {M}arton couplings and spectral methods},
  journal={Electronic Journal of Probability},
  volume={20},
  number={79},
  pages={1--32},
  year={2015},
  doi={10.1214/EJP.v20-4039},
  note={See also arXiv:1212.2015v5 for the corrected version}
}

@inproceedings{bietti2023birth,
  title={Birth of a Transformer: A Memory Viewpoint},
  author={Alberto Bietti and Vivien Cabannes and Diane Bouchacourt and Herve Jegou and Leon Bottou},
  booktitle={Thirty-seventh Conference on Neural Information Processing Systems},
  year={2023}
}

@inproceedings{makkuva2024local,
  title={Local to Global: Learning Dynamics and Effect of Initialization for Transformers},
  author={Ashok Vardhan Makkuva and Marco Bondaschi and Adway Girish and Alliot Nagle and Hyeji Kim and Michael Gastpar and Chanakya Ekbote},
  booktitle={The Thirty-eighth Annual Conference on Neural Information Processing Systems},
  year={2024},
}

@inproceedings{edelman2024evolution,
  title={The Evolution of Statistical Induction Heads: In-Context Learning Markov Chains},
  author={Ezra Edelman and Nikolaos Tsilivis and Benjamin L. Edelman and E. Malach and Surbhi Goel},
  booktitle={The Thirty-eighth Annual Conference on Neural Information Processing Systems},
  year={2024},
}

@inproceedings{makkuva2024attention,
  title={Attention with Markov: A Curious Case of Single-layer Transformers},
  author={Ashok Vardhan Makkuva and Marco Bondaschi and Adway Girish and Alliot Nagle and Martin Jaggi and Hyeji Kim and Michael Gastpar},
  booktitle={The Thirteenth International Conference on Learning Representations},
  year={2025},
}

@inproceedings{rajaraman2024transformersmarkovdataconstant,
  title={Transformers on Markov data: Constant depth suffices},
  author={Nived Rajaraman and Marco Bondaschi and Ashok Vardhan Makkuva and Kannan Ramchandran and Michael Gastpar},
  booktitle={The Thirty-eighth Annual Conference on Neural Information Processing Systems},
  year={2024},
}

@article{cai1,
  author={Cai, Changxiao and Verdú, Sergio},
  title={Conditional {R}ényi divergence saddlepoint and the maximization of $\alpha$-mutual information},
  journal={Entropy},
  volume={21},
  year={2019},
  number={10},
  doi={10.3390/e21100969}
}

@article{yagli1,
  author={Yagli, Semih and Altuğ, Yücel and Verdú, Sergio},
  journal=IT, 
  title={Minimax {R}ényi redundancy}, 
  year={2018},
  volume={64},
  number={5},
  pages={3715-3733},
  doi={10.1109/TIT.2018.2803070}
}

@article{verdu1,
  author={Verdú, Sergio},
  journal=ITA,
  title={$\alpha$-mutual information}, 
  year={2015},
  doi={10.1109/ITA.2015.7308959}
}

@article{esposito2024sibsonsalphamutualinformationvariational,
  author={Esposito, Amedeo Roberto and Gastpar, Michael and Issa, Ibrahim},
  journal={IEEE Transactions on Information Theory}, 
  title={Sibson $\alpha$-Mutual Information and Its Variational Representations}, 
  year={2026},
  volume={72},
  number={7},
  pages={4434-4467},
  doi={10.1109/TIT.2025.3587340}
}

@article{tomamichel1,
  author={Tomamichel, Marco and Hayashi, Masahito},
  journal=IT,
  title={Operational Interpretation of {R}ényi Information Measures via Composite Hypothesis Testing Against Product and Markov Distributions}, 
  year={2018},
  volume={64},
  number={2},
  pages={1064-1082},
  doi={10.1109/TIT.2017.2776900}
}

@article{esposito1,
  author={Esposito, Amedeo Roberto and Wu, Diyuan and Gastpar, Michael},
  journal=ISIT, 
  title={On conditional {S}ibson's $\alpha$-Mutual Information}, 
  year={2021},
  doi={10.1109/ISIT45174.2021.9517944}
}

@article{bondaschi1,
  author={Bondaschi, Marco and Gastpar, Michael},
  journal=IT,
  title={Alpha-{NML} Universal Predictors}, 
  year={2025},
  volume={71},
  number={2},
  pages={1171-1183},
  doi={10.1109/TIT.2024.3521221}
}

@inproceedings{bondaschi2024batch,
  title={Batch universal prediction},
  author={Bondaschi, Marco and Gastpar, Michael},
  booktitle={2024 IEEE International Symposium on Information Theory (ISIT)},
  pages={3552--3557},
  year={2024},
  organization={IEEE}
}

@inproceedings{bondaschi2025markov,
  title={From Markov to Laplace: How Mamba In-Context Learns Markov Chains},
  author={Marco Bondaschi and Ashok Vardhan Makkuva and Nived Rajaraman and Xiuying Wei and Razvan Pascanu and Caglar Gulcehre and Michael Gastpar},
  booktitle={The Fourteenth International Conference on Learning Representations},
  year={2026},
}

@inproceedings{bondaschi2025conditional,
  title={The Conditional Regret-Capacity Theorem for Batch Universal Prediction},
  author={Bondaschi, Marco and Gastpar, Michael},
  booktitle={2025 IEEE Information Theory Workshop (ITW)},
  pages={746--751},
  year={2025},
  organization={IEEE}
}

@book{laplace1814essaiprob,
  title={Essai philosophique sur les probabilit\'es},
  author={Laplace, Pierre Simon},
  year={1814},
  publisher={Courcier, Paris, France}
}

@article{Sibson:69,
  title={Information radius},
  author={Sibson, R.},
  year={1969},
  journal = {Z. Wahrscheinlichkeitstheorie verw. Gebiete},
  number = {14},
  pages={149-160}
}

@book{boyd2004convex,
  author={Stephen Boyd and Lieven Vandenberghe},
  title={Convex Optimization},
  publisher={Cambridge University Press},
  year={2004}
}

@article{witsenhausen1980convexity,
  author={Hans S. Witsenhausen},
  title={Some aspects of convexity useful in information theory},
  journal={IEEE Transactions on Information Theory},
  volume={26},
  number={3},
  pages={265--271},
  year={1980}
}

@inproceedings{barletta2024binomial,
  author={Luca Barletta and Ian Zieder and Antonino Favano and Alex Dytso},
  title={Binomial Channel: On the Capacity-Achieving Distribution and Bounds on the Capacity},
  booktitle={2024 IEEE International Symposium on Information Theory (ISIT)},
  pages={711--716},
  year={2024},
  doi={10.1109/ISIT57864.2024.10619601}
}

@article{haussler1997minimax,
  author={David Haussler},
  title={A General Minimax Result for Relative Entropy},
  journal={IEEE Transactions on Information Theory},
  volume={43},
  number={4},
  pages={1276--1280},
  month={July},
  year={1997},
  doi={10.1109/18.605594}
}

@article{bhaskar-song-descartes,
  author  = {Bhaskar, Anand and Song, Yun S.},
  title   = {Descartes' Rule of Signs and the Identifiability of Population Demographic Models from Genomic Variation Data},
  journal = {The Annals of Statistics},
  volume  = {42},
  number  = {6},
  pages   = {2469--2493},
  year    = {2014},
  doi     = {10.1214/14-AOS1264}
}

@book{levin-peres,
  author    = {David A. Levin and Yuval Peres},
  title     = {Markov Chains and Mixing Times},
  edition   = {2},
  publisher = {American Mathematical Society},
  address   = {Providence, RI},
  year      = {2017},
  doi       = {10.1090/mbk/107}
}

\end{document}